\documentclass[runningheads,a4paper,10pt]{llncs}
\usepackage{adjustbox}
\usepackage{algorithm}
\usepackage[noend]{algpseudocode}
\usepackage{amsmath}
\usepackage{array}
\usepackage{booktabs}
\usepackage{breqn}
\usepackage{caption}
\usepackage{enumitem}
\usepackage{graphicx}
\usepackage[hidelinks,colorlinks=true,allcolors=blue]{hyperref}
\usepackage[utf8]{inputenc}
\usepackage{siunitx}
\usepackage{subcaption}

\newcommand\secpar[1]{\vspace{0.4em}\noindent \textbf{#1.}}
\newcolumntype{H}{>{\setbox0=\hbox\bgroup}c<{\egroup}@{}}

\DeclareMathOperator*{\argmax}{arg\,max}

\graphicspath{{./figures/}}

\newcommand{\tsf}{\tilde{s}_f}
\newcommand{\tv}{\tilde{v}(t)}
\newcommand{\ec}{\epsilon_c}

\newcommand{\eh}[1][\epsilon]{\frac{#1+1}{2}}
\newcommand{\meh}[1][\epsilon]{\frac{2-#1}{2}}
\newcommand{\nin}[1][\epsilon]{N^{-1}\left(\eh[#1]\right)}
\newcommand{\ds}{\Delta\Sigma}
\newcommand{\vma}{V_{PK}^{Adv}}
\newcommand{\vrms}{V_{RMS}}
\newcommand{\vra}{\vrms^{Adv}}

\newcommand{\jw}{j\omega}

\begin{document}

\title{A Framework for Evaluating Security in the Presence of Signal Injection Attacks}
\titlerunning{A Security Framework for Signal Injection Attacks}

\author{Ilias Giechaskiel \and Youqian Zhang \and Kasper B. Rasmussen}
\institute{University of Oxford, Oxford, UK
\email{\{ilias.giechaskiel,youqian.zhang,kasper.rasmussen\}@cs.ox.ac.uk}
}

\maketitle

\begin{abstract}
Sensors are embedded in security-critical applications from medical
devices to nuclear power plants, but their outputs
can be spoofed through electromagnetic and other types of signals
transmitted by attackers at a distance.
To address the lack of a unifying framework for evaluating the effects of
such transmissions, we introduce a system and threat model for signal
injection attacks.
We further define the concepts of existential, selective, and universal security,
which address attacker goals from mere disruptions of the sensor readings to precise
waveform injections. Moreover, we
introduce an algorithm which allows circuit designers
to concretely calculate the security level of real systems. Finally, we apply
our definitions and algorithm in practice using measurements
of injections against a smartphone microphone, and analyze the demodulation
characteristics of commercial Analog-to-Digital Converters (ADCs).
Overall, our work highlights the importance of
evaluating the susceptibility of systems against signal injection attacks,
and introduces both the terminology and the methodology to do so.\footnote{
  This article is the extended technical report version of the paper presented at
  ESORICS 2019. DOI: \href{https://doi.org/10.1007/978-3-030-29959-0_25}{10.1007/978-3-030-29959-0\_25}.
}
\end{abstract}

\section{Introduction}
In our daily routine we interact with dozens of sensors: from motion detection
in home security systems and tire pressure monitors in cars, to
accelerometers in smartphones and heart rate monitors in smartwatches.
The integrity of these sensor outputs is crucial, as
many security-critical decisions are taken in response
to the sensor values. However,
specially-crafted adversarial signals can be used to remotely
induce waveforms into the outputs of sensors, thereby attacking
pacemakers~\cite{ghost}, temperature sensors~\cite{trusting_sensors},
smartphone microphones~\cite{smartphone_iemi}, and
car-braking mechanisms~\cite{abs}. These attacks cause a system to
report values which do not match the true sensor measurements, and trick it
into performing dangerous actions such as raising false alarms, or even
delivering defibrillation shocks.

The root cause of these vulnerabilities lies in the unintentional side-effects
of the physical components of a system. For example,
the wires connecting sensors to microcontrollers behave like low-power, low-gain
antennas, and can thus pick up high-frequency electromagnetic radiations.
Although these radiations are considered ``noise'' from an electrical point of view,
hardware imperfections in the subsequent parts of the circuit can transform
attacker injections into meaningful waveforms. Specifically, these
radiations are digitized along with the true sensor outputs, which represent
a physical property as an analog electrical quantity. This digitization process
is conducted by Analog-to-Digital Converters (ADCs), which, when used outside
of their intended range, can cause high-frequency signals to be
interpreted as meaningful low-frequency signals.

Despite the potential that signal injection attacks have to break security
guarantees, there is no unifying framework for evaluating
the effect of such adversarial transmissions. Our work fills this gap
through the following contributions:

\begin{enumerate}
  \item We propose a system model which abstracts away from
  engineering concerns associated with remote transmissions,
  such as antenna design (Section~\ref{sec:model}).
  \item We define security against adversarial signal injection attacks.
  Our definitions address effects ranging from mere disruptions of
  the sensor readings, to precise waveform injections of attacker-chosen values
  (Section~\ref{sec:definitions}).
  \item We introduce an algorithm to calculate the security
  level of a system under our definitions and demonstrate it in practice
  by injecting ``OK Google'' commands into a smartphone (Section~\ref{sec:case_study}).
  \item We investigate how vulnerable commercial ADCs are to malicious signal
        injection attacks by testing their demodulation properties (Section~\ref{sec:adcs}).
  \item We discuss how our model can be used to inform circuit design choices,
  and how to interpret defense mechanisms and other types
  of signal injection attacks in its context (Section~\ref{sec:discussion}).
\end{enumerate}

Overall, our work highlights the importance of testing systems against
signal injection attacks, and proposes a methodology to test the security of real devices.

\section{System and Adversary Model}
\label{sec:model}

Remote signal injection attacks pose new challenges
from a threat-modeling perspective, since the electrical properties of systems
suggest that adversaries cannot arbitrarily and precisely
change any sensor reading. To create a threat
model and define security in its context, we need to first abstract away from
specific circuit designs and engineering concerns related to remote
transmissions. To do so, we separate the behavior of a system
into two different transfer functions. The first function describes
circuit-specific behavior, including how adversarial signals
enter the circuit (e.g., through PCB wires acting as antennas), while the second one
is ADC-specific, and dictates how the signals
which have made it into the circuit are digitized.
We describe this model in greater detail in Section~\ref{sec:adc_model}, taking
a necessary detour into electrical engineering to show why our proposal
makes for a good system model. We then explain
some sources of measurement errors even in the absence of an
adversary in Section~\ref{sec:samp_errors} and finish by detailing the
capabilities and limitations of the adversary in Section~\ref{sec:adv_model}.
Both sub-sections are crucial in motivating the security definitions of
Section~\ref{sec:definitions}.

\subsection{Circuit Model}
\label{sec:adc_model}

\begin{figure*}[t]
  \centering
  \includegraphics[width=\textwidth]{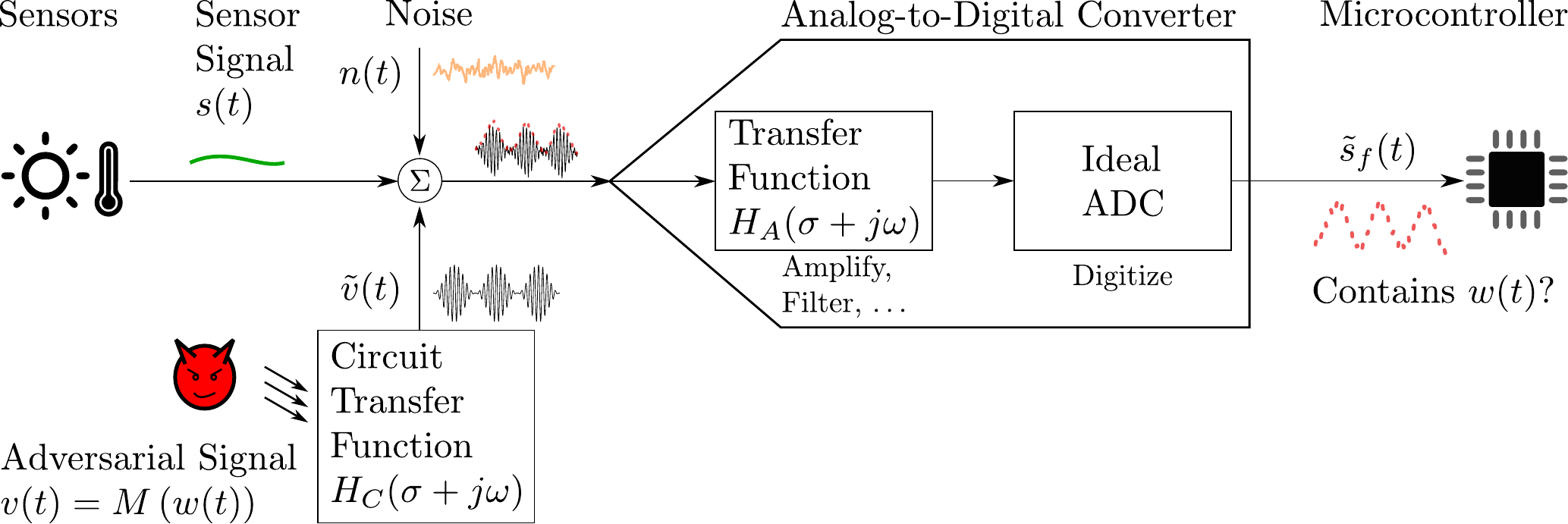}
  \caption{System model: an adversarial signal $v(t)$ enters the circuit
           and is transformed via the transfer function $H_C$.
           It is digitized along with the sensor signal
           $s(t)$ and the noise $n(t)$ through an ADC-specific transfer function $H_A$.
           In successful attacks, the digitized signal will contain the demodulated
           version $w(t)$ of the attacker signal $v(t)=M(w(t))$, where $M$
           is the modulation function (e.g., amplitude modulation over
           a high-frequency carrier).}
  \label{fig:adc_model}
  \vspace{-2em}
\end{figure*}

Analog-to-Digital Converters (ADCs) are central in the
digitization process of converting signals from the analog to the digital realm,
and our circuit block diagram (Figure~\ref{fig:adc_model}) reflects that.
In the absence of an adversary, the ADC digitizes the sensor signal $s(t)$
as well as the environmental noise $n(t)$, and transfers the digital bits
to a microcontroller. We model the ADC in two parts: an ``ideal'' ADC which simply
digitizes the signal, and a transfer function $H_A$. This transfer function
describes the internal behavior of the ADC, which includes effects such as
filtering and amplification.
The digitized version of the signal $\tsf(t)$ depends both on this
transfer function, and the sampling frequency $f$ of the ADC.
An adversarial signal can enter the system (e.g., through the wires
connecting the sensor to the ADC) and add to the sensor signal and the noise.
This process can be described by a second, circuit-specific transfer function $H_C$,
which transforms the adversarial signal $v(t)$ into $\tilde{v}(t)$.
Note that components such as external filters and amplifiers in the signal path between
the point of injection and the ADC can be included in either $H_A$ or $H_C$.
We include them in $H_A$ when they also affect the sensor signal $s(t)$,
but in $H_C$ when they are specific to the coupling effect.
$H_C$ and $H_A$ are discussed in detail below.

\secpar{Circuit Transfer Function $H_C$}
To capture the response of the circuit to external signal injections,
we introduce a transfer function $H_C$. This transfer function explains
why the adversarial waveforms must be modulated, and why it is helpful
to try and reduce the number of remote experiments to perform.
For electromagnetic interference (EMI) attacks,
the wires connecting the sensor to the ADC pick up signals by
acting as (unintentional) low-power and low-gain antennas,
which are resonant at specific
frequencies related to the inverse of the wire length~\cite{pcb_emi}.
Non-resonant frequencies are attenuated more, so for a successful attack
the adversary must transmit signals at frequencies with relatively low attenuation.
For short wires, these frequencies are in the $\si{\giga\hertz}$ range~\cite{pcb_emi},
so the low-frequency waveform $w(t)$ that the adversary wants
to inject into the output of the ADC $\tsf(t)$ may need to be modulated
over a high-frequency carrier using a function $M$. We denote this
modulated version of the signal by $v(t)=M(w(t))$.

$H_C$ is also affected by passive and active components on the path to the ADC,
and can also be influenced by inductive and capacitive coupling for small
transmission distances, as it closely depends on the circuit components and
their placement. Specifically, it is possible for 2 circuits with
``the same components, circuit topology and placement area''
to have different EMI behavior depending on the component placement
on the board~\cite{emi_predict}.
Despite the fact that it is hard to mathematically
model and predict the behavior of circuits in response to different
signal transmissions, $H_C$ can still be determined
empirically using frequency sweeps. It presents a useful abstraction,
allowing us to separate the behavior of the ADC (which need only be determined
once, for instance by the manufacturer) from circuit layout and
transmission details.

Note, finally, that $H_C$
can also account for distance factors between the adversary
and the circuit under test: due to the Friis transmission formula~\cite{friis},
as distance doubles, EMI transmission power needs to
quadruple. This effect can be captured by increasing
the attenuation of $H_C$ by $\SI{6}{\decibel}$, while defense mechanisms such
as shielding can be addressed similarly. This approach allows us to
side-step engineering issues of
remote transmissions and reduce the number
of parameters used in the security definitions we
propose in Section~\ref{sec:definitions}.

\begin{figure}[t]
  \centering
  \includegraphics[width=0.45\linewidth]{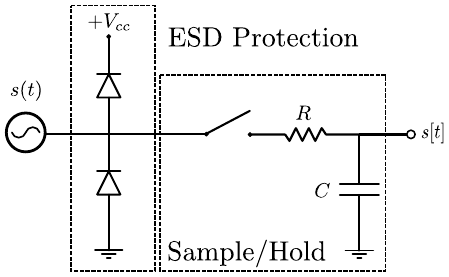}
  \caption{The sample-and-hold mechanism of an ADC is an $RC$ low-pass filter.
           Electrostatic Discharge (ESD) protection diodes can also introduce
           non-linearities.}
  \label{fig:sah_model}
  \vspace{-2em}
\end{figure}

\secpar{ADC Transfer Function $H_A$}
Every system with sensors contains one or more ADCs, which may even be integrated
into the sensor chip itself. ADCs are not perfect, but contain components
which may cause a mismatch between the ``true'' value at the ADC
input and the digitized output. In this section, we describe how these components
affect the digitization process.

Although there are many types of ADCs, every ADC contains
three basic components: a ``sample- or track-and-hold circuit where the
sampling takes place, the digital-to-analog
converter and a level-comparison mechanism''~\cite{adc_book}. The
sample-and-hold component acts as a low-pass filter, and makes it harder
for an adversary to inject signals modulated at high frequencies.
However, the level-comparison mechanism is essentially an amplifier with
non-linearities which induces DC offsets, and
allows low-frequency intermodulation products to pass through.
These ADC-specific transformations, modeled through $H_A$,
unintentionally demodulate high-frequency signals which are not attenuated by $H_C$.
They are explored in more detail in Section~\ref{sec:adcs} and Appendix~\ref{app:exp}.

\secpar{Sample-And-Hold Filter Characteristics}
A sample-and-hold (S/H) mechanism is an
$RC$ circuit connected to the analog input, with the resistor and
the capacitor connected in series (Figure~\ref{fig:sah_model}). The transfer
function of the voltage across the capacitor is
$H_{S/H}(\jw)=\frac{1}{1+\jw RC}$, and the magnitude of the gain
is $G_{S/H}=\frac{1}{\sqrt{1+\left(\omega R C\right)^2}}$. As the
angular frequency $\omega=2\pi f$ increases, the gain is reduced: the
S/H mechanism acts as a low-pass filter. The $\SI{-3}{\deci\bel}$ cutoff frequency
is thus $f_{cut}=\frac{1}{2\pi RC}$, which is often higher than the
ADC sampling rate (Section~\ref{sec:adcs}).
Hence, ``aliasing'' occurs when signals beyond the Nyquist frequency are
digitized by the ADC: high-frequency signals become indistinguishable
from low-frequency signals which the ADC can sample accurately.

\secpar{Amplifier Non-Linearities}
Every ADC contains amplifiers: a comparator, and possibly buffer and
differential amplifiers. Many circuits also contain additional external amplifiers
to make weak signals measurable. All these amplifiers have
harmonic and intermodulation non-linear distortions~\cite{emc_analog_book},
which an adversary can exploit.
Harmonics are produced when an amplifier transforms an input $v_{in}$ to an output $
v_{out}=\sum_{n=1}^\infty a_nv_{in}^n$. In particular, if $v_{in}=\hat{v}\cdot \sin(\omega t)$,
then:
\[
  v_{out} = \left(\frac{a_2\hat{v}^2}{2}+\frac{3a_4\hat{v}^4}{8}+\cdots\right)
          +\left(a_1\hat{v}+\cdots\right)\sin(\omega t)  
          - \left(\frac{a_2\hat{v}^2}{2}+\cdots\right)\cos(2\omega t) + \cdots
\]
This equation shows that ``the frequency
spectrum of the output contains a spectral component at the original (fundamental)
frequency, [and] at multiples of the fundamental
frequency (harmonic frequencies)''~\cite{emc_analog_book}.
Moreover, the output includes a DC component,
which depends only on the even-order non-linearities of the system.
Besides harmonics, intermodulation products arise when the input signal
is a sum of two sinusoids (for instance when the injected signal sums with
the sensor signal): $v_{in}=\hat{v}_1\cdot \sin(\omega_1 t)+\hat{v}_2\cdot \sin(\omega_2 t)$.
In that case, the output signal contains frequencies of the form
$n\omega_1 \pm m\omega_2$ for integers $n,m\neq 0$.
These non-linearities demodulate attacker waveforms,
even when they are modulated on high-frequency carriers.

\secpar{Diode Rectification} Figure~\ref{fig:sah_model} shows that the input
to an ADC can contain reverse-biased diodes to ground and $V_{cc}$ to protect
the input from Electrostatic Discharge (ESD). When
the input to the ADC is negative, or when it exceeds $V_{cc}$, the diodes
clamp it, causing non-linear behavior. When the sensor signal $s(t)$
is positive, this behavior is also asymmetric, causing a DC shift~\cite{emc_analog_book},
which compounds with the amplifier non-linearities.

\secpar{Conclusion} All ADCs contain the same basic building blocks,
modeled through $H_A$. Although the sample-and-hold mechanism should
attenuate high-frequency signals beyond the maximum sampling rate of the ADC,
non-linearities due to ESD diodes and amplifiers in the ADC cause
DC offsets and the demodulation of signals through harmonics and intermodulation
products. Section~\ref{sec:adcs} and Appendix~\ref{app:exp} exemplify these
effects through experiments on different ADCs.

\subsection{Sampling Errors in the Absence of an Adversary}
\label{sec:samp_errors}

\begin{figure}[t]
  \centering
  \includegraphics[width=0.45\linewidth]{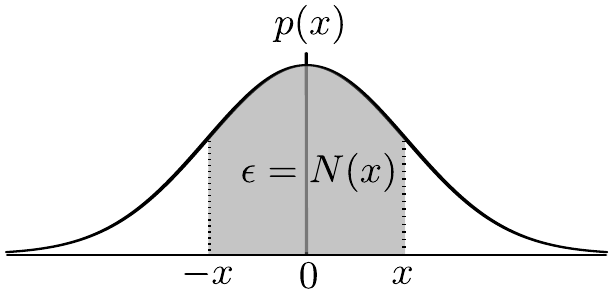}
  \caption{Noise probability distribution $p(x)$. The shaded area represents
           the probability $\epsilon=N(x)=\textrm{Pr}[|n(t)|\le x]$.}
  \label{fig:noise}
  \vspace{-2em}
\end{figure}

The digitization process through ADCs entails errors due to quantization
and environmental noise.
Quantization errors exist due to the inherent loss of accuracy
in the sampling process. An ADC can only
represent values within a range, say between $V_{min}$ and $V_{max}$ volts,
with a finite binary representation of $N$ bits, called
the {\em resolution} of the ADC. In other words, every value between $V_{min}$ and $V_{max}$
is mapped to one of the $2^N$ values that can be represented using $N$ bits.
As a result, there is a {\em quantization error} between the true sensor
analog value $s$ and the digitized value $\tilde{s}$. The maximum value
of this error is
\begin{equation}
\label{eq:quantization}
  Q=\frac{V_{max}-V_{min}}{2^{N+1}}\ge|s-\tilde{s}|
\end{equation}

The second source of error comes from environmental noise, which may affect
measurements. We assume that this noise, denoted by $n(t)$, is
independent of the signal being measured, and that it comes from a zero-mean
distribution, i.e., that the noise is {\em white}. The security definitions
we introduce in Section~\ref{sec:definitions} require an estimate of the
level of noise in the system, so we introduce some relevant notation
here. We assume that $n(t)$ follows a probability distribution function (PDF)
$p(x)$, and define $N(x)$ as the probability that the noise is between $-x$ and $x$,
as shown in Figure~\ref{fig:noise}, i.e.,
\[
  N(x)=\textrm{Pr}\left[|n(t)|\le x\right]=\int_{-x}^x p(u) du
\]
Note that typically the noise is assumed to come from a normal distribution, but
this assumption is not necessary in our models and definitions.

We are also interested in the inverse of this function, where given a
probability $0\le \epsilon < 1$, we want to find $x\ge 0$ such that $N(x)=\epsilon$.
For this $x$, the probability that the noise magnitude falls within $[-x,x]$
is $\epsilon$, as also shown in Figure~\ref{fig:noise}.
Because for some distributions there might be multiple $x$ for which $N(x)=\epsilon$,
we use the smallest such value:
\begin{equation}\label{eq:noise}
  N^{-1}(\epsilon) =\inf\{x\ge 0:N(x)=\epsilon\}
\end{equation}
Since $N(x)$ is an increasing function, so is $N^{-1}(\epsilon)$.

To account for repeated measurements, we introduce a short-hand for
sampling errors, which we denote by $E_s(t)$.
The sampling errors depend on the sensor input into the
ADC $s(t)$, the sampling rate $f$, the discrete output of the ADC
$\tsf(t)$ as well as the conversion delay $\tau$, representing the
time the ADC takes for complete a conversion:
\begin{equation}\label{eq:samp_error}
  E_s(t)=
  \begin{cases}
    \left|\tsf(t+\tau) - s(t)\right| & \text{if a conversion starts at $t$} \\
    0 & \text{otherwise} \\
  \end{cases}
\end{equation}

\subsection{Adversary Model}
\label{sec:adv_model}

Our threat model and definitions can capture a range of attacker goals, from
attackers who merely want to disrupt sensor outputs, to those who wish
to inject precise waveforms into a system. We define these notions
precisely in Section~\ref{sec:definitions},
but here we describe the attacker capabilities based on our model of
Figure~\ref{fig:adc_model}. Specifically, in our model, the adversary can only
alter the transmitted adversarial signal $v(t)$. He/she cannot directly influence the sensor
signal $s(t)$, the (residual) noise $n(t)$, or the transfer functions
$H_A$ and $H_C$. The adversary knows $H_A$, $H_C$, and the distribution
of the noise $n(t)$, although the true sensor signal $s(t)$
might be hidden from the adversary (see Section~\ref{sec:select_sec}).
The only constraint placed on the
adversarial signal is that the attacker is only allowed to transmit signals $v(t)$
whose peak voltage level is bounded by some constant $\vma$, i.e.,
$|v(t)|\le \vma$ for all $t$. We call this adversary a {\em $\vma$-bound adversary},
and all security definitions are against such bounded adversaries.

We choose to restrict voltage rather than restricting power or distance, as
it makes for a more powerful adversarial model. Our model gives the adversary
access to any physical equipment necessary (such as powerful amplifiers and
highly-directional antennas), while reducing the number of parameters
needed for our security definitions of Section~\ref{sec:definitions}.
Distance and power effects can be compensated directly through altering $\vma$,
or indirectly by integrating them into $H_C$, as discussed in Section~\ref{sec:adc_model}.

\section{Security Definitions}
\label{sec:definitions}

Using the model of Figure~\ref{fig:adc_model}, we can
define security in the presence of signal injection attacks. The
$\vma$-bound adversary is allowed to transmit any
waveform $v(t)$, provided that $|v(t)|\le \vma$ for all $t$: the adversary is only
constrained by the voltage budget. Whether or not the adversary succeeds
in injecting the target waveform $w(t)$ into the output of the system depends on
the transfer functions $H_C$ and $H_A$. For a given system described by $H_A$
and $H_C$, there are three outcomes against an adversary whose only restriction is
voltage:

\begin{enumerate}
  \itemsep0em
  \item The adversary can disturb the sensor readings, but cannot precisely control
        the measurement outputs, an attack we call {\em existential injection}.
        The lack of existential injections can be considered {\em universal security}.
  \item The adversary can inject a target waveform $w(t)$ into
        the ADC outputs with high fidelity, performing a {\em selective injection}.
        If the adversary is unable to succeed, the system is {\em selectively secure}
        against $w(t)$.
  \item The adversary can {\em universally inject} any waveform $w(t)$. If there
        is any non-trivial waveform for which he/she fails, the system is
        {\em existentially secure}.
\end{enumerate}

This section sets out to precisely define the above security notions by
accounting for noise and quantization error (Equation~\eqref{eq:quantization}).
Our definitions capture the intuition that
systems are secure when there are no adversarial transmissions, and are
``monotonic'' in voltage, i.e., systems are more vulnerable
against adversaries with access to higher-powered transmitters. Our definitions
are also monotonic in noise: in other words, in
environments with low noise, even a small disturbance of the output is sufficient
to break the security of a system. Section~\ref{sec:exist_sec}
evaluates whether an adversary can disturb the ADC
output away from its correct value sufficiently. Section~\ref{sec:select_sec} then
formalizes the notion of selective security against target waveforms $w(t)$.
Finally, Section~\ref{sec:univ_sec} introduces universal injections by defining
what a non-trivial waveform is. The three types of signal injection attacks,
the corresponding security properties, and the ensuing ADC errors (injected
waveforms) are summarized in Table~\ref{table:sec_defs}.\footnote{
  The terminology chosen was inspired by attacks against signature
  schemes, where how broken a system is depends on what types of messages an attacker
  can forge~\cite{sig_scheme}.
}

\begin{table}[t]
  \centering
  \caption{Correspondence between security properties of a sensor system,
           adversarial injection attacks, and the resulting ADC waveform errors (signals).}
  \begin{tabular}{llr}
  \toprule
  \textbf{Security}   & \textbf{Injection}  & \textbf{ADC Error $E_s(t)$}  \\
  \midrule
  Universal           & Existential         & Bounded away from $0$             \\
  Selective           & Selective           & Target waveform $w(t)$            \\
  Existential         & Universal           & Non-trivial waveforms $w(t)$      \\
  \bottomrule
  \end{tabular}
  \label{table:sec_defs}
\end{table}

\subsection{Existential Injection, Universal Security}
\label{sec:exist_sec}
The most primitive type of signal injection attack is a simple disruption of
the sensor readings. There are two axes in which this notion can be evaluated:
adversarial voltage and probability of success (success is probabilistic,
as noise is a random variable). For a fixed probability of success,
we want to determine the smallest voltage level for which an attack is successful.
For a fixed voltage level, we want to find the probability of a successful attack.
Alternatively, if we fix both the voltage and the probability of success, we want to
determine if a system is secure against disruptive signal injection attacks.

The definition for universal security is a formalization of the
above intuition, calling a system secure when, even in the presence of
injections (bounded by adversarial voltage), the true analog sensor value and the ADC digital output
do not deviate by more than the quantization error and the noise, with sufficiently
high probability. Mathematically:

\begin{definition}[Universal Security, Existential Injection]
For $0\le\epsilon< 1$, and $\vma \ge 0$,
we call a system \textbf{universally $(\epsilon, \vma)$-secure} if
\begin{equation}
  Pr\left[E_s(t) \ge Q+\nin{}\right]\le \eh{} \label{eq:ex_sec}
\end{equation}
for every adversarial waveform $v(t)$, with $|v(t)|\le\vma$ for all $t$.
$Q$ is the quantization error of the system, $N^{-1}$ is the noise distribution inverse
defined in Equation~\eqref{eq:noise}, and $E_s$ is the sampling error as defined
by Equation~\eqref{eq:samp_error}.
The probability is taken over the duration of the attack, i.e.,
at each sampling point within the interval $t_{start}\le t \le t_{end}$.
We call a successful attack an \textbf{existential injection}, and simply call
a system universally $\epsilon$-secure, when $\vma$ is implied.
\end{definition}

We first show that in the absence of injections,
the system is universally $\epsilon$-secure for all $0\le \epsilon < 1$.
Indeed, let $x=\nin{}$, so that $Pr\left[|n(t)|\le x\right]=\eh{}$.
Then, in the absence of injections,
\begin{align*}
  &Pr\left[E_s(t) \ge Q+\nin{}\right]=Pr\left[|n(t)|\ge x\right]=\\
  &=1-\eh{}=\frac{1-\epsilon}{2}\le\eh{}
\end{align*}
which holds for all $0\le \epsilon<1$, as desired. This proof is precisely the
reason for requiring a noise level and probability of at least 50\% in the definition:
the proof no longer works if $(1+\epsilon)/2$ is replaced by just $\epsilon$.
In other words, mere noise would be classified as an attack by the modified
definition.

\secpar{Voltage} We now show
that a higher adversarial voltage budget can only make a system more vulnerable.
Indeed, if a system is universally $(\epsilon, V_1)$-secure, then
it is universally $(\epsilon, V_2)$-secure for $V_2\le V_1$. For this,
it suffices to prove the contrapositive, i.e., that if a system is
not universally $(\epsilon, V_2)$-secure,
then it is not universally $(\epsilon, V_1)$-secure. For the proof,
let $v(t)$ be an adversarial waveform with $|v(t)| \le V_2$ such that
Equation~\eqref{eq:ex_sec} does not hold, which exists by the assumption that
the system is not universally $(\epsilon, V_2)$-secure. Then, by the transitive property,
$|v(t)| \le V_1$, making $v(t)$ a valid counterexample for universal
$(\epsilon, V_1)$ security.

\secpar{Probability} The third property we show is probability monotonicity,
allowing us to define a ``critical threshold'' for $\epsilon$, above which
a system is universally secure (for a fixed $\vma$), and below which
a system is not universally secure. Indeed, for fixed $\vma$,
if a system is universally $(\epsilon,\vma)$-secure, then it is
universally $(\epsilon+\delta, \vma)$-secure for $0\le \delta < 1-\epsilon$, as
\begin{align*}
  &Pr\left[E_s(t) \ge Q+\nin[\epsilon+\delta]\right]\le \\
  &Pr\left[E_s(t) \ge Q+\nin{}\right]\le \eh{} \le \eh[\epsilon+\delta]
\end{align*}
because $N^{-1}$ is increasing. The contrapositive is, of course, also true:
if a system is not universally secure for a given $\epsilon$, it is also not
universally secure for $\epsilon-\delta$ with $0\le \delta \le \epsilon$.

\secpar{Thresholds} For a given security level $\epsilon$,
then, we can talk about the maximum (if any) $\vma$
such that a system is universally $(\epsilon,\vma)$-secure, or conversely
the minimum (if any) $\vma$ such that a system is not
universally $(\epsilon,\vma)$-secure. This is the \textbf{critical universal
voltage level $V_c$} for the given $\epsilon$.
Moreover, for any $\vma$, there is a unique \textbf{critical universal
security threshold $\ec$} such that the system is universally
$(\epsilon,\vma)$-secure for $\ec<\epsilon<1$ and not universally
$(\epsilon,\vma)$-secure for $0\le\epsilon<\ec$. By convention we take
$\ec = 0$ if the system is secure for all $\epsilon$, and
$\ec=1$ if there is no $\epsilon$ for which the system is secure.
This critical threshold indicates the security level of a system: the
lower $\ec$ is, the better a system is protected against signal injection attacks.

\subsection{Selective Injection and Security}
\label{sec:select_sec}

The second definition captures the notion of security against specific
target waveforms $w(t)$:
we wish to find the probability that a $\vma$-bounded adversary can make
$w(t)$ appear in the output of the ADC. Conversely, to define security
in this context, we must make sure that
the digitized signal $\tsf(t)$ differs from the waveform $s(t)+w(t)$ with high
probability, even if plenty of noise is allowed. There are two crucial points to
notice about the waveform $w(t)$. First, $w(t)$ is not the raw signal $v(t)$
the adversary is transmitting, as this signal undergoes two transformations
via $H_C$ and $H_A$. Instead, $w(t)$ is the signal
that the adversary wants the ADC to think that it is seeing, and is usually
a demodulated version of $v(t)$ (see Figure~\ref{fig:adc_model}).
Second, $w(t)$ does not necessarily cancel out or overpower $s(t)$,
because that would require predictive modeling of the sensor signal $s(t)$.
However, if the adversary
can predict $s(t)$ (e.g., by monitoring the output of the ADC, or by
using identical sensors), we can then ask about security against the
waveform $w'(t)=w(t)-s(t)$ instead. Given this intuition, we can define
selective security as follows:

\begin{definition}[Selective Security, Selective Injection]
  For $0\le\epsilon< 1$, and $\vma \ge 0$, a system is called
  \textbf{selectively $(\epsilon, w(t), \vma)$-secure} if
  \begin{equation}
    Pr\left[E_{s+w}(t)\ge Q+\nin[(1-\epsilon)]\right]> \meh{}\label{eq:sel_sec}
  \end{equation}
  for every adversarial waveform $v(t)$, with $|v(t)|\le\vma$ for all $t$,
  where the probability is taken over the duration of the attack.
  $Q$ is the quantization error of the system, $N^{-1}$ is the noise distribution inverse
  defined in Equation~\eqref{eq:noise}, and
  $E_{s+w}(t)=\left|\tsf(t+\tau) - s(t) - w(t)\right|$ during
  sampling periods, and $0$ otherwise.
  We call a successful attack a \textbf{selective injection}, and simply call
  a system selectively $\epsilon$-secure, when $\vma$ and $w(t)$ are clear from context.
\end{definition}

This definition is monotonic in voltage and
the probability of success, allowing us to talk about ``the'' probability of success
for a given waveform:

\secpar{Voltage} A similar argument 
shows that increasing $\vma$ can only make a secure
system insecure, but not vice versa, i.e., that if
a system is selectively $(\epsilon, w(t), V_1)$-secure, then
it is selectively $(\epsilon, w(t), V_2)$-secure for $V_2\le V_1$. We can
thus define the \textbf{critical selective voltage level $V_c^w$} for a given
$\epsilon$ and $w(t)$.

\secpar{Probability} If a system is selectively $\epsilon$-secure (against a target waveform and
voltage budget), then it is selectively $(\epsilon+\delta)$-secure for $0\le\delta<1-\epsilon$, because
\begin{align*}
P&=Pr\left[E_{s+w}(t)\ge Q+\nin[1-(\epsilon+\delta)]\right]\\
 &\ge Pr\left[E_{s+w}(t)\ge Q+\nin[1-\epsilon]\right] > \meh{}\ge \meh[(\epsilon+\delta)]
\end{align*}
If the system is not selectively $\epsilon$-secure,
then it is not selectively $(\epsilon-\delta)$-secure.

Given the above, for a given waveform $w(t)$ and fixed $\vma$,
we can define a waveform-specific \textbf{critical selective security threshold} $\ec^w$
such that the system is vulnerable for all $\epsilon^w$ with
$0\le\epsilon^w<\ec^w$ and secure for all $\epsilon^w$ with $\ec^w<\epsilon^w<1$.
By convention we take $\ec^w = 0$ if there is no $\epsilon$ for which
the system is vulnerable, and $\ec^w=1$ if there is no $\epsilon$ for which
the system is secure.

\secpar{Threshold Relationship} The critical universal threshold of a system
$\ec$ is related to the critical selective threshold $\ec^0$ against the zero waveform $w(t)=0$
through the equation $\ec^0=1-\ec$. Indeed, if a system is not universally
$\epsilon$-secure, then
$P=Pr\left[E_s(t) \ge Q+\nin{}\right]> \eh{}$, so
$\meh[(1-\epsilon)]=\eh{}<P=Pr\left[E_{s+0}(t) \ge Q+\nin[(1-(1-\epsilon))]\right]$,
making the system selectively $(1-\epsilon)$-secure for the zero waveform. Conversely, if a system
is selectively $(1-\epsilon)$-secure for the zero waveform,
then it is not universally $\epsilon$-secure.
The fact that a low critical universal threshold results in a high critical selective
threshold for the zero threshold is not surprising: it is
easy for an adversary to inject a zero signal by simply not transmitting anything.

\subsection{Universal Injection, Existential Security}
\label{sec:univ_sec}

The final notion of security is a weak one, which requires that the adversary
cannot inject at least one ``representable'' waveform into the system, i.e.,
one which is within the ADC limits. We can express this more precisely as follows:

\begin{definition}[Representable Waveform]
  A waveform $w(t)$ is called \textbf{representable}
  if it is within the ADC voltage levels, and has a maximum frequency component
  bounded by the Nyquist frequency of the ADC. Mathematically,
  $V_{min} \le w(t) \le V_{max}$ and $f_{max} \le f_s/2$.
\end{definition}
\noindent Using this, we can define security against at least one
representable waveform:

\begin{definition}[Existential Security, Universal Injection]
  For $0\le\epsilon< 1$, and $\vma \ge 0$, a system is called
  \textbf{existentially $(\epsilon, \vma)$-secure} if there exists
  a representable waveform $w(t)$ for which the system is
  selectively $(\epsilon, w(t), \vma)$-secure.
  We call a system existentially $\epsilon$-secure
  when $\vma$ is clear. If there is no such $w(t)$,
  we say that the adversary can perform any \textbf{universal injection}.
\end{definition}

\noindent As above, voltage and probability are monotonic in
the opposite direction.

\secpar{Voltage} If a system is existentially $(\epsilon, V_1)$-secure,
then it is $(\epsilon, V_2)$-secure for $V_2\le V_1$. By assumption, there is a representable
$w(t)$ such that the system is selectively $(\epsilon, w(t), V_1)$-secure.
By the previous section, this system is $(\epsilon, w(t), V_2)$-secure,
concluding the proof.

\secpar{Probability}  If a system is existentially $(\epsilon_1, V)$-secure,
then it is $(\epsilon_2, V)$-secure for $\epsilon_1\le \epsilon_2$.
By assumption, there is a representable
$w(t)$ such that the system is selectively $(\epsilon_1, w(t), V)$-secure.
By the previous section, the system is also $(\epsilon_2, w(t), V)$-secure, as desired.

\secpar{Thresholds} Extending the definitions of the previous sections, for fixed
$\epsilon$ we can define a \textbf{critical existential voltage level $V_c^{exist}$} below which
a system is existentially $\epsilon$-secure, and above which the system
is existentially $\epsilon$-insecure. Similarly, for a fixed adversarial voltage
we can define the
\textbf{critical existential security threshold $\epsilon_c^{exist}$}, above
which the system is existentially secure, and below which the system is insecure.

In some cases,
security designers may wish to adjust the definitions to restrict target waveforms (and
existential security counterexamples) even further.
For instance, we might wish to check whether an adversary can inject all
waveforms which are sufficiently bounded away from 0, periodic waveforms, or
waveforms of a specific frequency. The proofs for voltage and probability monotonicity
still hold, allowing us to talk about universal security against $\mathcal{S}$-representable
waveforms: waveforms which are representable and also in a set $\mathcal{S}$.

\section{Security Evaluation of a Smartphone Microphone}
\label{sec:case_study}

\begin{table}[t]
  \centering
  \caption{The adversary can
           easily disturb the smartphone output (existential injection), and inject
           human speech (universal injection). Selective injections of sines are less
           precise than exponentials of the same frequency.}
  \begin{tabular}{llr}
  \toprule
  \textbf{Injection} & \textbf{Resulting Signal}    & \textbf{Crit. Thres.}\\
  \midrule
  Existential        & $w(t) \ne 0$                 & $0.892$ \\
  Selective          & $w(t)=e^{\sin(2\pi f_m t)}$  & $0.747$ \\
  Selective          & $w(t)=\sin(2\pi f_m t)$      & $0.562$ \\
  Universal          & ``OK Google'' commands       & $\le0.562$\\
  \bottomrule
  \end{tabular}
  \vspace{-1em}
  \label{table:results}
\end{table}

In this section, we illustrate how our security definitions can be used to
determine the security level of a commercial, off-the-shelf smartphone microphone.
We first introduce an algorithm to calculate the critical selective security
threshold $\ec^w$ against a target waveform $w(t)$ in Section~\ref{sec:algorithm}.
We then use the algorithm to calculate the critical thresholds
of a smartphone in Section~\ref{sec:phone_security}.
Finally, we comment on universal security in Section~\ref{sec:phone_sine},
where we show that we are able to inject complex ``OK Google'' commands.
We summarize our results in Table~\ref{table:results}, while
Appendix~\ref{app:exp} contains additional experiments for further
characterization of the smartphone's ADC.

\subsection{Algorithm for Selective Security Thresholds}
\label{sec:algorithm}

In this section, we introduce an algorithm to calculate the critical
selective security threshold $\ec^w$ of a system against a target waveform $w(t)$,
using a transmitted signal $v(t)$.
The first step in the algorithm (summarized in
pseudocode as Algorithm~\ref{alg:adc:sec_calc}) is to determine the noise
distribution. To that end, we collect $N$ measurements of
the system output $\tsf(t)$ during the injection and pick one as the
{\em reference} signal. We then pick $1\le k\le N-2$ of them to calculate
the noise ({\em estimation} signals), while the remaining are used to verify
our calculations ({\em validation} signals).

\begin{algorithm}[t]
  \small
  \begin{algorithmic}[1]
  \Procedure{FindCriticalEpsilon}{$measured$, $ideal$, $sigma$}
  \State $errors \gets |measured - ideal|/sigma$ \Comment{Calculate normalized absolute errors}
  \State $lo \gets 0.5$ \Comment{Probabilities need to be between $0.5$ and $1$}
  \State $hi \gets 1$
  \While {$lo < hi$}
  \State $mid \gets (lo + hi)/2$ \Comment{$mid$ represents $(2-\epsilon)/2$}
  \State $n_{inv} \gets \text{ppf}((1+mid)/2) $ \Comment{Percentile point function}
  \State $p_{error} \gets \text{length}([x\ge n_{inv}:x\in errors])/\text{length}(errors)$

  \If{$|p_{error} - mid| < \delta \And p_{error} \le mid$} \Comment{Threshold $\delta=10^{-4}$}
    \State \Return $2-2*mid$ \Comment{Break out if sufficiently close}
  \ElsIf{$p_{error} < mid$}
    \State $hi \gets mid$
  \Else
    \State $lo \gets mid$
  \EndIf
  \EndWhile

  \EndProcedure

  \Procedure{Compare}{$\boldsymbol{measurements}$, $ideal$} \Comment{Repeated $\boldsymbol{measurements}$}
  \State $ref \gets \text{detrend}(\boldsymbol{measurements}[0])$ \Comment{Pick first as reference, remove DC}
  \State $\boldsymbol{estimating} \gets \text{align}(\text{scale}(\text{detrend}(\boldsymbol{measurements}[1:]), ref), ref)$
  \State $\boldsymbol{errors} \gets (measured - ref)\ \forall measured \in \boldsymbol{estimating}$
  \State $\sigma_{noise} \gets \text{std\_deviation}(\boldsymbol{errors})$ \Comment{Calculate noise from estimations}
  \State $ideal \gets \text{align}(\text{scale}(\text{detrend}(ideal), ref), ref)$
  \State \Return FindCriticalEpsilon($ref$, $ideal$, $\sigma_{noise}$)
  \EndProcedure
  \end{algorithmic}
  \caption{Determining the Critical Selective Security Threshold}
  \label{alg:adc:sec_calc}
\end{algorithm}

Our algorithm first removes any DC offset and re-scales the measurements
so that the root-mean-square (RMS) voltages of the signals are the same. The
repeated measurements are then phase-aligned, and we calculate the distance
between the reference signal and the estimation signals. The average of this
distance should be very close to 0, as the signals are generated in the same
way. However, the standard deviation $\sigma$ is non-zero, so we can model
noise as following a zero-mean normal distribution $n(t)\sim N(0, \sigma^2)$.
We can then find the critical threshold between the reference signal
and any target {\em ideal} waveform $w(t)$ as follows: we first detrend, scale, and align
the ideal signal to the reference waveform, as with the estimation signals.
Then, we calculate the errors (distance)
between the ideal and the reference signal. Finally, we perform a binary search
for different values of $\epsilon$, in order to find the largest $\epsilon$
for which Equation~\eqref{eq:sel_sec} does not hold: this is the critical
threshold $\ec^w$. To calculate the inverse of the noise, we use the percentile point
function $ppf(\epsilon)$, which is the inverse of the cumulative distribution
function, and satisfies $N^{-1}(\epsilon)=ppf((1+\epsilon)/2)$.
Note that since the critical universal threshold $\ec$ is related to the
selective critical threshold of the zero waveform $\ec^0$ through $\ec=1-\ec^0$
(Section~\ref{sec:select_sec}), the same algorithm can be used to calculate the
critical universal security threshold $\ec$.

\subsection{Existential and Selective Injections into a Smartphone}
\label{sec:phone_security}

\begin{figure*}[t]
  \centering
  \begin{subfigure}[b]{0.49\textwidth}
  \centering
          \includegraphics[width=\textwidth]{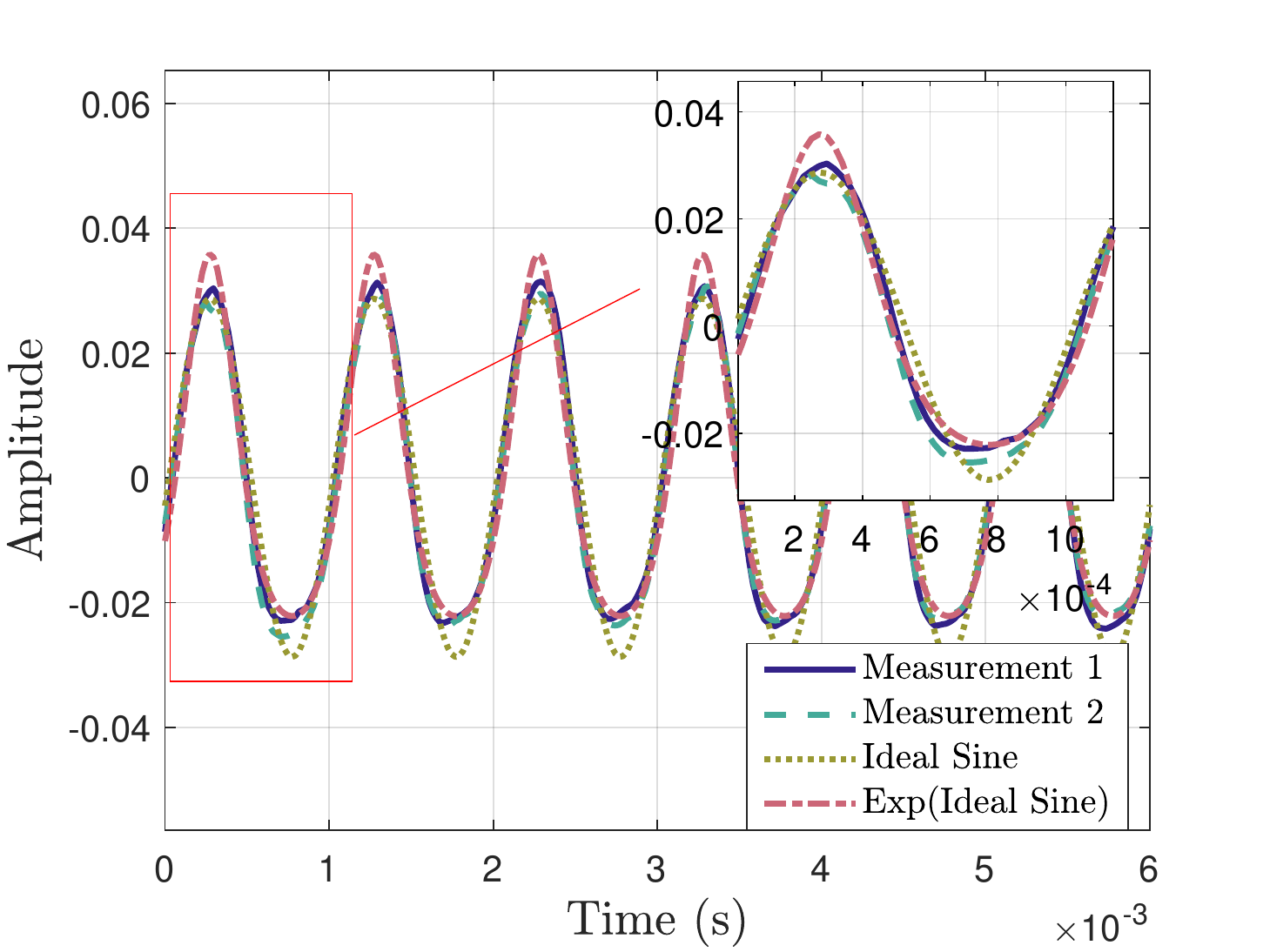}
          \caption{$\vra=\SI{0.2}{\volt}$, $f_c=\SI{200}{\mega\hertz}$}
          \label{fig:clean_waveform}
  \end{subfigure}\hfill
  \begin{subfigure}[b]{0.49\textwidth}
  \centering
          \includegraphics[width=\textwidth]{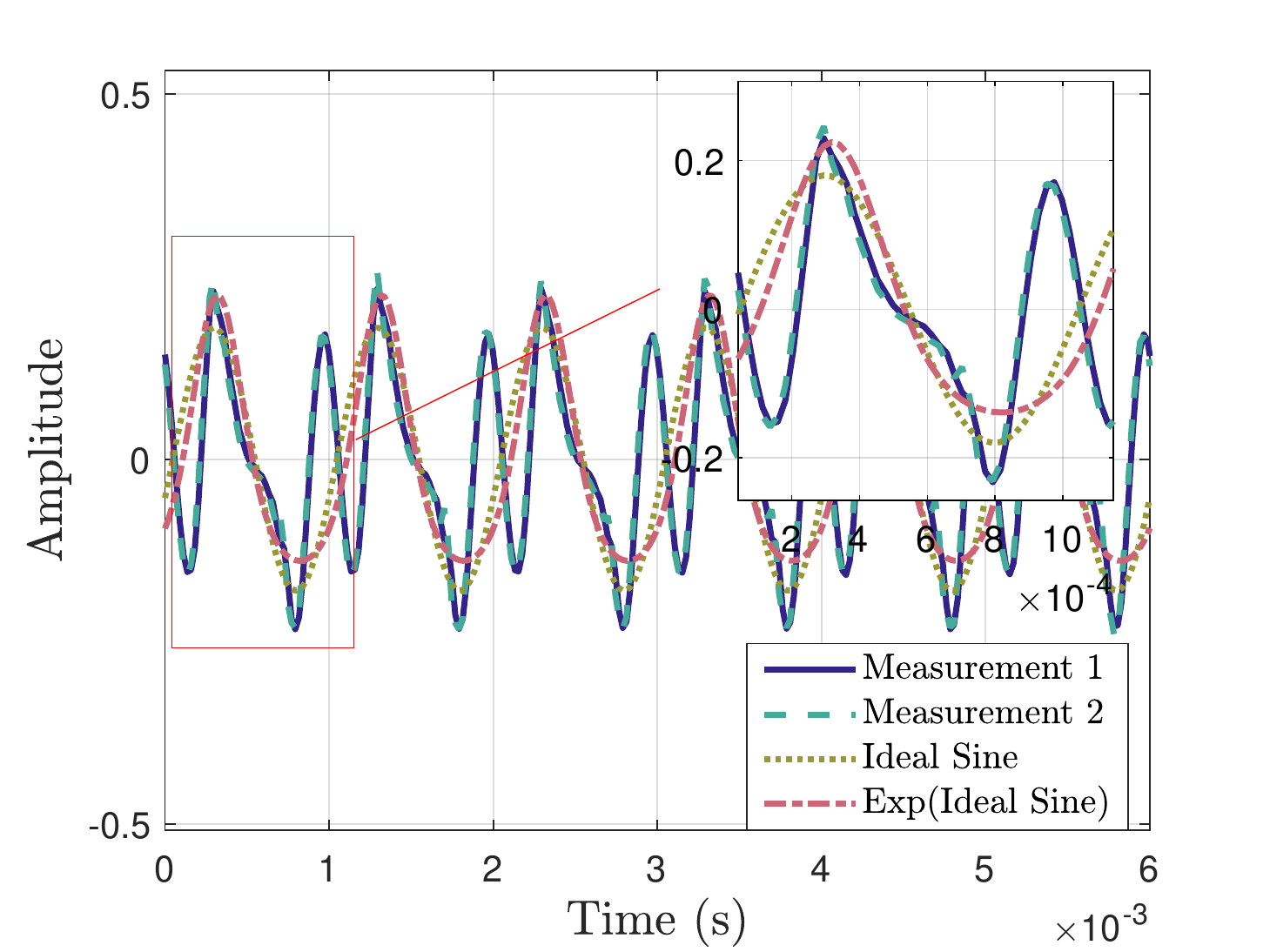}
          \caption{$\vra=\SI{0.9}{\volt}$, $f_c=\SI{25}{\mega\hertz}$}
          \label{fig:distorted_waveform}
  \end{subfigure}\hfill
  \caption{Clean (\subref{fig:clean_waveform}) and Distorted (\subref{fig:distorted_waveform}) waveforms injected into the smartphone, with ideal sine and exponential sine functions for comparison.}
  \label{fig:phone_waveform}
  \vspace{-1em}
\end{figure*}

We demonstrate how our algorithm can be used in a realistic setup
using a Motorola XT1541 Moto G3 smartphone. We inject amplitude-modulated
$f_m=\SI{1}{\kilo\hertz}$ signals using a Rohde \& Schwarz SMC100A/B103
generator into the headphone jack of the phone, following direct
power injection (DPI) methodology~\cite{sok}. We collect $N=10$ measurements of
$2^{15}$ sample points per run using an ``Audio Recorder'' app,
and record the data
at a frequency of $f_s=\SI{44.1}{\kilo\hertz}$ in a $\left[-1,1\right]$
dimensionless range (AAC encoding). We first amplitude-modulate $f_m$ over
$f_c=\SI{200}{\mega\hertz}$ using an output level of $\vra=\vma/\sqrt2=\SI{0.2}{\volt}$.
This injection is demodulated well by the smartphone and has a ``similarity''
(see Appendix~\ref{app:exp})
of over $0.98$ compared to a pure $\SI{1}{\kilo\hertz}$ tone. We call
this example the ``clean'' waveform.
The second injection, which we call the ``distorted'' waveform,
uses $f_c=\SI{25}{\mega\hertz}$, $\vra=\SI{0.9}{\volt}$, and has a similarity
of less than $0.55$ to the ideal tone.
Example measurements of these signals and ``ideal'' signals (see below) are
shown in Figure~\ref{fig:phone_waveform}.

The algorithm first calculates the noise level using the reference signals.
As expected, the error average is very close to 0 (usually less than $10^{-6}$),
while the standard deviation $\sigma$ is noticeable at around $0.0015$.
Taking the reference signals as the target signal $w(t)$, the critical selective thresholds
are close to $1$. In other words, even if the injected waveforms do not
correspond to ``pure'' signals, the adversary can inject them
with high fidelity: the system is not selectively secure against them with
high probability.

We also tried two signals as the signal $w(t)$ that the adversary is trying
to inject: a pure $\SI{1}{\kilo\hertz}$ sine wave,
and an exponential of the same sine wave. The averages and standard deviations for
the calculated thresholds over all combinations of $k$
and reference signals are shown in
Table~\ref{table:selective_thresholds}. As we would expect, the thresholds
for the distorted waveform are much lower than the values for the clean waveform:
the signal is distorted, so it is hard to inject an ideal signal. We also find
that the exponential function is a better fit for the signal we are seeing,
and can better explain the harmonics. Table~\ref{table:selective_thresholds}
also includes the critical universal injection threshold based on the two
waveform injections. This
threshold is much higher for both waveforms, as injections disturb the ADC output
sufficiently, even when the demodulated signal is not ideal.

\begin{table}[t]
  \centering
  \caption{Mean and std. deviation $(\mu, \sigma)$ of critical selective
            thresholds $\ec^w$ for different target signals $w(t)$.
            Injections using the clean waveform are always more successful than
            with the distorted waveform. Validation signals are injected
            with high fidelity, and are better modeled by an exponential
            rather than a pure sine.}
  \begin{tabular}{lrrrr}
    \toprule
    \textbf{Waveform}  & \textbf{Validation} & \textbf{Ideal Sine} & \textbf{$e^{\text{Ideal Sine}}$} & \textbf{$w(t) \ne 0$}\\
    \midrule
    Clean       & $(0.98, 0.03)$ & $(0.56, 0.04)$ & $(0.75, 0.06)$ & $(0.89, 0.01)$\\
    Distorted   & $(0.95, 0.09)$ & $(0.31, 0.05)$ & $(0.34, 0.05)$ & $(0.71, 0.04)$ \\
    \bottomrule
  \end{tabular}
  \vspace{-1em}
  \label{table:selective_thresholds}
\end{table}

\subsection{Universal Injections on a Smartphone}
\label{sec:phone_sine}

In this section, we demonstrate that the smartphone is vulnerable to the
injection of arbitrary
commands, which cause the smartphone to behave as if the user initiated an action.
Using the same setup of direct power injection (Section~\ref{sec:phone_security}),
we first inject a modulated recording of ``OK Google, turn on the flashlight''
into the microphone port,
checking both whether the voice command service was activated in response to
``OK Google'', and whether the desired action
was executed. We repeat measurements 10 times, each time amplitude-modulating
the command at
a depth of $\mu=1.0$ with $\vra=\SI{0.6}{\volt}$ on 26 carrier frequencies
$f_c$: $\SI{25}{\mega\hertz}$, $\SI{50}{\mega\hertz}$, and $\SIrange{100}{2400}{\mega\hertz}$
at a step of $\SI{100}{\mega\hertz}$.
The voice-activation feature (``OK Google'') worked with 100\% success
rate (10/10 repetitions) for all frequencies, while the full command was successfully executed
for 23 of the 26 frequencies we tested (all frequencies
except $f_c\in\{1.3, 2.0, 2.4\si{\giga\hertz}\}$). Increasing the output level
to $\vra=\SI{0.9}{\volt}$, increased success rate to 25/26 frequencies.
Only $f_c=\SI{2.4}{\giga\hertz}$ did not
result in a full command injection, possibly because the Wi-Fi disconnected in
the process.

We repeated the above injections, testing 5 further commands to (1) call
a contact; (2) text a contact; (3) set a timer; (4) mute the volume; and
(5) turn on airplane mode. The results remained identical, regardless of
the actual command to be executed. As a result, all carrier frequencies which are not severely
attenuated by $H_C$ (e.g., when coupling to the user's headphones) are vulnerable
to injections of complex waveforms such as human speech.

\section{Commercial ADC Response $H_A$ to Malicious Signals}
\label{sec:adcs}

\begin{table}[t]
  \centering
  \caption{The ADCs used in our experiments cover a range of different properties.}
  \begin{tabular}{llllrrHHHr}
  \toprule
  \textbf{ADC} & \textbf{Manufacturer} & \textbf{Package}  & \textbf{Type}  & \textbf{Bits}& \textbf{Max $f_s$} & \textbf{Eff. $f_s$} & \textbf{$R$} & \textbf{$C$} & \textbf{$f_{cut}$}\\
  \midrule
  TLC549 & Texas Instruments & DIP & SAR & 8  & \SI{40}{\kilo\hertz} & \SI{29}{\kilo\hertz} & \SI{1}{\kilo\ohm} & \SI{60}{\pico\farad} & \SI{2.7}{\mega\hertz}\\
  ATmega328P & Atmel & Integrated & SAR  & 10 & \SI{76.9}{\kilo \hertz} & \SI{8.3}{\kilo\hertz}
                  & \SIrange{1}{100}{\kilo\ohm} & \SI{14}{\pico\farad} & \SIrange{0.1}{11.4}{\mega\hertz} \\
  Artix7 & Xilinx & Integrated & SAR & 12& \SI{1}{\mega \hertz} & \SI{198}{\kilo\hertz} & \SI{10}{\kilo\ohm} & \SI{3}{\pico\farad}  & \SI{5.3}{\mega\hertz} \\
  AD7276 & Analog Devices & TSOT & SAR  & 12& \SI{3}{\mega \hertz} & \SI{35}{\kilo\hertz} & \SI{75}{\ohm} & \SI{32}{\pico\farad} & \SI{66.3}{\mega\hertz}\\
  AD7783 & Analog Devices & TSSOP & $\ds$ & 24 & \SI{19.79}{\hertz} & \SI{19.71}{\hertz} & N/A &  N/A & [$50$,\SI{60}{\hertz}] \\
  AD7822 & Analog Devices & DIP & Flash & 8 & \SI{2}{\mega \hertz} & \SI{84}{\kilo\hertz} & \SI{310}{\ohm} & \SI{4}{\pico\farad} & \SI{128.4}{\mega\hertz}\\
  \bottomrule
  \end{tabular}
  \label{table:adcs}
  \vspace{-.5em}
\end{table}

\vspace{-0.5em}
As explained in Section~\ref{sec:model}, an adversary trying to inject signals
remotely into a system typically needs to transmit modulated signals over
high-frequency carriers. As $H_C$ is unique to each circuit and needs to be re-calculated even
for minor changes to its components and layout~\cite{emi_susceptibility},
the first step to determine the
system vulnerability is to understand the behavior $H_A$ of the ADC used.

\begin{figure}[t]
  \centering
  \includegraphics[width=0.5\textwidth]{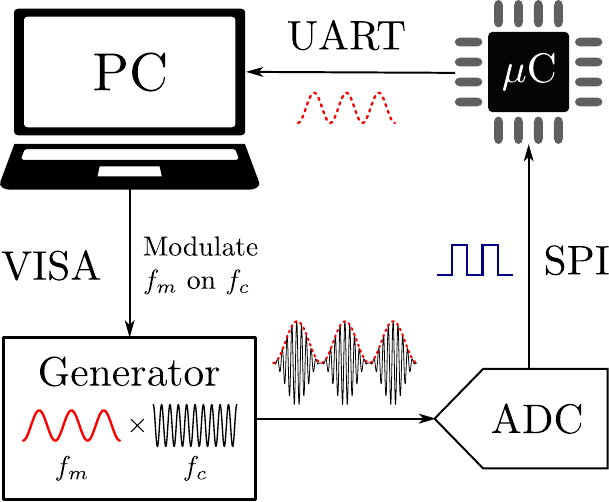}
  \caption{Amplitude-modulated signals are directly injected into ADCs using a
          signal generator controlled over the VISA interface. Measurements are
          transferred to a computer for analysis via a microcontroller's
          UART interface.}
  \label{fig:setup}
  \vspace{-.5em}
\end{figure}

To do so, we inject signals into the ADCs to determine their
demodulation characteristics. As shown in Figure~\ref{fig:setup},
the output of the Rohde \& Schwarz signal generator is directly connected to the
ADC under test,
while additional experiments with an amplifier or with remote transmissions are
performed in Appendix~\ref{app:exp}. An Arduino Uno (ATmega328P microcontroller)
interfaces with the ADC over the appropriate protocol, while a computer
collects the measurements from the microcontroller over the UART,
and controls the signal generator over the VISA interface.

Experiments are conducted with six ADCs from four manufacturers (Texas
Instruments, Analog Devices, Atmel, and Xilinx) in different packages: some are part
of the silicon in other ICs, while others are standalone surface-mount
or through-hole chips. Delta-Sigma ($\ds$), half-flash, and successive
approximation (SAR) ADCs
are tested, with sampling rates $f_s$ ranging from a few $\si{\hertz}$
to several $\si{\mega\hertz}$, and resolutions between $8$ and $24$ bits.
Table~\ref{table:adcs} shows these properties
along with the $\SI{-3}{\decibel}$ cutoff frequency $f_{cut}$, calculated using
the $R,C$ parameters in the ADCs' datasheets.

We use sinusoidals of frequencies $f_m$
that have been amplitude-modulated on carrier frequencies
$f_c$. In other words, we consider the intended
signal to be $w(t)=\sin(2\pi f_m t)$, the sensor signal to be absent ($s(t)=0$),
and evaluate how ``close'' $w(t)$ is to the ADC output $\tsf(t)$.
We summarize typical results for each ADC here, and present more details
in Appendix~\ref{app:exp}.

\begin{figure*}[t]
  \centering
  \begin{subfigure}[b]{0.45\textwidth}
  \centering
          \includegraphics[width=\textwidth]{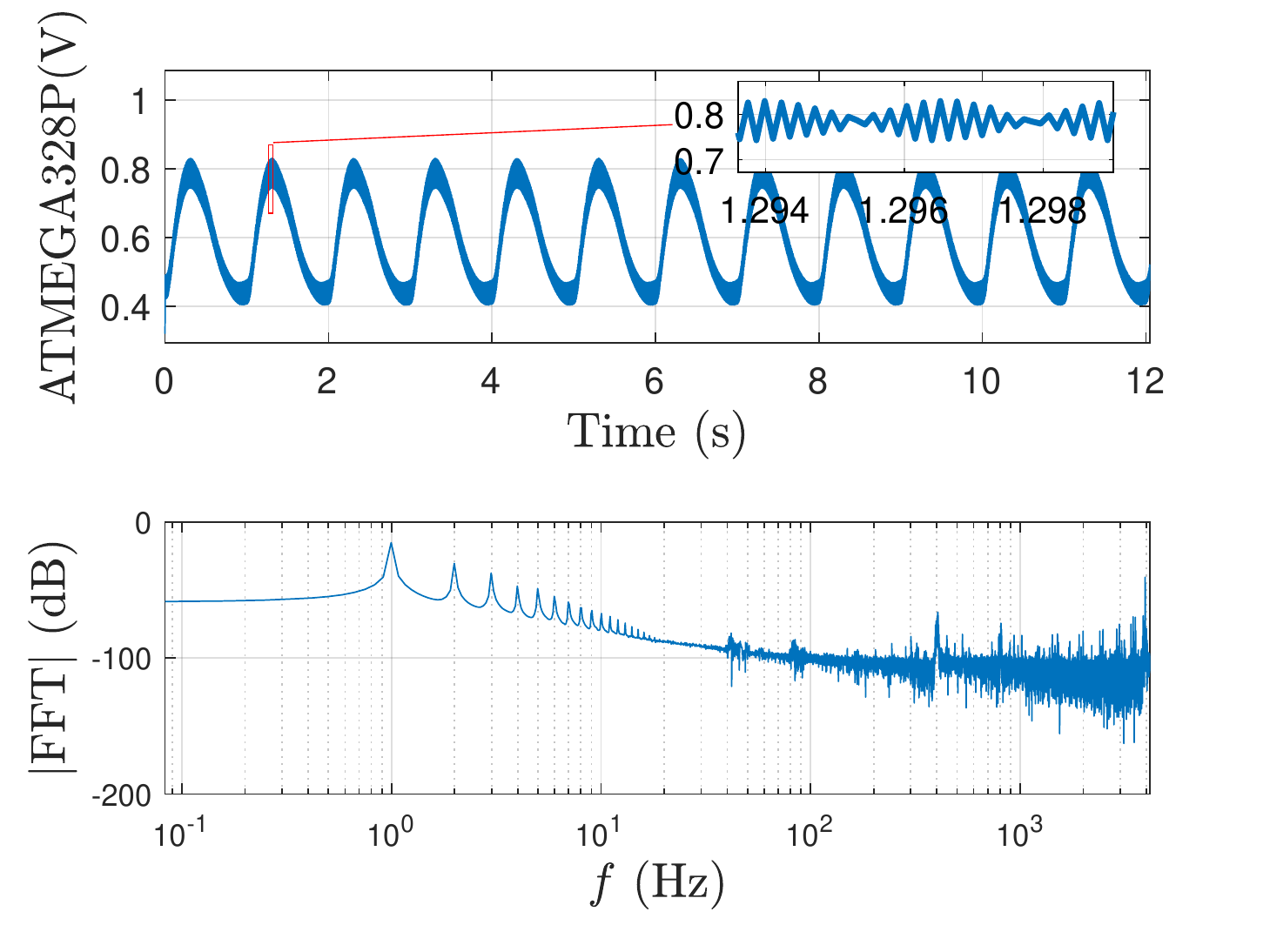}
          \caption{$f_c=\SI{10}{\mega\hertz}$}
  \end{subfigure}\hfill
  \begin{subfigure}[b]{0.45\textwidth}
  \centering
          \includegraphics[width=\textwidth]{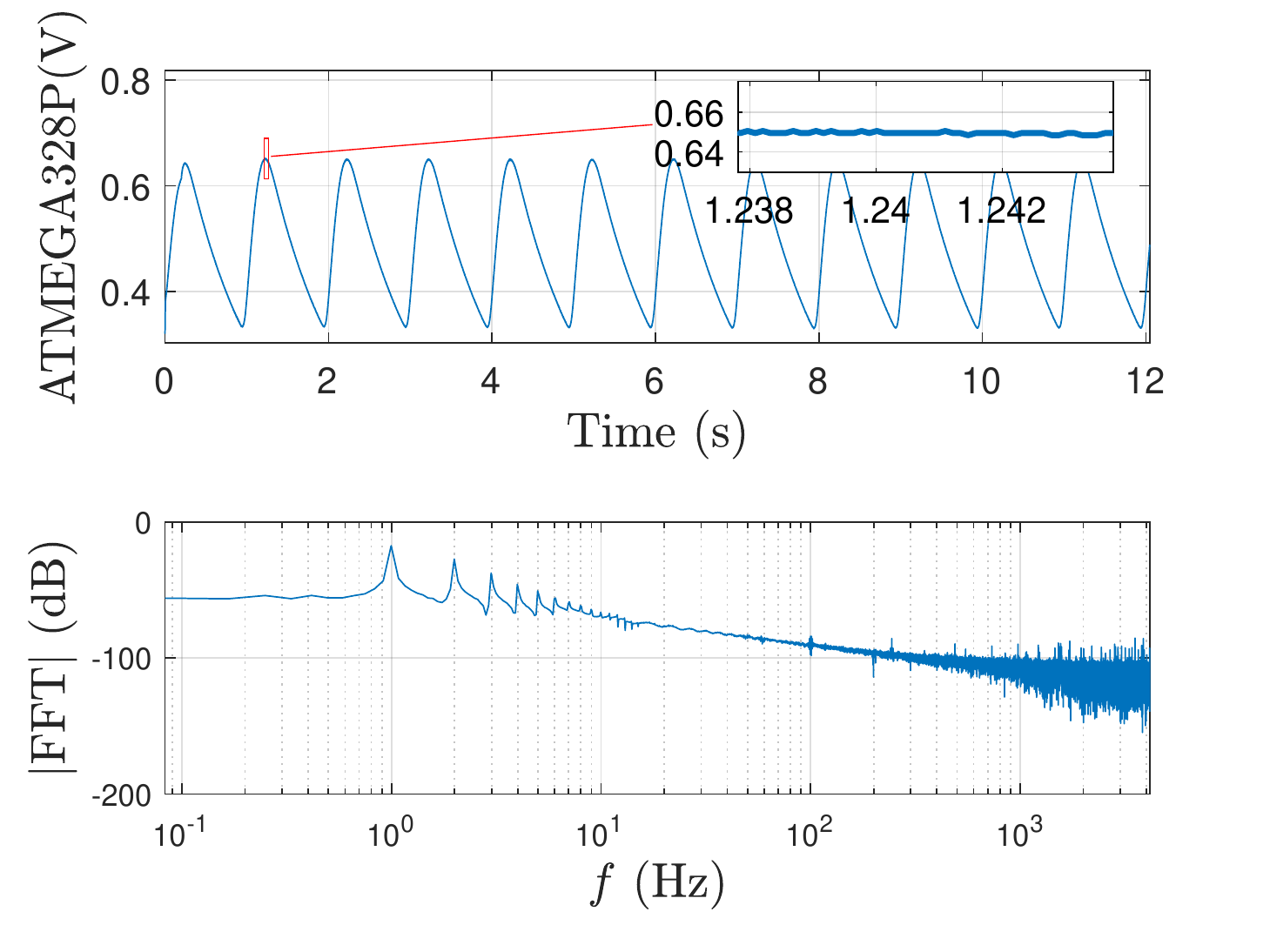}
          \caption{$f_c=\SI{80}{\mega\hertz}$}
  \end{subfigure}\hfill
  \caption{Example ATmega328P output for power $P=\SI{0}{\decibel m}$, signal frequency $f_m=\SI{1}{\hertz}$,
          and modulation depth $\mu=0.5$. The signal exhibits the correct fundamental frequency, but also
          contains strong harmonics and a high-frequency component, which is attenuated as the carrier frequency $f_c$ increases.}
  \label{fig:atmega_raw}
  \vspace{-1em}
\end{figure*}

\secpar{ATmega328P} Figure~\ref{fig:atmega_raw} presents two example measurements
of outputs of the ATmega328P, both in the time domain and in the frequency
domain. The input to the ADC is a $f_m=\SI{1}{\hertz}$ signal modulated over
different high-frequency carriers. As shown in the frequency
domain (bottom of Figure~\ref{fig:atmega_raw}), the fundamental frequency $f_m$
dominates all other frequencies, so the attacker is able to inject a signal
of the intended frequency into the output of the ADC.
However, the output at both carrier frequencies has strong harmonics
at $2f_m,3f_m,\ldots\si{\hertz}$,
which indicates that the resulting signal is not pure. Moreover, there is
a residual high-frequency component, which is attenuated as the carrier
frequency $f_c$ increases. Finally, there is a frequency-dependent
DC offset caused, in part, by the ESD diodes, while the peak-to-peak
amplitude of the measured signal decreases as the carrier frequency increases.
This is due to the low-pass filtering behavior of the sample-and-hold mechanism,
which also explains why we are only able to demodulate signals
for carrier frequencies until approximately $\SI{150}{\mega\hertz}$.

\begin{figure*}[!t]
  \centering
  \begin{subfigure}[b]{0.45\textwidth}
  \centering
          \includegraphics[width=\textwidth]{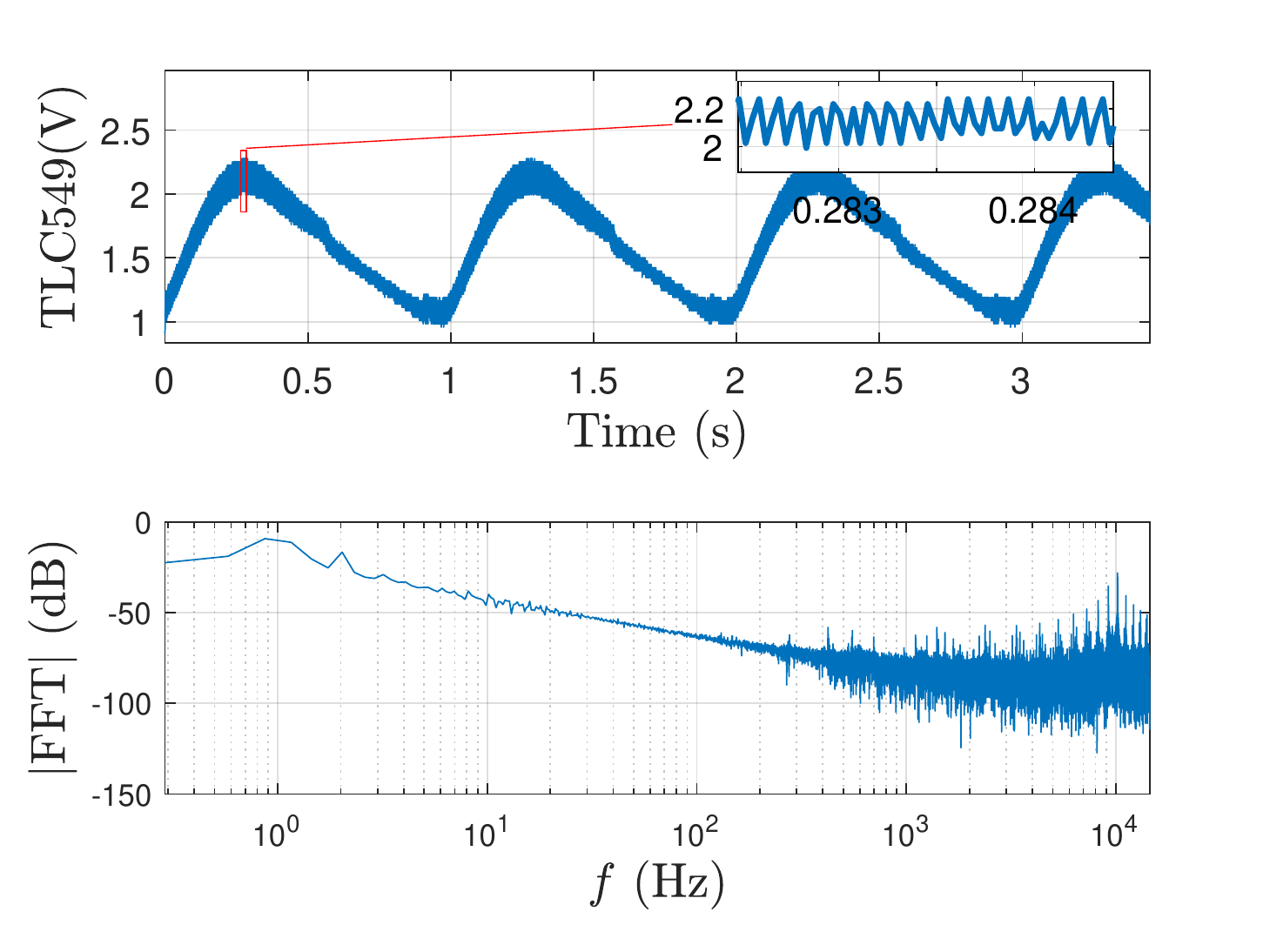}
          \caption{TLC549: $f_c=\SI{80}{\mega\hertz}$, $\mu=0.5$, $f_m=\SI{1}{\hertz}$}
  \label{fig:tlc549_raw}
  \end{subfigure}\hfill
  \begin{subfigure}[b]{0.45\textwidth}
    \centering
            \includegraphics[width=\textwidth]{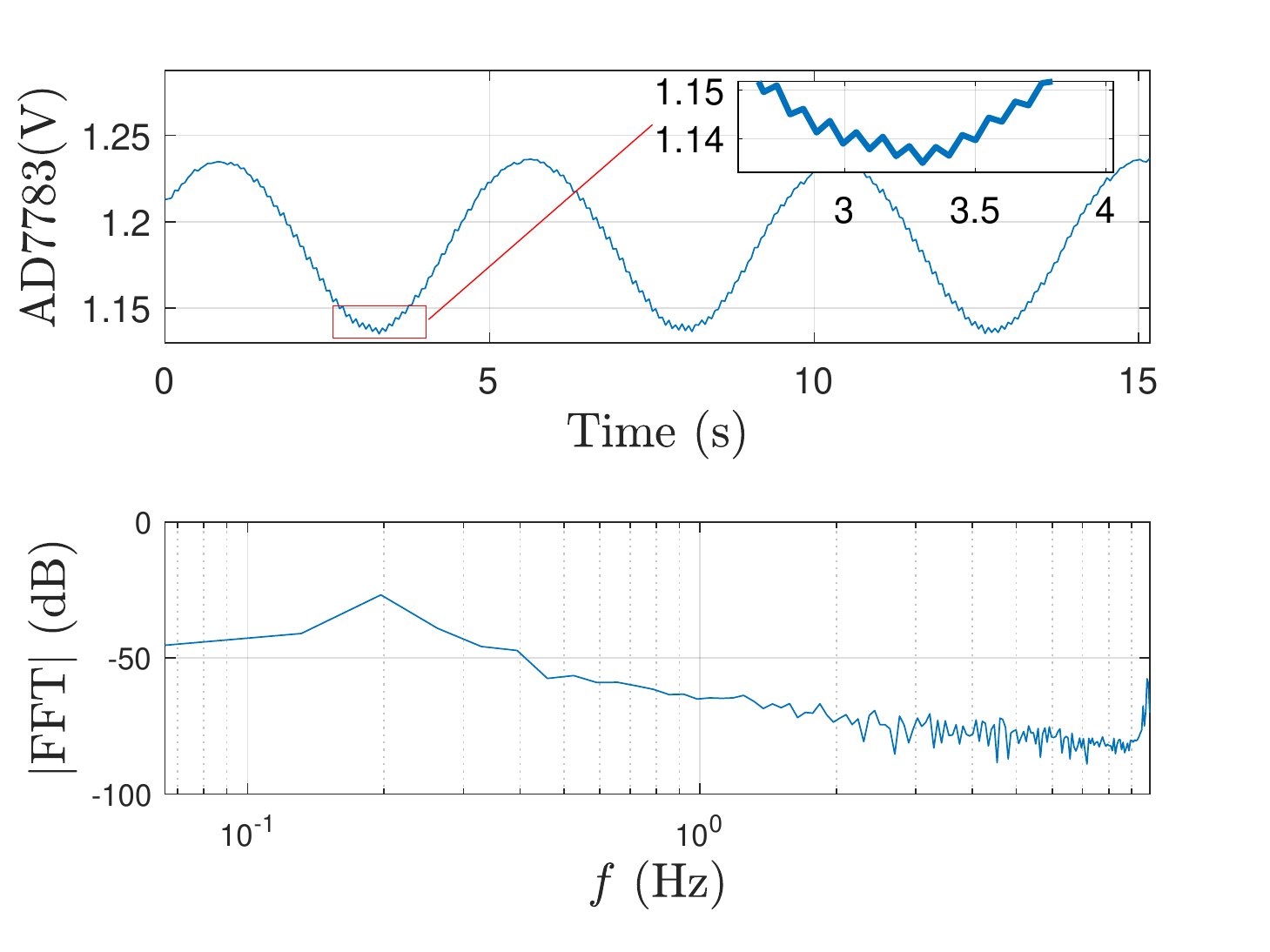}
            \caption{AD7783: $f_c=\SI{40}{\mega\hertz}$, $\mu=1.0$, $f_m=\SI{10}{\hertz}$}
    \label{fig:ad7783_raw}
  \end{subfigure}\hfill
  \caption{Example TLC549 (\subref{fig:tlc549_raw}) and AD7783 (\subref{fig:ad7783_raw})
           outputs for a transmission power of $P=\SI{5}{\decibel m}$.
           Both ADCs demodulate the injected signal, but present harmonics
           and some high-frequency components. The AD7783 signal is aliased.}
  \vspace{-1em}
\end{figure*}

\secpar{TLC549} The TLC549 (Figure~\ref{fig:tlc549_raw}) also demodulates the
injected signal, but still
contains harmonics and a small high-frequency component.

\begin{figure*}[t]
  \centering
  \begin{subfigure}[b]{0.33\textwidth}
  \centering
          \includegraphics[width=\textwidth]{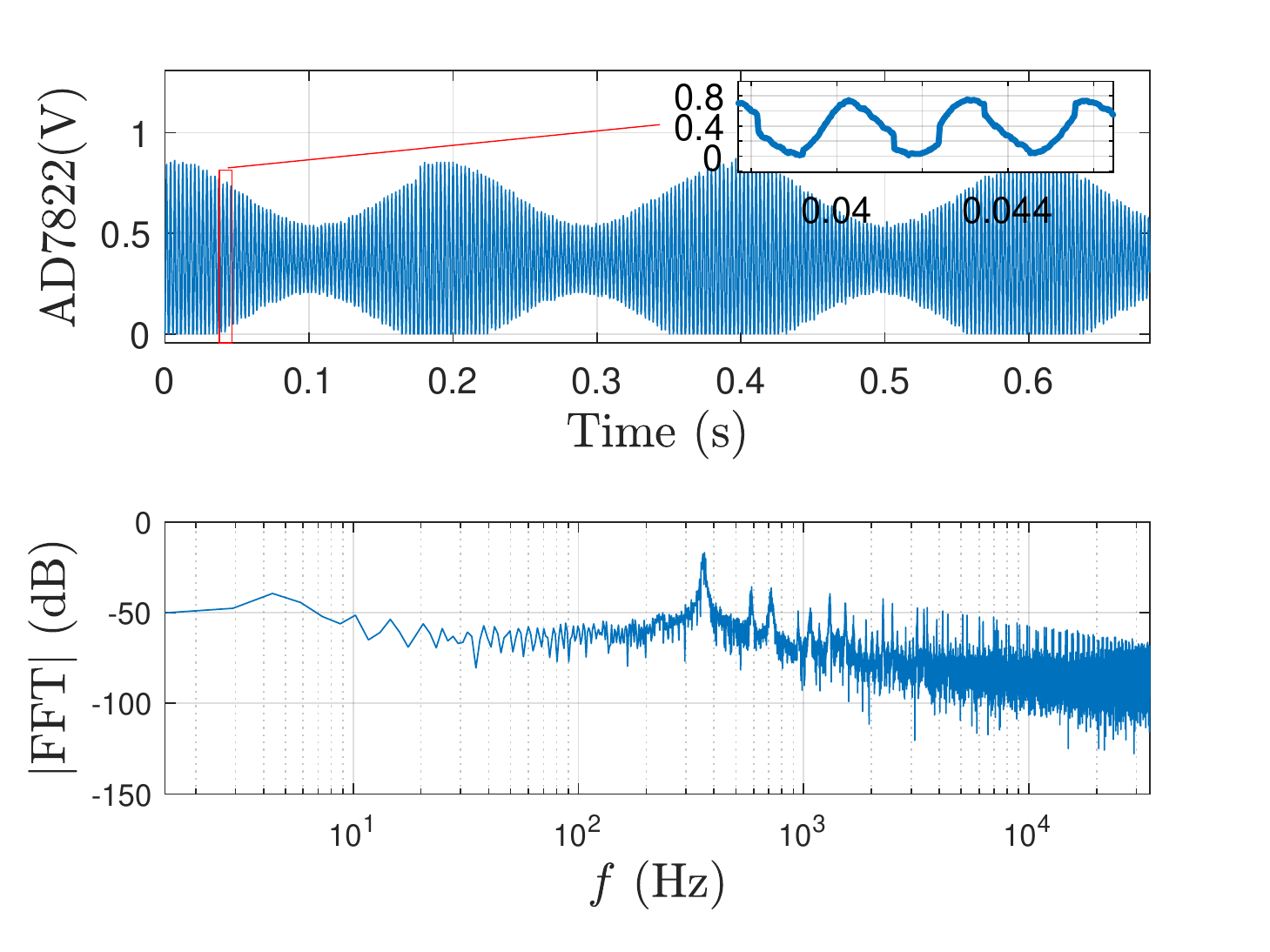}
          \caption{$f_c=\SI{79.4933}{\mega\hertz}$}
  \end{subfigure}\hfill
  \begin{subfigure}[b]{0.33\textwidth}
    \centering
            \includegraphics[width=\textwidth]{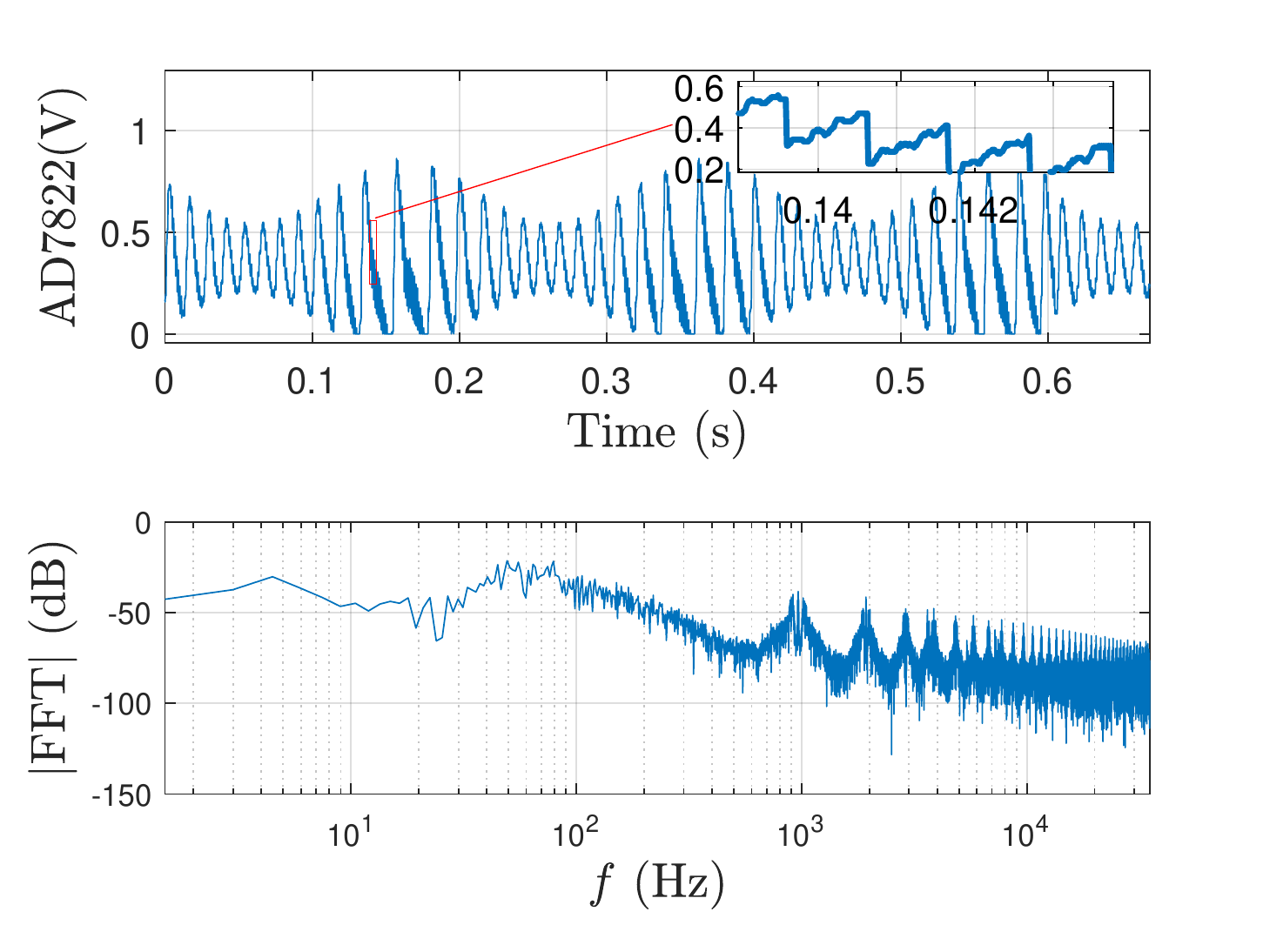}
            \caption{$f_c=\SI{79.4936}{\mega\hertz}$}
  \end{subfigure}\hfill
  \begin{subfigure}[b]{0.33\textwidth}
  \centering
          \includegraphics[width=\textwidth]{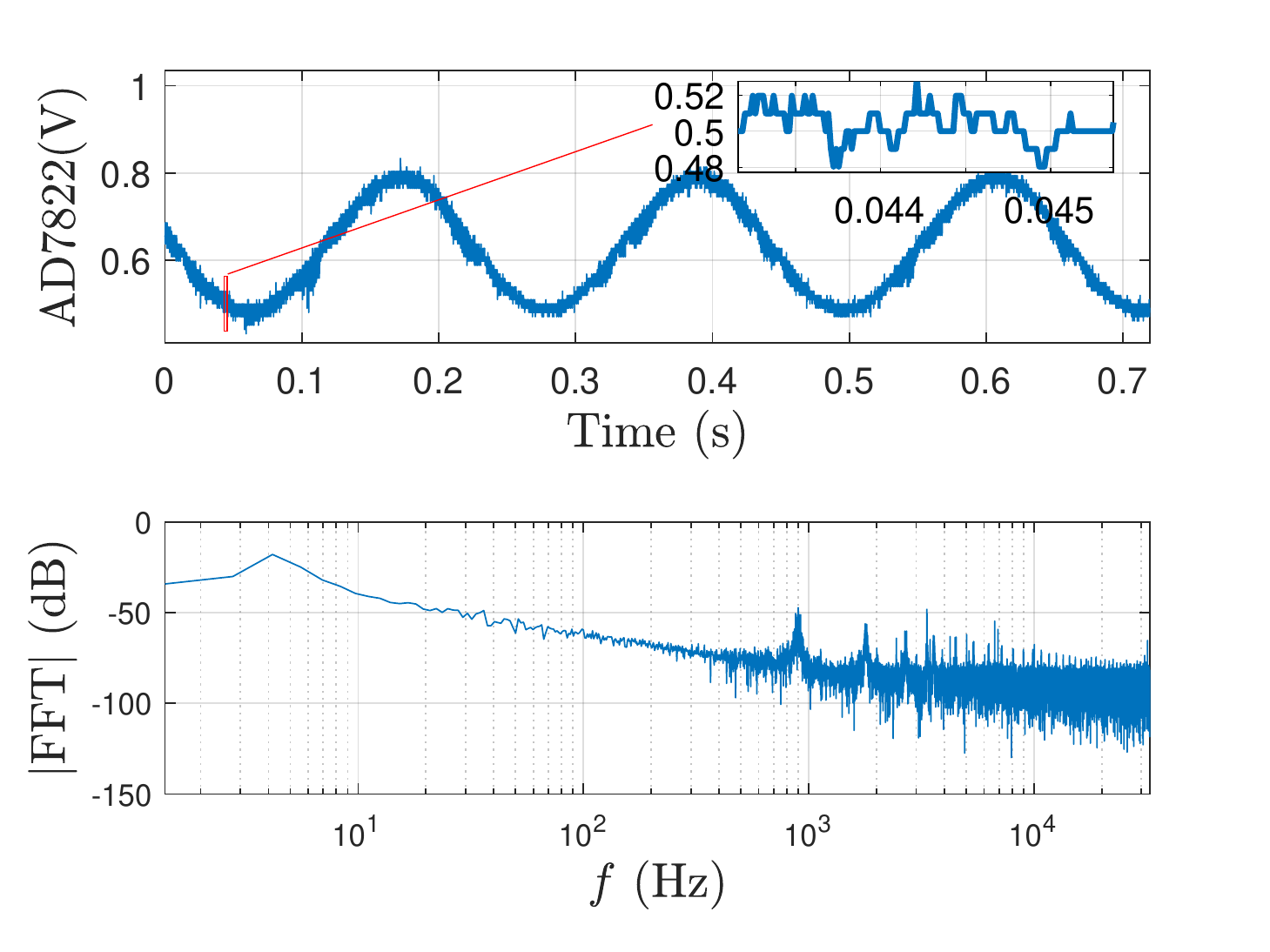}
          \caption{$f_c=\SI{79.4937}{\mega\hertz}$}
  \end{subfigure}\hfill
  \caption{Example AD7822 output for power $P=\SI{-5}{\decibel m}$, signal frequency $f_m=\SI{5}{\hertz}$,
          and depth $\mu=0.5$. Signal demodulation requires a fine-tuned $f_c$.}
  \label{fig:ad7822_tuning}
  \vspace{-2em}
\end{figure*}

\secpar{AD7783} As the AD7783 (Figure~\ref{fig:ad7783_raw}) only has a sampling frequency of
$f_s=\SI{19.79}{\hertz}$, aliasing occurs when the baseband signal exceeds
the Nyquist frequency $f_s/2$. For example, when the baseband frequency is
$f_m=\SI{10}{\hertz}$,
the fundamental frequency dominating the measurements is of frequency
$2f_m-f_s=20-19.79=\SI{0.21}{\hertz}$, with a high-frequency component of
$f_s-f_m=\SI{9.79}{\hertz}$.

\secpar{AD7822, AD7276, Artix7}
The three remaining ADCs contain strong high-frequency components which dominate
the low-frequency signal. Their outputs appear to be AM-modulated,
but at a carrier frequency which is below the ADC's Nyquist frequency.
However, with manual tuning of the carrier frequency, it is possible to
remove this high frequency component, causing the ADC to demodulate
the input. This is shown for the Flash ADC AD7822 in
Figure~\ref{fig:ad7822_tuning}, where we change the carrier frequency $f_c$
in steps of $\SI{100}{\hertz}$.

\secpar{Conclusion} The results of our experiments lead to the following observations:
\begin{enumerate}
  \vspace{-0.5em}
  \item {\bf Generality} -- All 6 ADCs tested are
  vulnerable to signal injections at multiple carrier frequencies, as they
  demodulate signals, matching the theoretical expectations of
  Section~\ref{sec:adc_model}. As the ADCs are of all
  major types and with a range of different resolutions and sampling frequencies,
  the conclusions drawn should also be valid for other ADC chips.
  \item {\bf Low-Pass Filter} -- Although all ADCs exhibited low-pass
  filtering characteristics,
  the maximum vulnerable carrier frequency for a given power level was multiple
  times the cut-off frequency of the $RC$ circuit at the input of the ADC.
  This extended the frequency range that an attacker could use for
  transmissions to attack the system.
  \item {\bf Power} -- The adversary needs to select the power level of
  transmissions carefully: too much power in the input of the ADC can cause
  saturation and/or clipping of the measured signal. Too little power, on the
  other hand, results in output that looks like noise or a zero signal.
  \item {\bf Carrier Frequency} -- Some ADCs were vulnerable at any
  carrier frequency that is not severely attenuated by the sample-and-hold
  mechanism. For others, high-frequency components dominated the intended
  baseband signal of frequency $f_m$ in the ADC output for most frequencies.
  Even then, carefully-chosen carrier frequencies
  resulted in a demodulated ADC output.
  \vspace{-1.5em}
\end{enumerate}

\section{Discussion}
\label{sec:discussion}

We now discuss how our work can inform design choices.
To start, choosing the right ADC directly impacts
the susceptibility to signal injection attacks. As shown in Section~\ref{sec:adcs},
some ADCs distort the demodulated output and result in more sawtooth-like
output, making them more resilient to clean sinusoidal injections. Moreover,
other ADCs require fine-grained control over the carrier frequency of
injection. As the adversarial signal is transformed
through the circuit-specific transfer function $H_C$, the adversary may
not have such control, resulting in a more secure system.

Having chosen the appropriate ADC based on cost, performance, security, or
other considerations, a designer needs to assess the impact of $H_C$.
Prior work has shown that even small layout or component changes
affect the EMI behavior of a
circuit~\cite{emi_susceptibility,emi_predict,statistics_demodulation_rfi}.
Since the ADC behavior can be independently determined through direct power injections,
fewer experiments with remote transmissions are required to evaluate the
full circuit behavior and how changes in the circuit's topology
influence the system's security.

Our selective security definition and algorithm
address how to determine the vulnerability of a system against specific waveforms.
Universal security, on the other hand, allows us to directly compare
the security of two systems for a fixed adversarial voltage budget through their critical universal
security thresholds. Moreover, given a probability/threshold
$\epsilon$, we can calculate the critical universal voltage level, which is
the maximum output level for which a system is still universally $\epsilon$-secure.

Our smartphone case study showed that our framework can be used in practice
with real systems, while our ``OK Google'' experiments
demonstrated that less-than-perfect injections of
adversarial waveforms can have the same effect as perfect injections.
This is because there is a mismatch between the true noise level of a system
and the worst-case noise level that the system expects. In other
words, injections worked at all carrier frequencies, even when
the demodulated output was noisy or distorted.
This is a deliberate, permissive design decision, which allows the
adversary to succeed with a range of different and noisy waveforms $w(t)$,
despite small amplitudes and DC offsets.

Although not heavily discussed in this paper, our model and definitions are
general enough to capture alternative signal
injection techniques. For instance,
electro-mechanical sensors have resonant frequencies which allow acoustic
injection attacks~\cite{drones,injected_actuation}. $H_C$ can
account for such imperfections in the sensors themselves, attenuating
injection frequencies which are not close to the resonant frequencies.
Our system model also makes it easy to evaluate countermeasures and
defense mechanisms in its context. For example, shielding increases
the attenuation factor of $H_C$, thereby
increasing the power requirements for the adversary
(Section~\ref{sec:adc_model}).
Alternatively, a low-pass
filter (LPF) before the ADC and/or amplifier changes $H_A$, and attenuates the
high-frequency components which would induce non-linearities.
Note, however, that even
moving the pre-amplifier, LPF and ADC into the same IC package does not fully
eliminate the vulnerability to signal injection attacks
(Section~\ref{sec:related}) as the channel between
the analog sensor and the ADC cannot be fundamentally authenticated.

\section{Related Work}
\label{sec:related}

Ever since a 2013 paper by Kune et al. showed that
electromagnetic (EM) signals can be used to cause medical devices to
deliver defibrillation shocks~\cite{ghost},
there has been a rise in EM, acoustic, and
optical signal injection attacks against sensor and actuator
systems~\cite{sok}. Although some papers have focused on vulnerabilities caused by
the ADC sampling process itself~\cite{ics_adc_attack}, others have focused
on exploiting the control algorithms that make use of the digitized signal.
For example, Shoukry et al. showed how to force the Anti-Lock Braking
Systems (ABS) to model the real input signal as a disturbance~\cite{abs}.
Selvaraj et al. also used the magnetic field to perform attacks on actuators,
but further explored the relationship between frequency and the average injected voltage
into ADCs~\cite{induction_embedded}. By contrast, our paper primarily focused on
a formal mathematical framework to understand security in the context of
signal injection attacks, but also investigated the demodulation properties of
different ADCs.

Our work further highlighted how to use the introduced algorithm and definitions to
investigate the security of a smartphone, complementing earlier work which
had shown that AM-modulated electromagnetic transmissions can be picked up
by hands-free headsets to trigger voice commands in smartphones~\cite{smartphone_iemi}.
Voice injection attacks can also be achieved
by modulating signals on ultrasound frequencies~\cite{dolphin}, or by playing two tones at
different ultrasound frequencies and exploiting non-linearities in
components~\cite{backdoor_microphone}.
Acoustic transmissions at a device's resonant frequencies can also
incapacitate~\cite{drones} or precisely control~\cite{walnut} drones, with
attackers who account for sampling rate drifts being able to control the outputs
of accelerometers for longer periods of time~\cite{injected_actuation}.
Moreover, optical attacks can be used to spoof medical infusion pump
measurements~\cite{pump}, and cause autonomous cars and unmanned aerial
vehicles (UAVs) to drift or fail~\cite{uavs,lidar,autonomous_vehicles}.

It should be noted that although the literature has primarily focused on signal
injection attacks, some works have also proposed countermeasures. These defense
mechanisms revolve around better sampling techniques, for example by adding
unspoofable physical and computational delays~\cite{pycra}, or by oversampling
and selectively turning the sensors off using a secret
sequence~\cite{zhang_detection_2020}.

Overall, despite the extensive literature on signal injection attacks and defenses,
the setup and effectiveness of different works is often reported in an inconsistent way,
making their results hard to compare~\cite{sok}. Our work, recognizing this gap,
introduced a formal foundation to define and quantify security against signal injection attacks,
working towards unifying the reporting methodology for competing works.

\section{Conclusion}
\label{sec:conlusion}

Sensors guide many of our choices, and we often blindly
trust their values. However, it is possible to spoof
their outputs through electromagnetic or other signal injection attacks. To
address the lack of a unifying framework
describing the susceptibility of devices to such attacks, we defined a system and
adversary model for signal injections. Our model is the first to abstract away from
specific environments and circuit designs and presents a strong adversary
who is only limited by transmission power.
It also makes it easy to discuss and evaluate countermeasures in its
context and covers different types of signal injection attacks.

Within our model, we defined existential, selective, and universal security,
capturing effects ranging from mere
disruptions of the ADC outputs to precise injections of all waveforms.
We showed that our definitions can be used to evaluate the 
security level of an off-the-shelf smartphone, and introduced an algorithm to
calculate ``critical'' thresholds, which express how close an
injected signal is to the ideal signal. Finally, we characterized
the demodulation characteristics of commercial ADCs to malicious injections.
In response to the emerging signal injection threat,
our work paves the way towards a future where security can be quantified
and compared through our methodology and security definitions.

\appendix

\section{Additional Experiments with ADCs}
\label{app:exp}

This appendix contains further measurements on the demodulation properties
of ADCs. Section~\ref{sec:app:metric} precisely defines the similarity metric
of Section~\ref{sec:case_study}, and validates the experimental setup.
Section~\ref{sec:app:mic} and~\ref{sec:app:atmega} then
conduct further characterization experiments of the smartphone microphone
and ATmega328P ADC respectively. Finally, Section~\ref{sec:app:others}
contains additional examples of the demodulation characteristics of the
remaining ADCs.

\subsection{Similarity Metric and Setup Validation}
\label{sec:app:metric}

\begin{figure}[!t]
  \centering
  \begin{subfigure}[b]{0.33\textwidth}
    \centering
      \includegraphics[width=\textwidth]{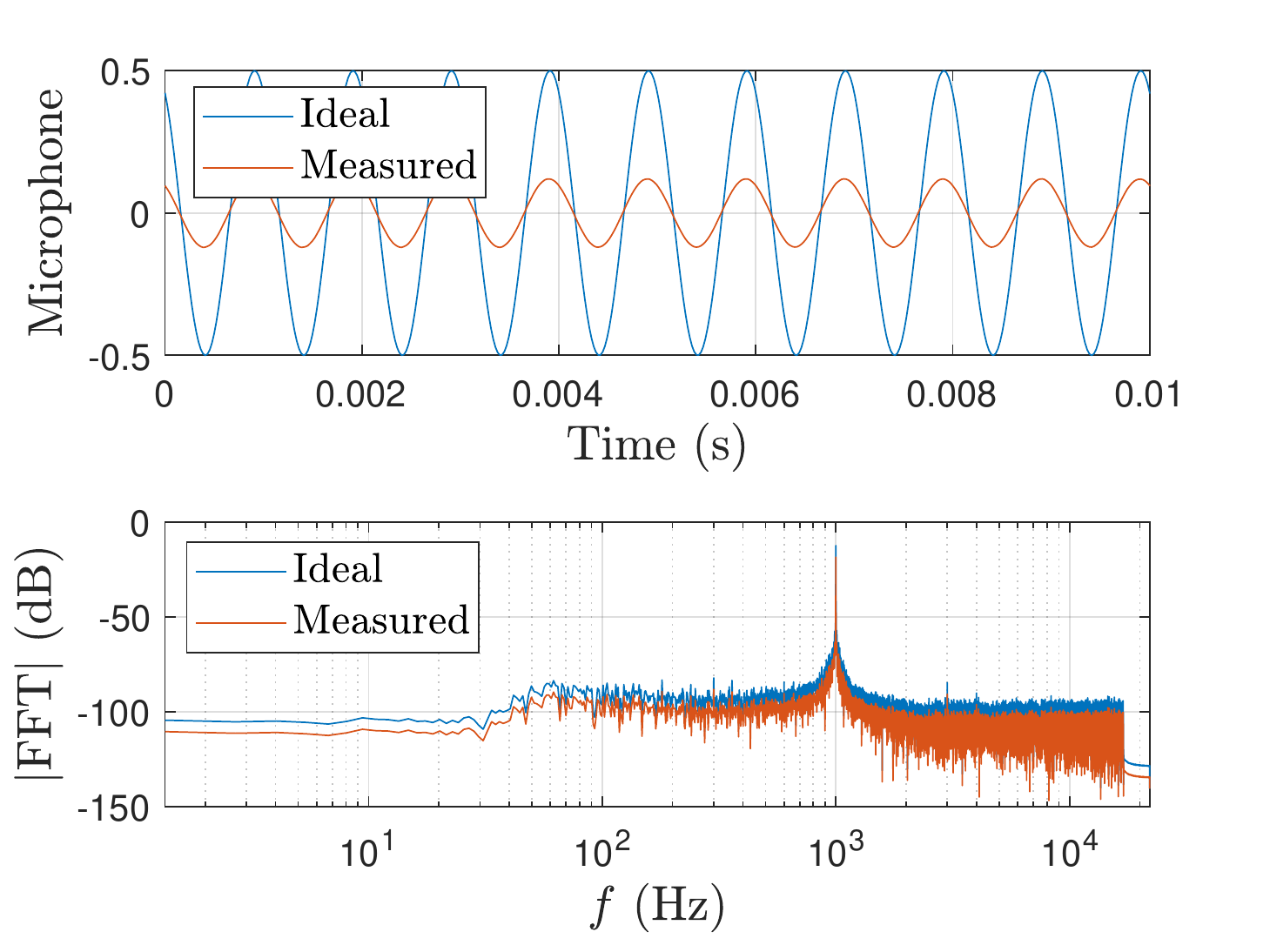}
      \caption{$f_m$ (Ideal \& Microphone)}
  \label{fig:app:ideal_ref}
\end{subfigure}\hfill
  \begin{subfigure}[b]{0.33\textwidth}
    \centering
      \includegraphics[width=\textwidth]{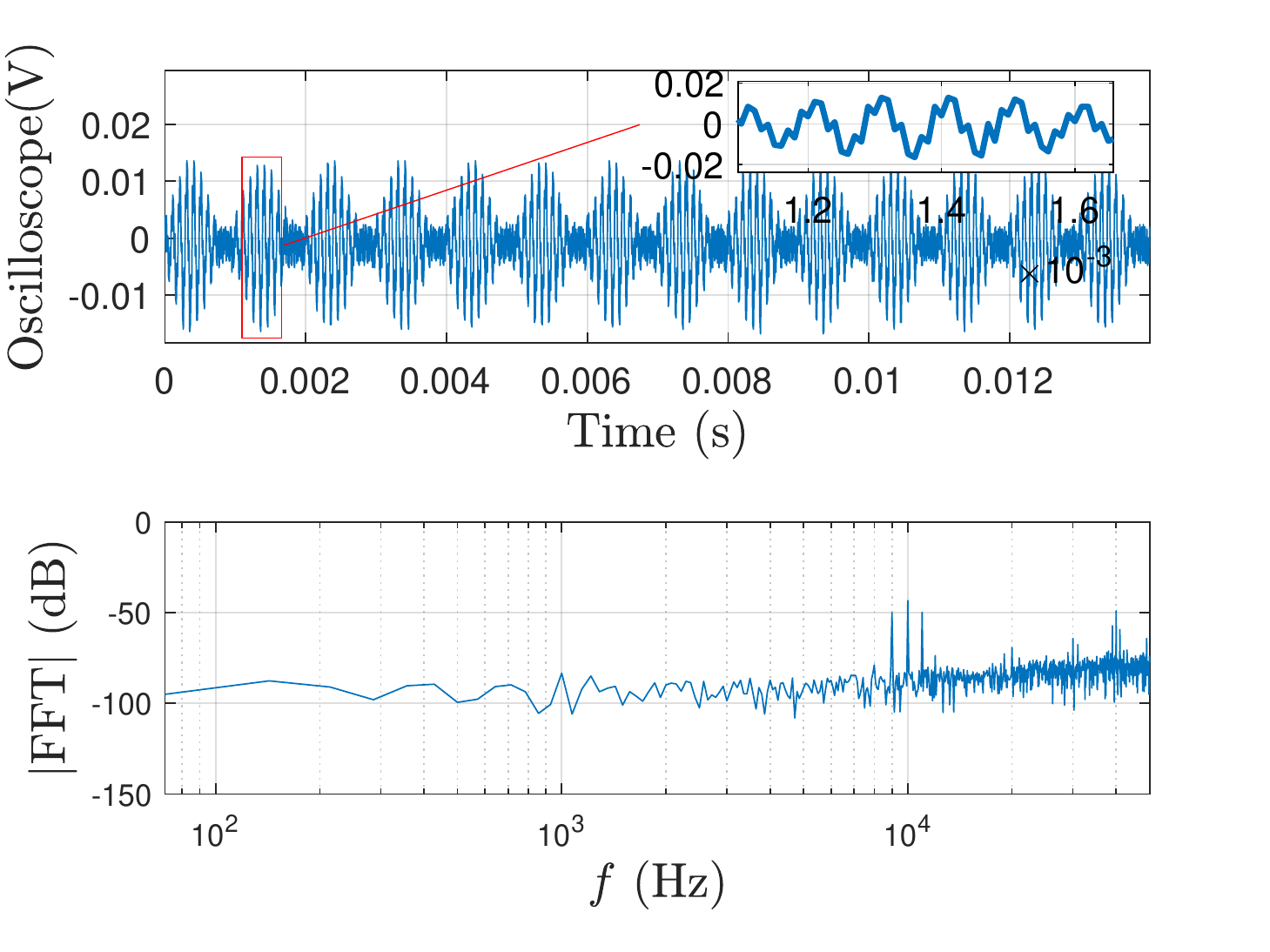}
      \caption{$f_m \times f_c$ (Oscilloscope)}
      \label{fig:app:osc_ref}
  \end{subfigure}\hfill
  \begin{subfigure}[b]{0.33\textwidth}
  \centering
    \includegraphics[width=\textwidth]{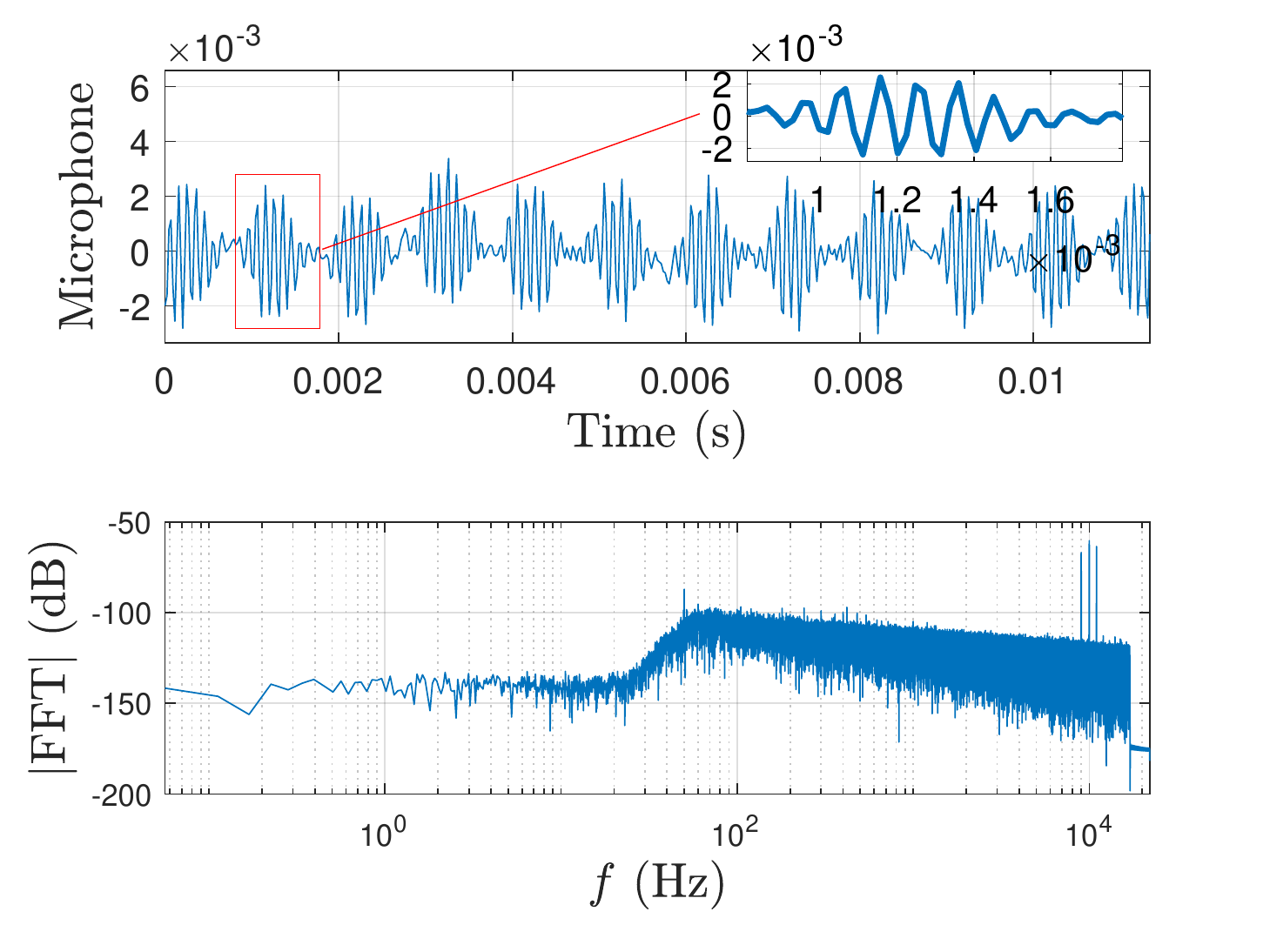}
    \caption{$f_m \times f_c$ (Microphone)}
  \label{fig:app:mic_ref}
  \end{subfigure}\hfill
  \caption{Unmodulated and modulated injections measured by
           an oscilloscope and a smartphone microphone, with RMS voltage
           $\vra=\SI{20}{\milli\volt}$, signal frequency $f_m=\SI{1}{\kilo\hertz}$,
           carrier frequency $f_c=\SI{10}{\kilo\hertz}$, and modulation depth $\mu=1.0$.}
  \label{fig:app:ref}
\end{figure}

The experiments of Section~\ref{sec:phone_security} required
an independent metric to evaluate how ``similar'' two signals are
as a way of independently validating the security definitions of Section~\ref{sec:definitions}.
The metric proposed for this task is based on the Pearson Correlation Coefficient (PCC),
which is commonly found in signal-alignment and optimization
applications~\cite{pearson_noise,pearson_alignment}. It is defined
as the covariance of two variables divided by the product of their
standard deviations:
\begin{equation}\label{eq:app:pcc}
  \rho\left(X,Y\right) = \frac{\text{cov}\left(X,Y\right)}{\sigma_X\sigma_Y}=\frac{\sum_{i=1}^{n}\left(x_i-\mu_x\right)\left(y_i-\mu_y\right)}{\sqrt{\sum_{i=1}^{n}\left(x_i-\mu_x\right)^2}{\sqrt{\sum_{i=1}^{n}\left(y_i-\mu_y\right)^2}}}
\end{equation}
It is a suitable metric because it removes the mean value of the signals
(DC shift), as well as the effects of scaling (related to transmission power).
In other words, $\rho\left(X, aX+b\right)=1$ for a variable $X$ and scalars $a,b$.
However, the PCC is sensitive to signal alignment.
To overcome this issue, the phase (time) offset between two signals $s_a$, $s_b$
can be found using cross-correlation. Specifically, the signals are
aligned when the cross-correlation coefficient is maximized:
\begin{equation}\label{eq:app:lag}
  \text{lag}\left(s_a, s_b\right) = \argmax\limits_{n} \left(\left(s_a \star s_b\right)\left(n\right)\right)
\end{equation}
Using Equations~\eqref{eq:app:pcc} and~\eqref{eq:app:lag}, the
similarity metric between the measured signal $\tsf(t)$ and the ideal
signal $w(t)$ can be defined as follows:
\begin{equation}\label{eq:app:similarity}
  \text{similarity}\left(\tsf, w\right) = \rho\left(\tsf, w^{lag}\right)
\end{equation}
To sanity-check this metric and the experimental setup, an unmodulated \SI{20}{\milli\volt}
$f_m=\SI{1}{\kilo\hertz}$ signal is generated. Figure~\ref{fig:app:ideal_ref} shows this waveform as
measured by the smartphone of Section~\ref{sec:phone_security} along
with an ideal $\SI{1}{\kilo\hertz}$ signal. Even though the amplitudes are different,
the frequency responses of the measured and the ideal signal are almost identical,
with the two signals having a similarity of $0.9991$ according to
Equation~\eqref{eq:app:similarity}.

Figures~\ref{fig:app:osc_ref} and~\ref{fig:app:mic_ref} additionally show
the same $f_m=\SI{1}{\kilo\hertz}$ signal modulated on a carrier frequency of
$f_c=\SI{10}{\kilo\hertz}$ at a depth of $\mu=1.0$. This carrier frequency was
chosen as it is within the Nyquist range of the smartphone ADC (sampling frequency
$f_s=\SI{44.1}{\kilo\hertz}$). Figure~\ref{fig:app:osc_ref} contains
measurements taken by a Rigol DS2302A oscilloscope with a timescale
division of $\SI{500}{\micro\second}$, while Figure~\ref{fig:app:mic_ref} uses
the smartphone microphone.

Unlike the examples of Section~\ref{sec:case_study}, the measurements shown in
Figure~\ref{fig:app:ref} do not exhibit harmonics, but rather high-frequency
components at $f_c$ and $f_c\pm f_m$, as expected. Consequently, the demodulation
characteristics are due to non-linearities in amplifiers and ADCs when
used outside of their intended range, instead of the experimental setup.

\subsection{Smartphone Microphone Properties}
\label{sec:app:mic}

\begin{figure}[!t]
  \centering
  \begin{subfigure}[!t]{0.49\textwidth}
  \centering
          \includegraphics[width=\textwidth]{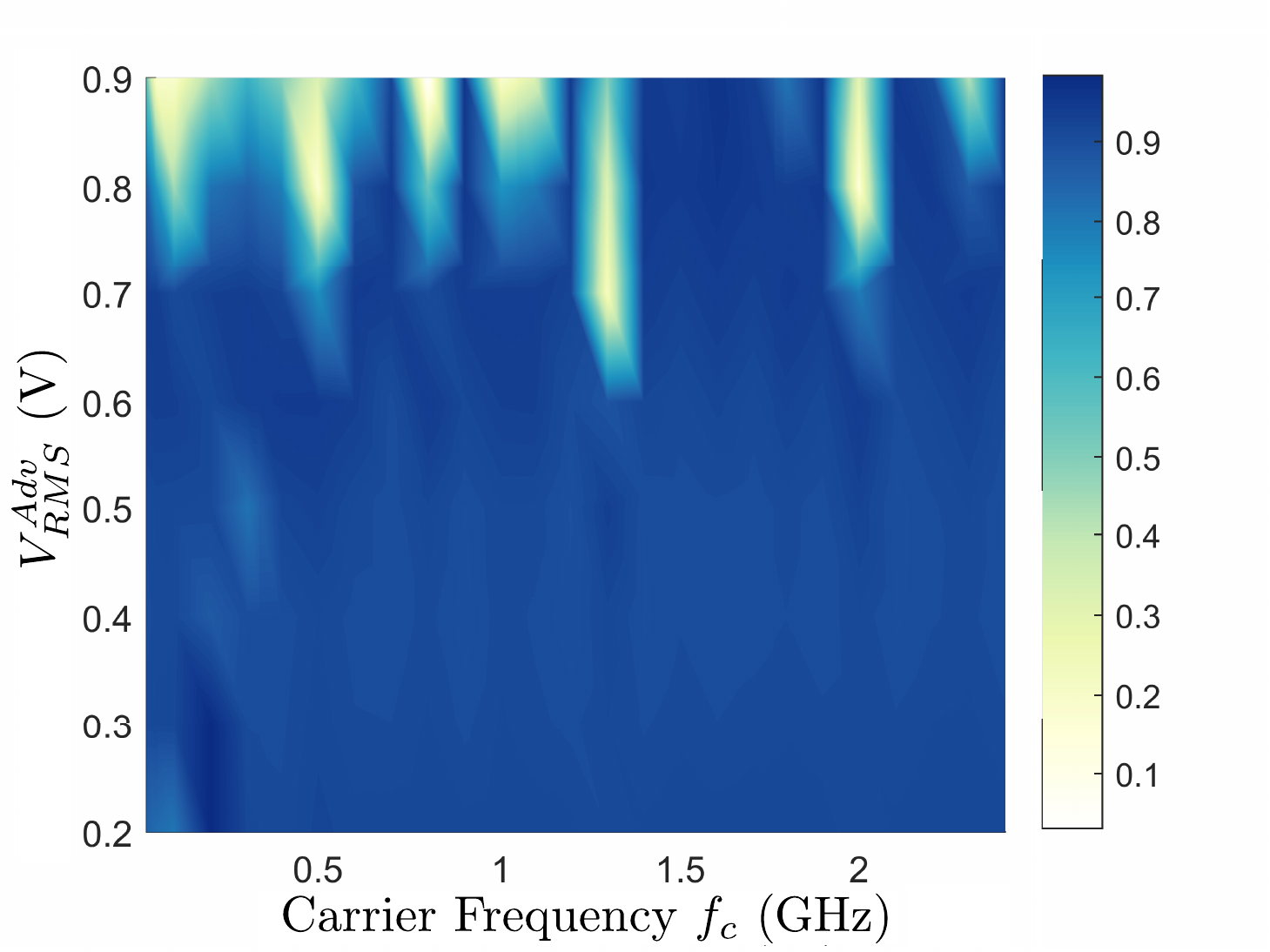}
          \caption{Similarity to the ideal signal}
          \label{fig:app:phone_similarity}
  \end{subfigure}\hfill
  \begin{subfigure}[!t]{0.49\textwidth}
    \centering
            \includegraphics[width=\textwidth]{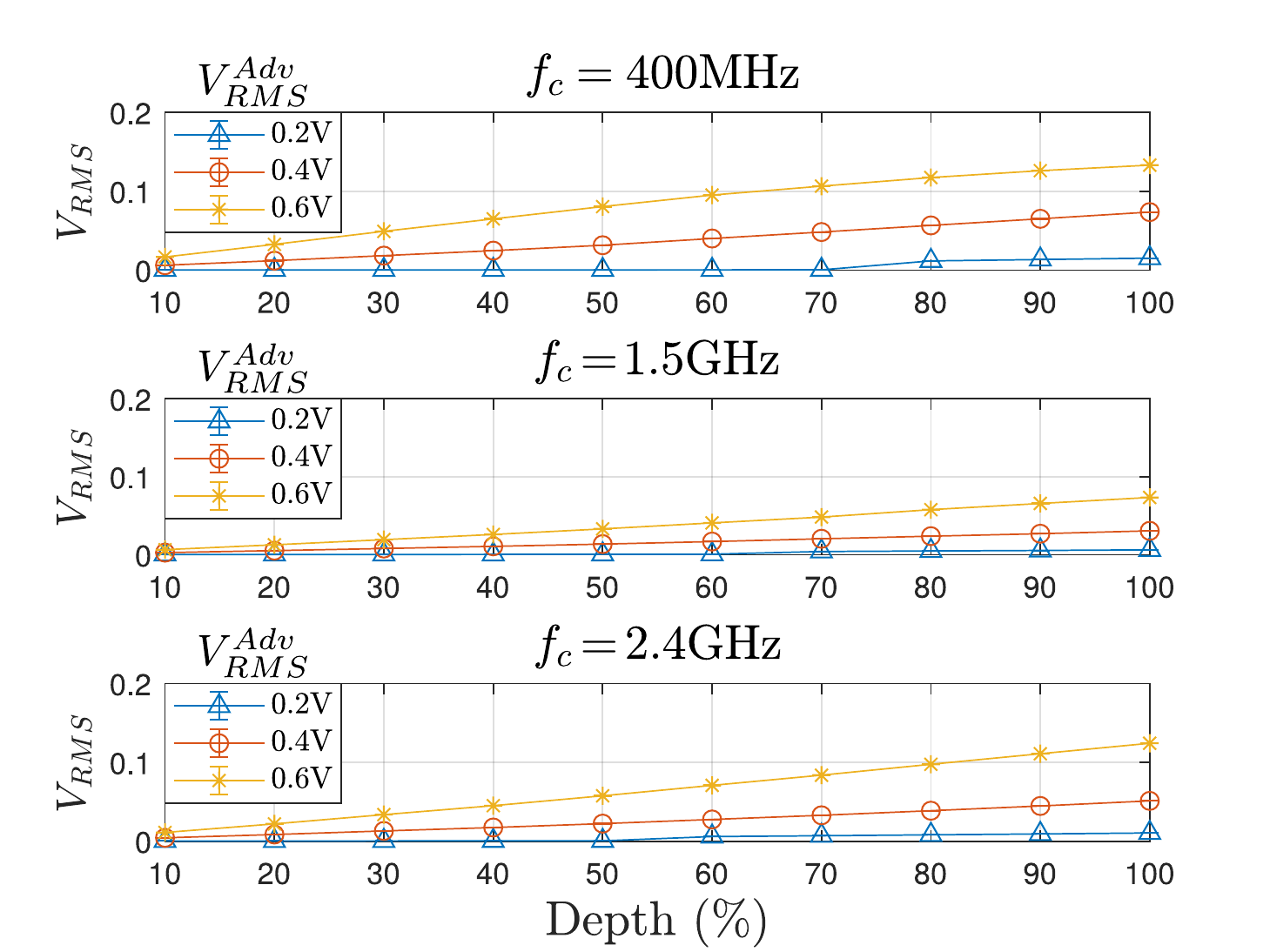}
            \caption{$\vrms$ for different $\mu$}
            \label{fig:app:phone_depth}
    \end{subfigure}\hfill
    \begin{subfigure}[!t]{0.49\textwidth}
      \centering
              \includegraphics[width=\textwidth]{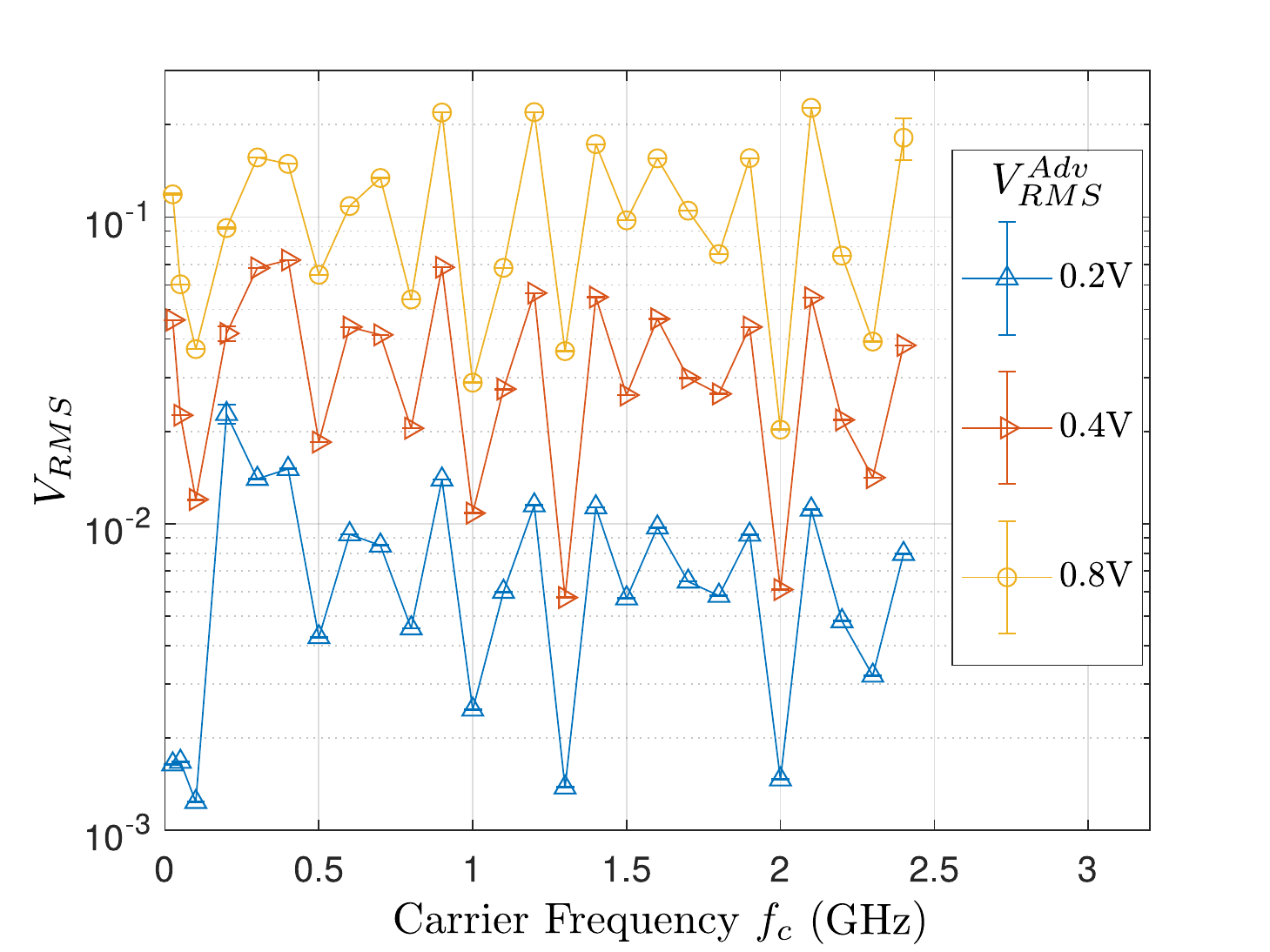}
              \caption{$\vrms$ for different $\vra$}
              \label{fig:app:phone_absolute}
    \end{subfigure}\hfill
  \begin{subfigure}[!t]{0.49\textwidth}
  \centering
          \includegraphics[width=\textwidth]{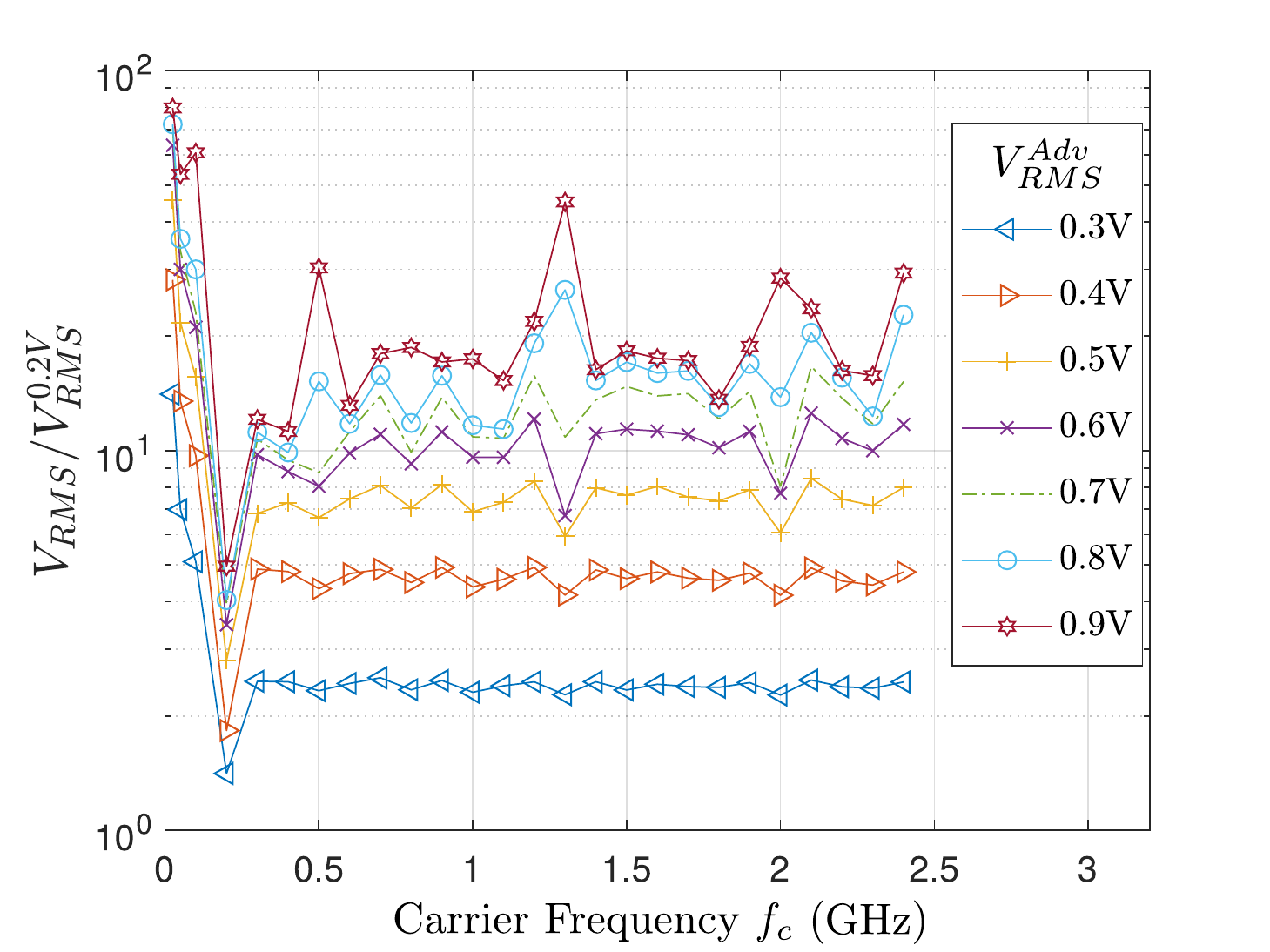}
          \caption{$\vrms$ relative to $\vra=\SI{0.2}{\volt}$}
          \label{fig:app:phone_relative}
  \end{subfigure}\hfill
  \caption{Results of amplitude-modulated $f_m=\SI{1}{\kilo\hertz}$ injections
           into the smartphone microphone for different carrier frequencies $f_c$,
           depths $\mu$, and output voltage levels $\vra$. (\subref{fig:app:phone_similarity})
           shows the similarity of the measured output compared to the ideal signal.
           (\subref{fig:app:phone_depth}) and~(\subref{fig:app:phone_absolute})
           illustrate the received RMS voltages $\vrms$ for different modulation
           depths $\mu$ and $\vra$ respectively. Finally, (\subref{fig:app:phone_relative})
           shows $\vrms$ relative to $\vra=\SI{0.2}{\volt}$.}
  \label{fig:app:phone}
\end{figure}

This section characterizes the smartphone microphone through
the direct injection methodology of Section~\ref{sec:adcs}.
An $f_m=\SI{1}{\kilo\hertz}$ tone is amplitude-modulated with a depth
of $\mu=1.0$ on the following carrier frequencies $f_c$:
$\SI{25}{\mega\hertz}$, $\SI{50}{\mega\hertz}$, and $\SIrange{0.1}{2.4}{\giga\hertz}$
at a step of $\SI{100}{\mega\hertz}$. The RMS output level
$\vra=\vma/\sqrt{2}$ of the signal generator is also varied
between $\SIrange{0.2}{0.9}{\volt}$ at a step of $\SI{100}{\milli\volt}$.

The similarity of the measured to the ideal signal for various
carrier frequencies $f_c$ and output levels $\vra$ is shown in
Figure~\ref{fig:app:phone_similarity}. It is consistently high for all
frequencies when $\SI{0.2}{\volt}\le \vra\le\SI{0.6}{\volt}$, while higher voltage
levels lead to more pronounced harmonics and clipping, reducing the similarity.
The results are consistent across measurements: the $99\%$ confidence interval
of the similarity is always below $\pm0.0005$, except for the $(f_c, \vra)$ pairs
$(\SI{300}{\mega\hertz},\SI{0.5}{\volt})$ and
$(\SI{2.4}{\giga\hertz},\SI{0.9}{\volt})$, where it reaches $0.035$.

According to work on amplifier properties~\cite{statistics_demodulation_rfi},
the measured RMS voltage level $\vrms$,
the input voltage $\vra$, the modulation depth $\mu$, the signal frequency $f_m$,
and the carrier frequency $f_c$ satisfy the following relationship:
\begin{equation}\label{eq:app:amp_demod}
  \vrms = \frac{\sqrt{2}}{2}\mu\left(\vra\right)^2\left| H_2\left(f_c, -\left(f_c-f_m\right)\right)\right|
\end{equation}
where $H_2$ is a second-order transfer function. Figure~\ref{fig:app:phone}
mostly confirms this equation for the microphone and ADC subsystem
of the smartphone.

Specifically, fixing $\mu$ and $f_m$ in Equation~\eqref{eq:app:amp_demod}
suggests that $\vrms^{V_1}/\vrms^{V_2}=(V_1/V_2)^2$ across all carrier
frequencies $f_c$. Indeed, Figure~\ref{fig:app:phone_depth} verifies that the received
RMS voltage $\vrms$ is linear in the modulation depth $\mu$ for fixed
$\vra\in\{0.4,0.6\}\si{\volt}$ with $R^2>0.97$. For $\SI{0.2}{\volt}$,
the relationship becomes linear after $\mu\approx0.8$, as the measured $\vrms$
is approximately $0$ below that.

Figure~\ref{fig:app:phone_absolute} illustrates that the transfer function for the
$\vrms$ response is frequency-dependent. Finally, Figure~\ref{fig:app:phone_relative} shows
$\vrms$ relative to $\vra=\SI{0.2}{\volt}$. For $\SI{0.3}{\volt}\le \vrms \le\SI{0.6}{\volt}$
and $f_c\ge\SI{100}{\mega\hertz}$, there is a linear relationship between
the carrier frequency and $\vrms/\vrms^{\SI{0.2}{\volt}}$, as
predicted by Equation~\eqref{eq:app:amp_demod}.

\begin{figure}[!t]
    \centering
    \begin{subfigure}[!t]{0.49\textwidth}
    \centering
            \includegraphics[width=\textwidth]{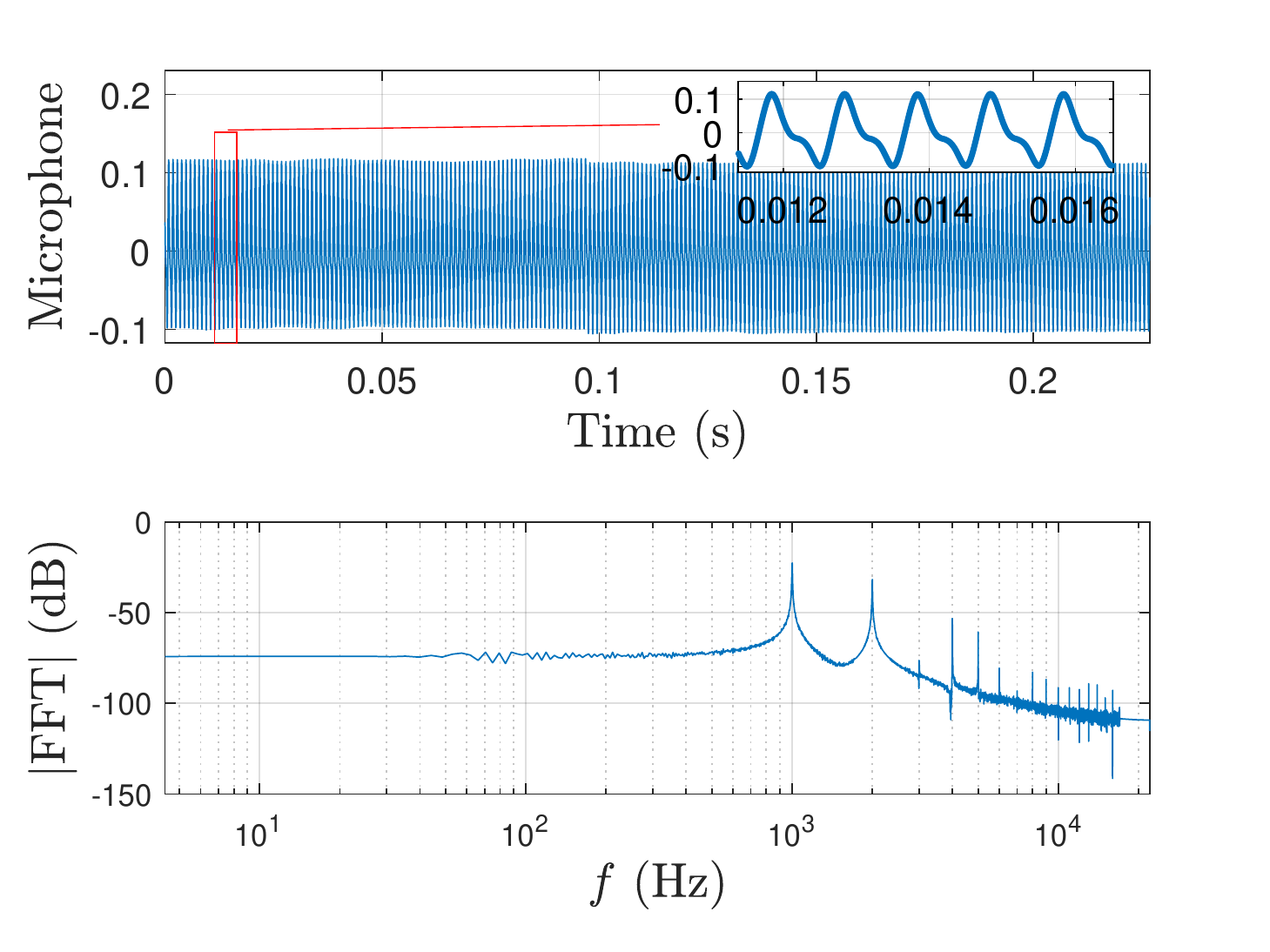}
            \caption{$\vra=\SI{0.6}{\volt}$, $f_c=\SI{1.5}{\giga\hertz}$}
    \label{fig:app:mic_06_15}
    \end{subfigure}\hfill
    \begin{subfigure}[!t]{0.49\textwidth}
      \centering
              \includegraphics[width=\textwidth]{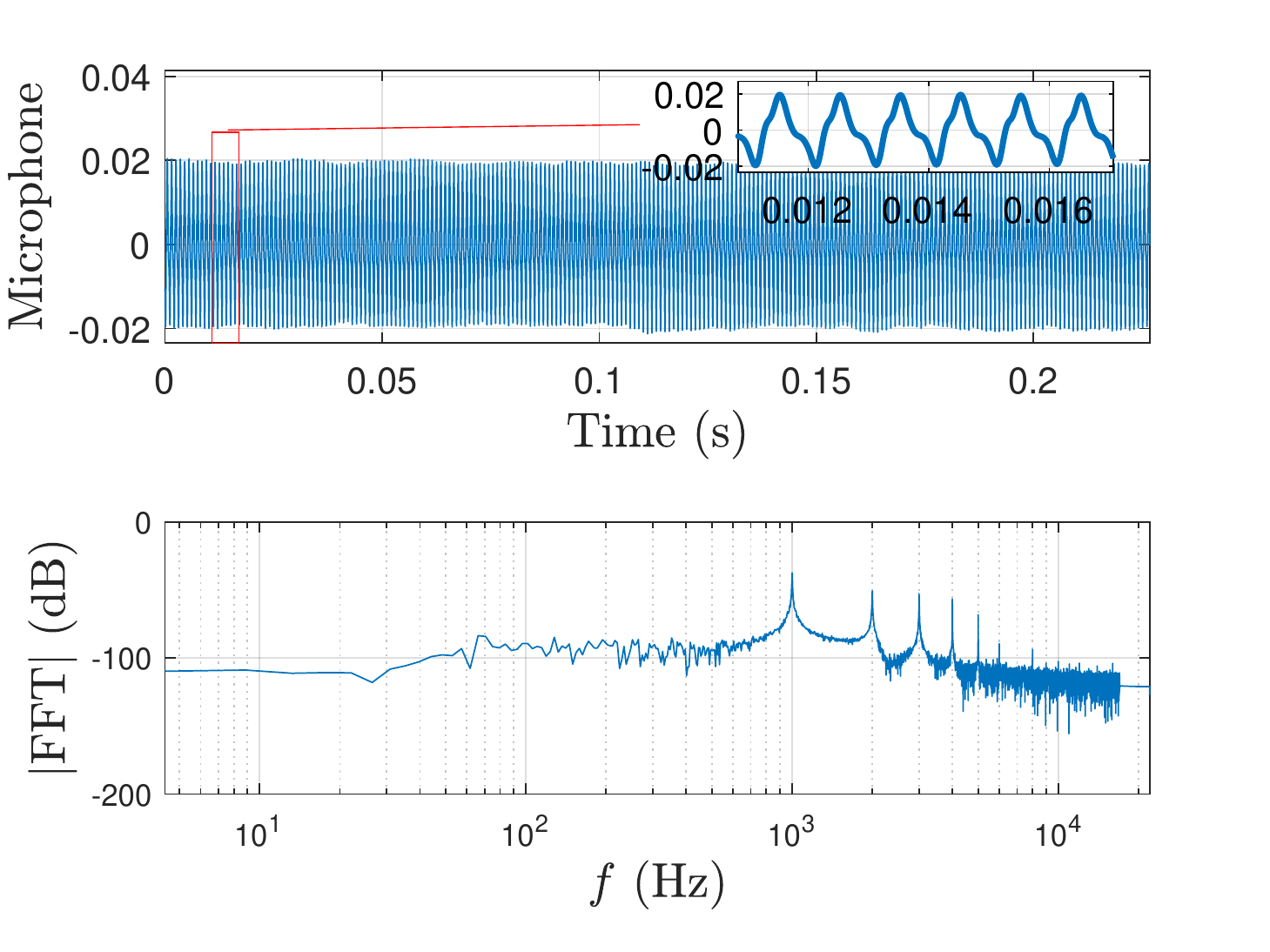}
              \caption{$\vra=\SI{0.6}{\volt}$, $f_c=\SI{2.0}{\giga\hertz}$}
    \label{fig:app:mic_06_20}
    \end{subfigure}\hfill
  \begin{subfigure}[!t]{0.49\textwidth}
      \centering
              \includegraphics[width=\textwidth]{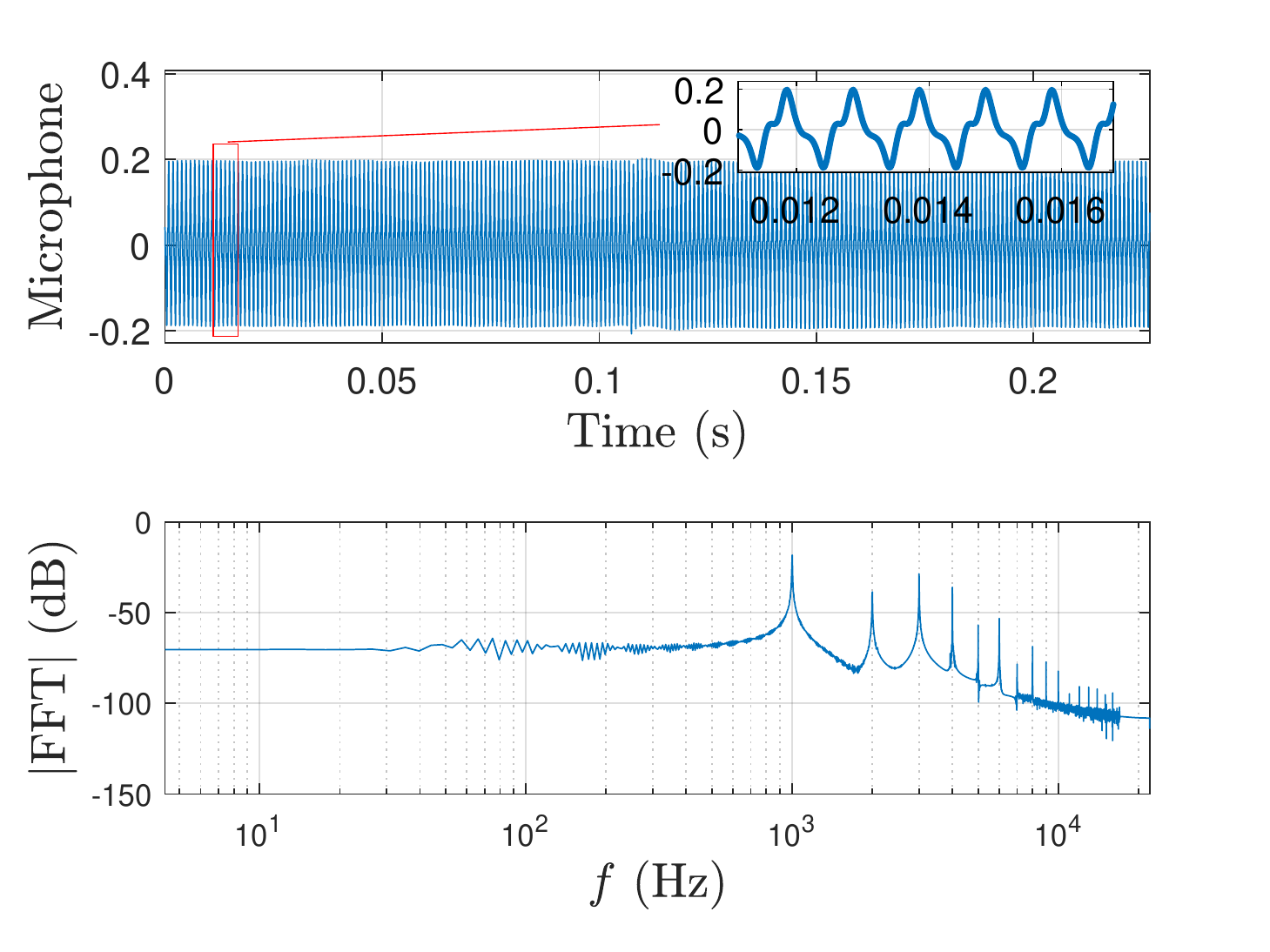}
              \caption{$\vra=\SI{0.9}{\volt}$, $f_c=\SI{1.5}{\giga\hertz}$}
    \label{fig:app:mic_09_15}
    \end{subfigure}\hfill
    \begin{subfigure}[!t]{0.49\textwidth}
    \centering
            \includegraphics[width=\textwidth]{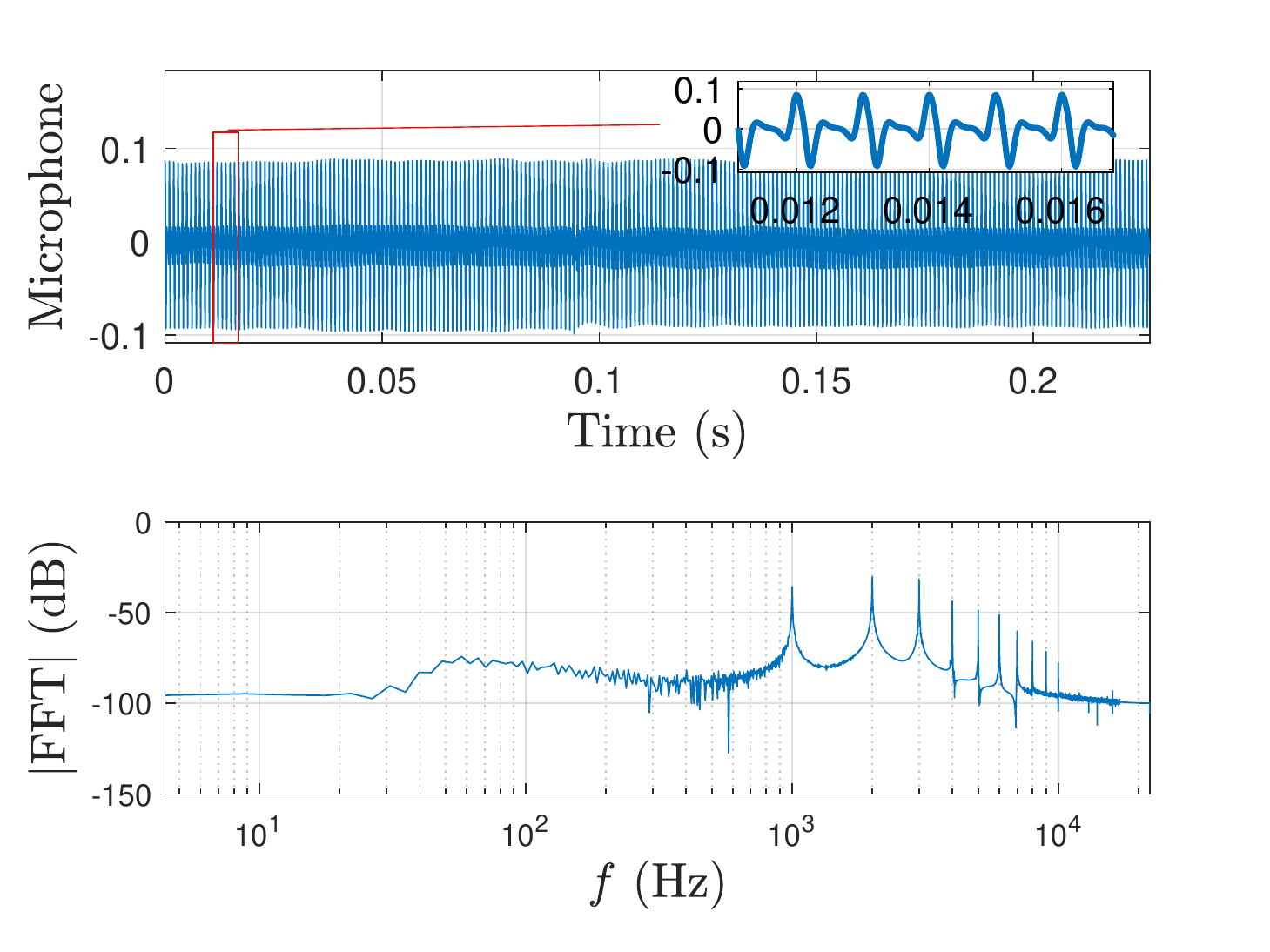}
            \caption{$\vra=\SI{0.9}{\volt}$, $f_c=\SI{2.0}{\giga\hertz}$}
    \label{fig:app:mic_09_20}
    \end{subfigure}\hfill
    \caption{Microphone output for $f_m=\SI{1}{\kilo\hertz}$ and $\mu=1.0$.}
    \label{fig:app:mic}
  \end{figure}

Figure~\ref{fig:app:mic} shows example microphone outputs for different
carrier frequencies $f_c$ and output voltages $\vra$. They all exhibit high harmonics,
but the similarity compared to the ideal signal for
Figures~\ref{fig:app:mic_06_15}--\ref{fig:app:mic_09_15}
is still over $0.9$. However, the injection
of Figure~\ref{fig:app:mic_09_20} contains more
pronounced distortions, and the similarity drops to less than $0.3$.

Overall, the results of this section show that
a higher-order transfer function may be needed to more accurately
predict the ADC output, both in terms of its RMS voltage,
and in terms of the harmonics it produces.

\subsection{ATmega328P Characterization}
\label{sec:app:atmega}

\begin{figure}[t]
  \centering
  \begin{subfigure}[!t]{0.7\linewidth}
  \includegraphics[width=\textwidth]{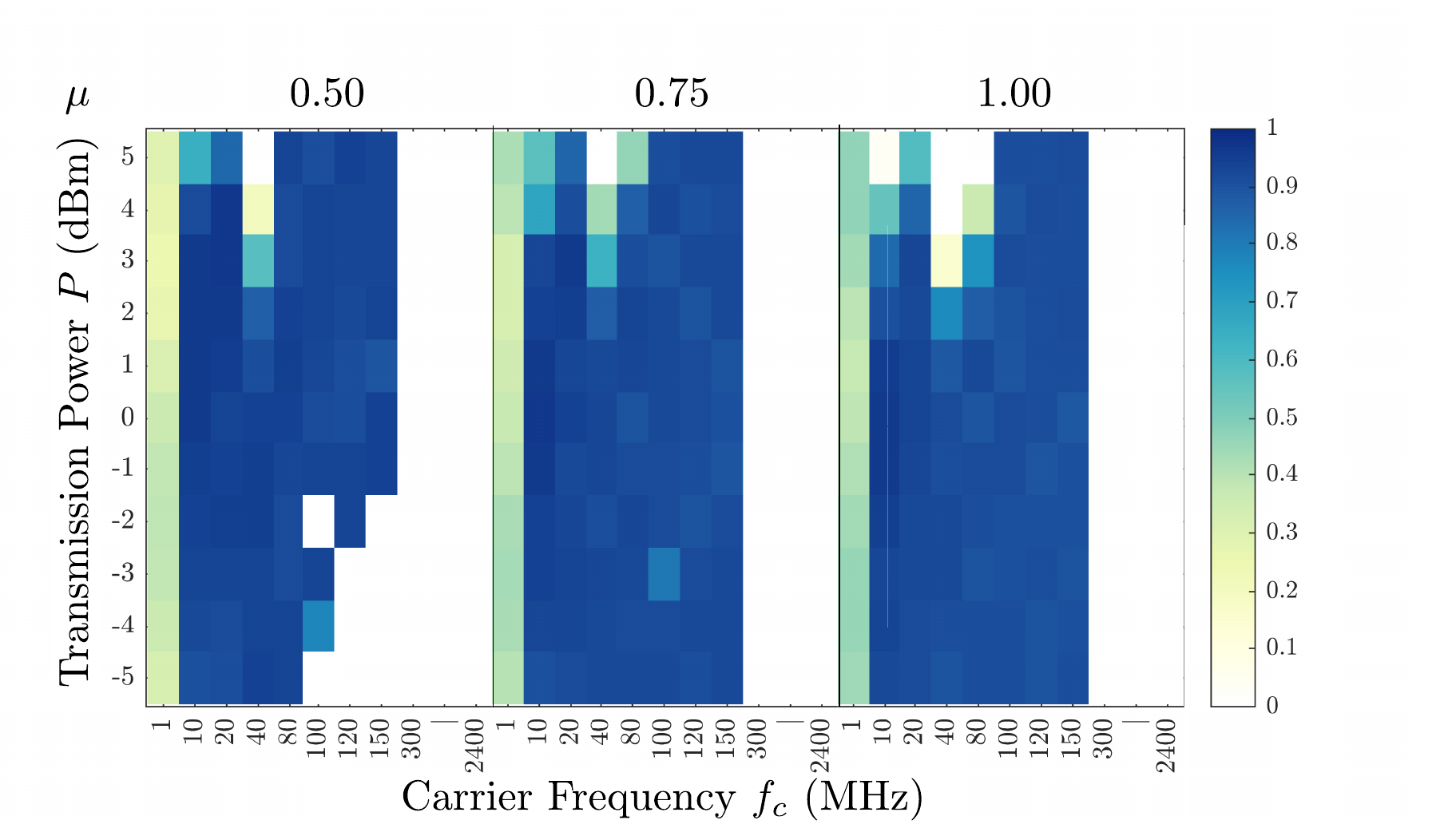}
  \caption{ADC only}
   \label{fig:app:atmega}
  \end{subfigure}\hfill
\end{figure}
\begin{figure}[!t]\ContinuedFloat
  \centering
  \begin{subfigure}[!t]{0.49\textwidth}
  \centering
  \includegraphics[width=\textwidth]{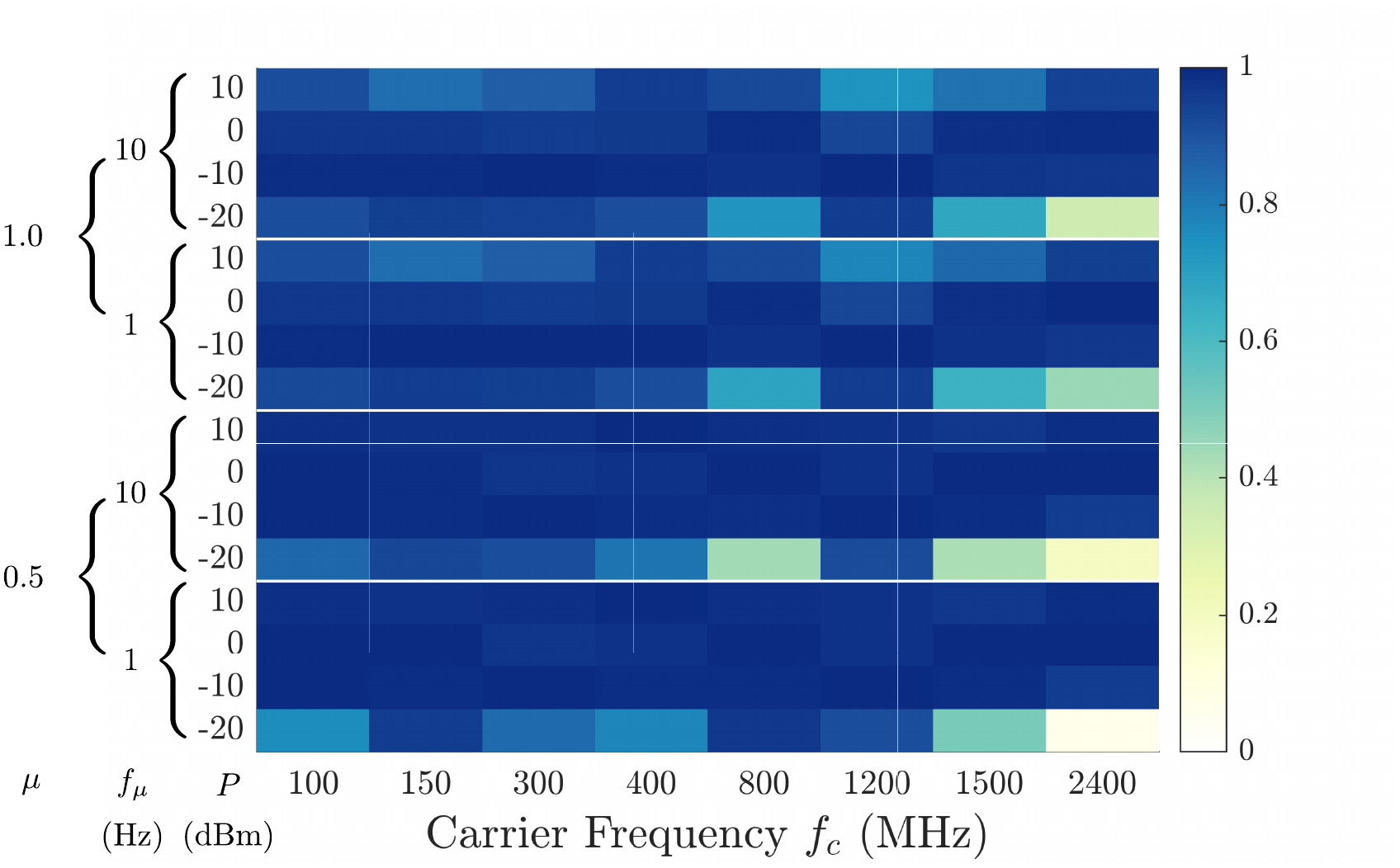}
  \caption{ADC + Amplifier}
  \label{fig:app:amp}
  \end{subfigure}\hfill
  \begin{subfigure}[!t]{0.49\textwidth}
    \centering
    \includegraphics[width=\textwidth]{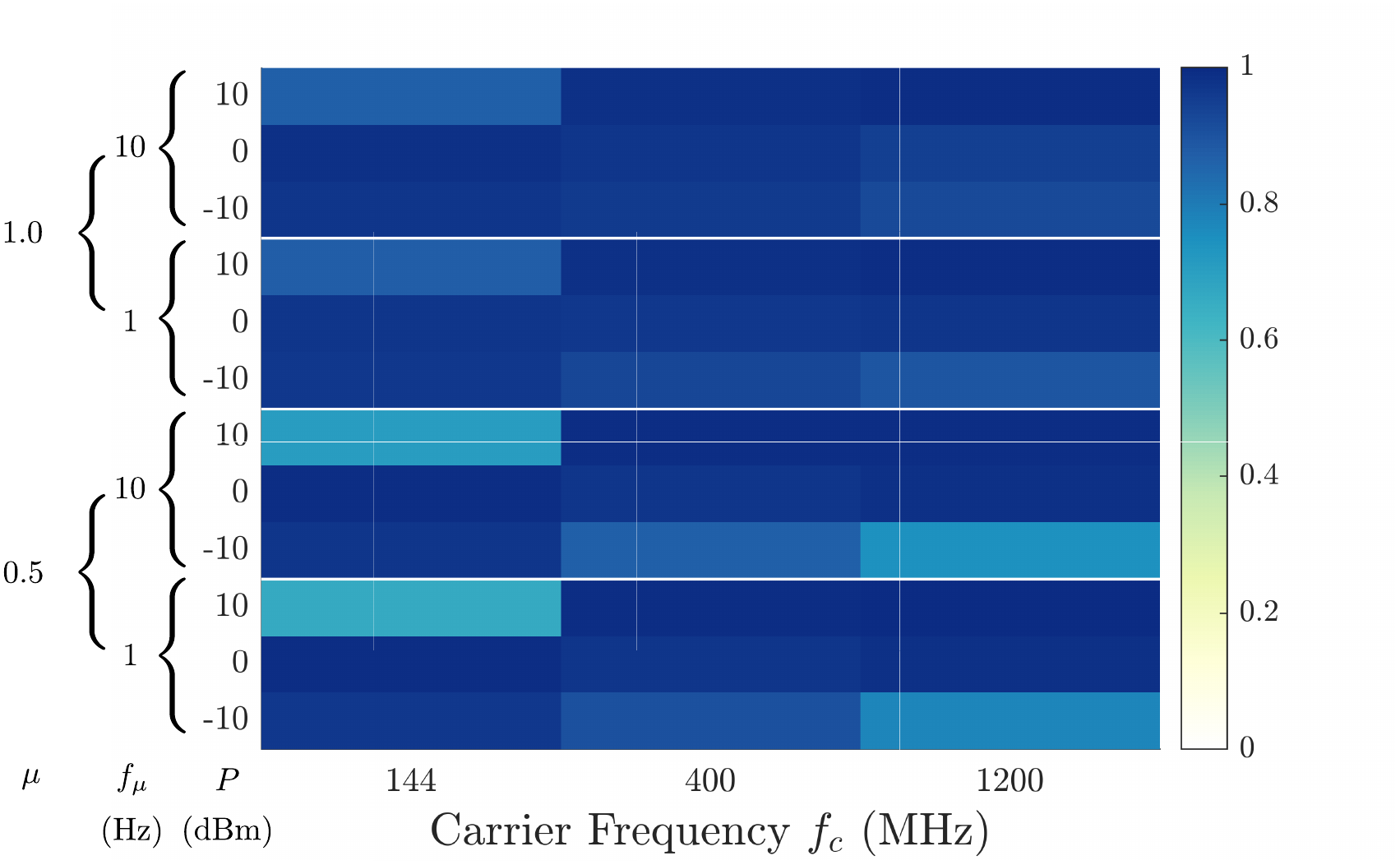}
    \caption{ADC + Amplifier + Antenna}
  \label{fig:app:ant}
  \end{subfigure}\hfill
  \caption{Similarity metrics for injections into the ATmega328P
           for different transmission powers $P$,
           modulation depths $\mu$, and carrier frequencies
           $f_c$. The amplifier increases the vulnerable frequencies to
           the $\si{\giga\hertz}$ range, allowing remote attacks.}
\end{figure}

This section contains detailed results for injections into the ATmega328P
ADC in three different arrangements: (a) the ADC on its own;
(b) the ADC with an amplifier; and (c) the ADC with an
amplifier and an antenna. The experimental results
support the theoretical model of Section~\ref{sec:model} and show that,
despite its low-pass filtering behavior, the ADC demodulates
signals carried at frequencies multiple times the cut-off frequency of
the sample-and-hold mechanism. Moreover, external amplifier non-linearities
increase the vulnerable frequency band into the $\si{\giga\hertz}$ range.

\secpar{ATmega328P Only} The first experiment targets the ATmega328P
directly, without using any additional components. The similarity
of the demodulated signal to the ideal signal is calculated for injections
at different powers $P$, modulation depths $\mu$, carrier frequencies $f_c$,
and a signal frequency of $f_m=\SI{1}{\hertz}$. As Figure~\ref{fig:app:atmega}
illustrates, the similarity for $f_c=\SI{1}{\mega\hertz}$ is always low due to aliasing.
However, similarity is high for $f_c$ between $\SIrange{10}{150}{\mega\hertz}$,
but signals are severely attenuated for $f_c\ge\SI{300}{\mega\hertz}$.
Small modulation depths and powers do not
result in demodulated outputs, while too much power causes the ADC to
be saturated. This leads to partial clipping of the signal, or
induces a DC offset which is beyond the range of the ADC. Overall,
the adversary has a range of choices for $P$ and $\mu$, and can use
carrier frequencies which are multiple times the cutoff frequency of the ADC,
provided these are not attenuated by the circuit-specific transfer function $H_C$.

\secpar{ATmega328P + Amplifier} A low-cost, off-the-shelf wideband LNA
is added before the input to the ADC
to change the transfer function $H_A$. The amplifier works between
$\SIrange{1}{2000}{\mega\hertz}$ and can perform a maximum amplification of
$\SI{32}{\decibel}$. It can output at most $\SI{10}{\decibel m}$,
and has a noise figure of approximately $\SI{2}{\decibel}$. Sinusoidals
of $f_m=\SI{1}{\hertz}$ and $f_m=\SI{10}{\hertz}$ are modulated at depths
of $\mu\in\{0.5,1.0\}$ on carrier frequencies $f_c$ from $\SI{100}{\mega\hertz}$
to $\SI{2.4}{\giga\hertz}$ at transmission powers $P$ between $\SI{-20}{\decibel m}$
and $\SI{10}{\decibel m}$. As can be seen in Figure~\ref{fig:app:amp},
the similarity is high across all frequencies, provided
the transmission power is above a minimum threshold.

The amplifier thus both reduces the power requirements for the adversary, and
increases the vulnerable frequencies to the $\si{\giga\hertz}$ range. This allows
an attacker to target systems with short wires between the  ADC and the
sensor with a lower power budget: short wires are not a sufficient defense
against electromagnetic out-of-band signal injection attacks. Moreover, it should
be noted that an adversary gains an advantage by not obeying the amplifier
constraints: abusing the amplifier by transmitting higher-powered signals or
by driving frequency signals outside of the intended range still results in
recognizable output. In other words, although the signal may be distorted,
non-linearities produce outputs within the range of the ADC.

\secpar{ATmega328P + Amplifier + Antenna} The final set of experiments
changes the circuit-specific transfer function $H_C$ by using a transmitting antenna
at the signal generator output and a receiving antenna at the amplifier input.
The antennas used are Ettus Research omnidirectional VERT400 antennas, which are
resonant at $\SI{144}{\mega\hertz}$, $\SI{400}{\mega\hertz}$, and $\SI{1.2}{\giga\hertz}$.
The antennas are placed in parallel at a distance of $\SI{5}{\centi\meter}$
to one another, and the results for sine signals of $f_m=\SI{1}{\hertz}$ and
$f_m=\SI{10}{\hertz}$ are presented in Figure~\ref{fig:app:ant}.
Although the minimum power required for successful injections
is higher due to transmission losses, the system remains vulnerable
for all three frequencies due to the amplifier and ADC non-linearities.
In other words, results are reproducible across multiple setups,
whether through remote transmissions, or through direct injections with an
identity transfer function
$H_C(\sigma+\jw)=\mathcal{L}\{\tv\}/\mathcal{L}\{v(t)\}=1$.

\begin{figure}[!t]
  \centering
  \begin{subfigure}[!t]{0.49\textwidth}
    \centering
            \includegraphics[width=\textwidth]{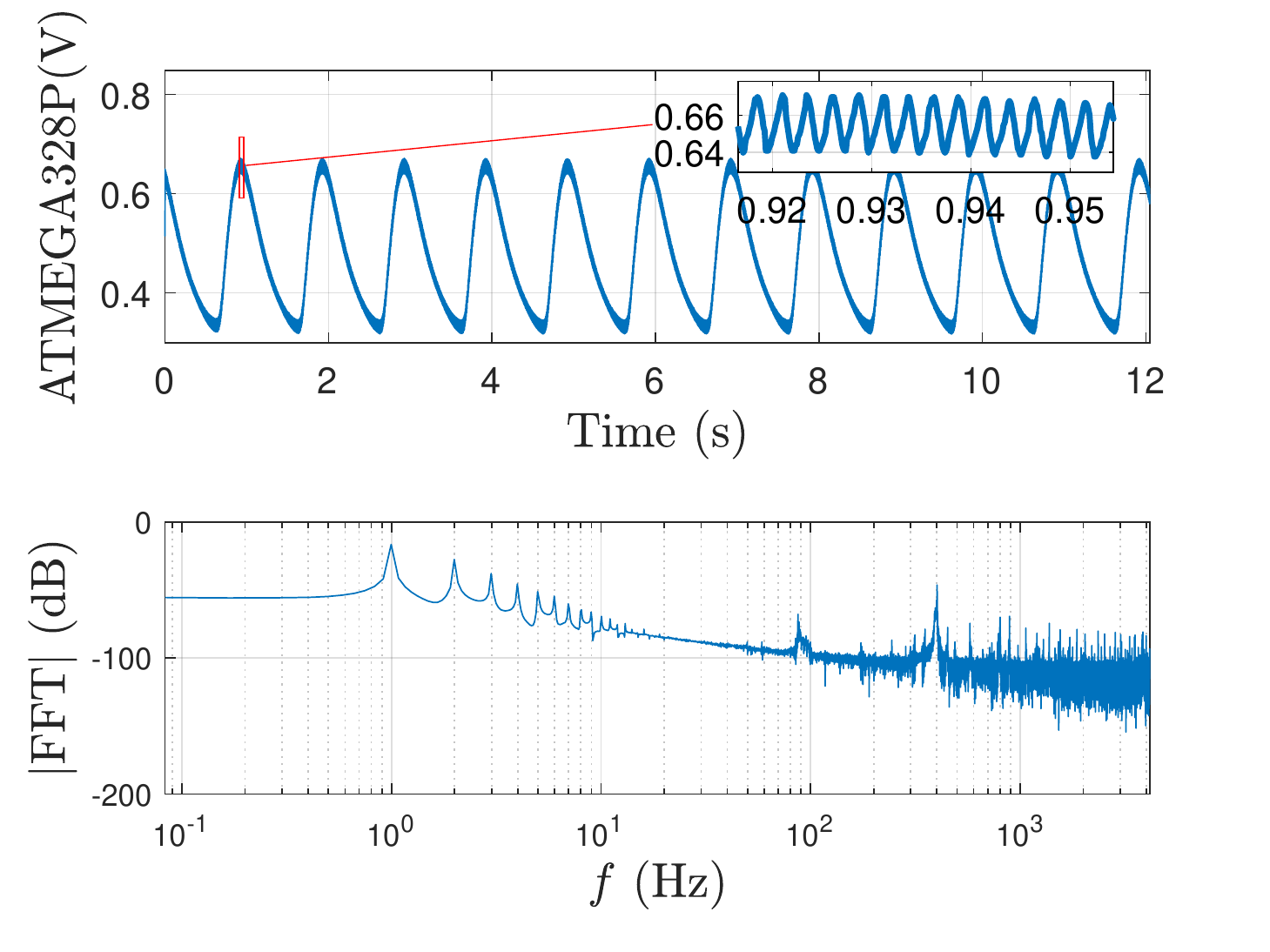}
            \caption{$f_c=\SI{20}{\mega\hertz}$}
  \end{subfigure}\hfill
  \begin{subfigure}[!t]{0.49\textwidth}
    \centering
            \includegraphics[width=\textwidth]{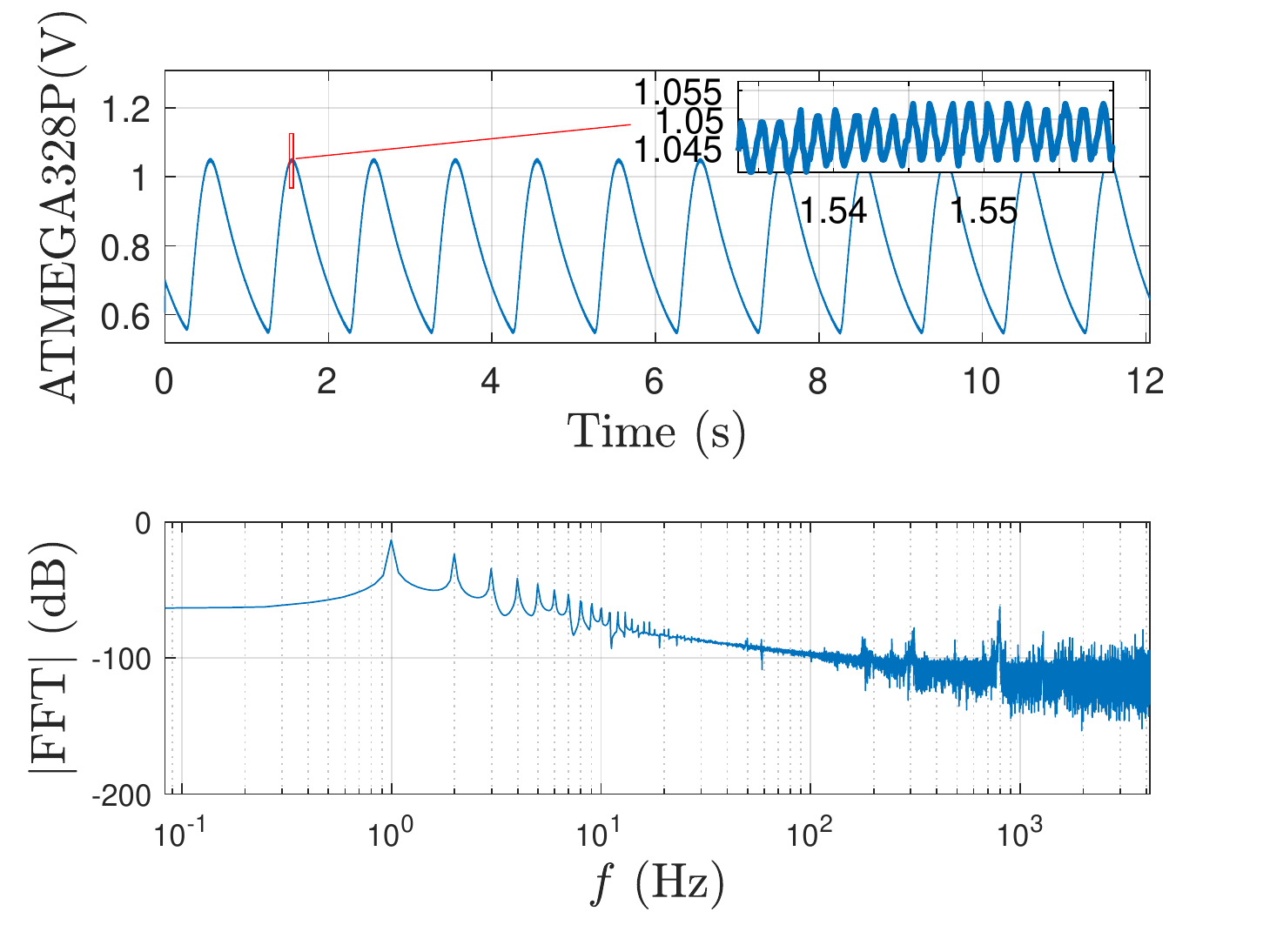}
            \caption{$f_c=\SI{40}{\mega\hertz}$}
  \end{subfigure}\hfill
  \caption{ATmega328P-only output for power $P=\SI{0}{\decibel m}$, signal frequency $f_m=\SI{1}{\hertz}$,
          and modulation depth $\mu=0.5$.}
  \label{fig:app:atmega_full}
\end{figure}

\begin{figure}[!t]
  \centering
  \begin{subfigure}[!t]{0.49\textwidth}
  \centering
          \includegraphics[width=\textwidth]{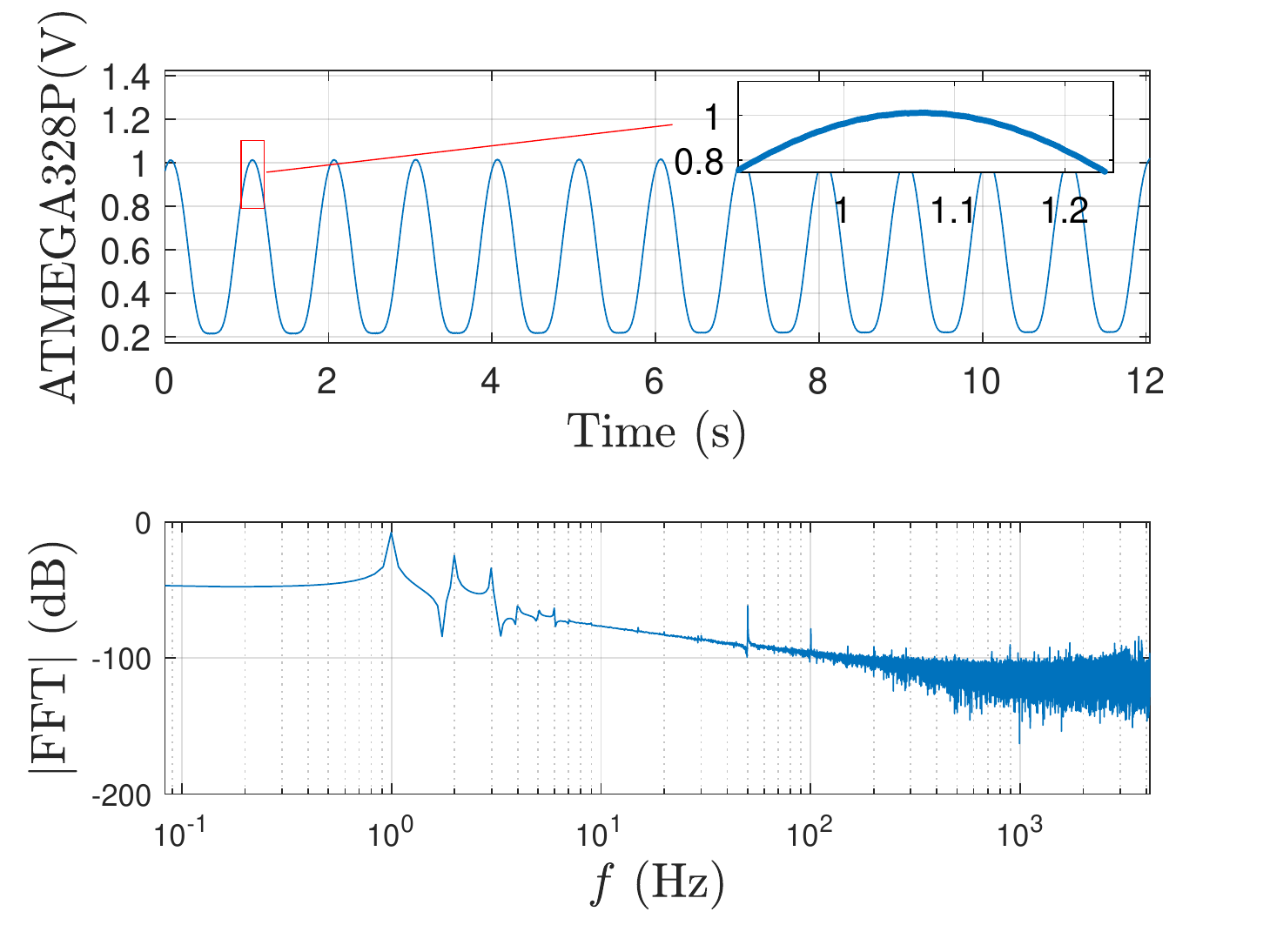}
          \caption{$f_c=\SI{1.5}{\giga\hertz}$}
  \end{subfigure}\hfill
  \begin{subfigure}[!t]{0.49\textwidth}
    \centering
            \includegraphics[width=\textwidth]{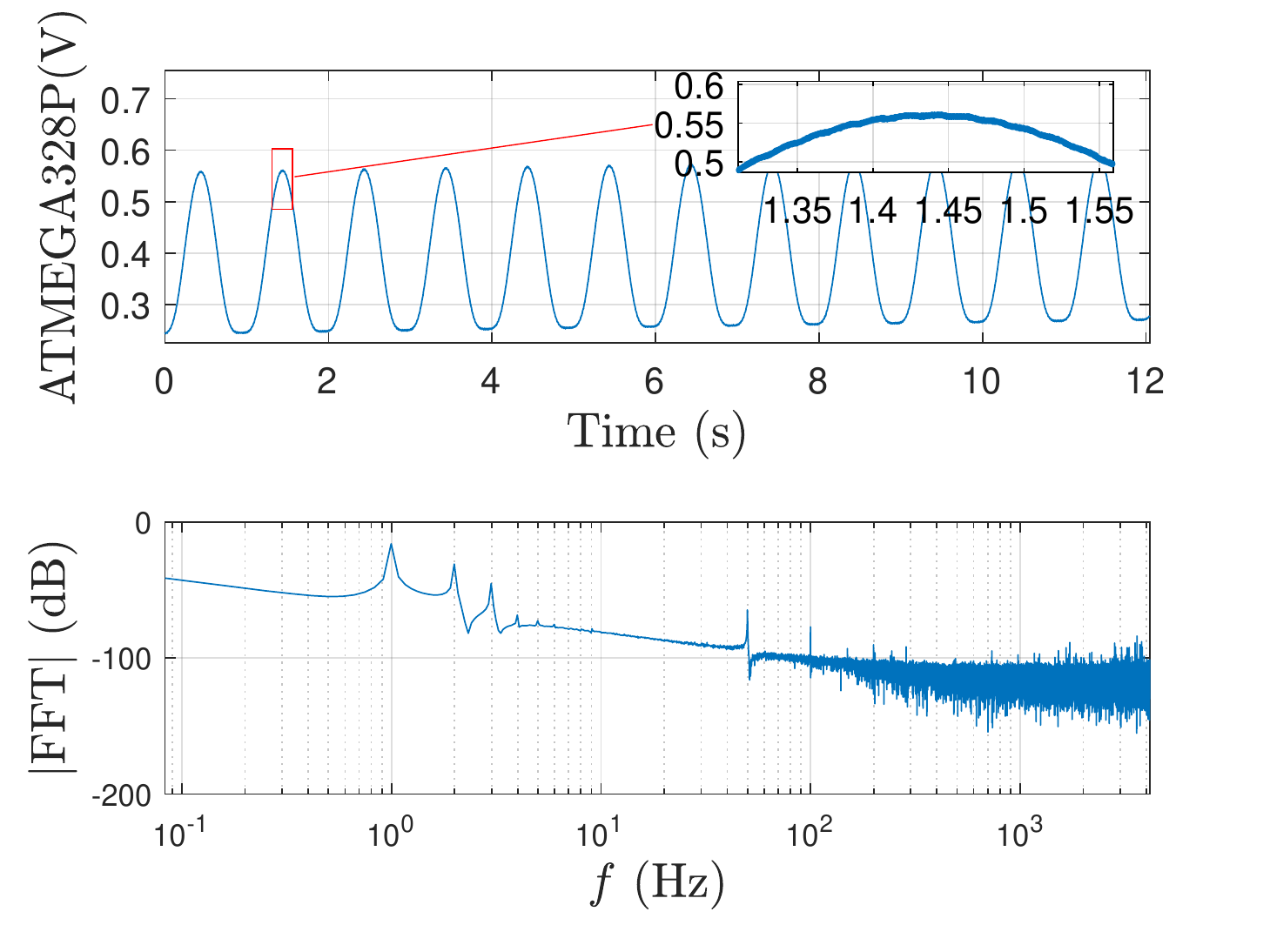}
            \caption{$f_c=\SI{2.4}{\giga\hertz}$}
  \end{subfigure}\hfill
  \caption{ATmega328P with amplifier output for power $P=\SI{-5}{\decibel m}$, signal frequency $f_m=\SI{1}{\hertz}$,
          and modulation depth $\mu=1.0$.}
  \label{fig:app:atmega_amp}
\end{figure}

\begin{figure}[!t]
  \centering
  \begin{subfigure}[!t]{0.33\textwidth}
  \centering
          \includegraphics[width=\textwidth]{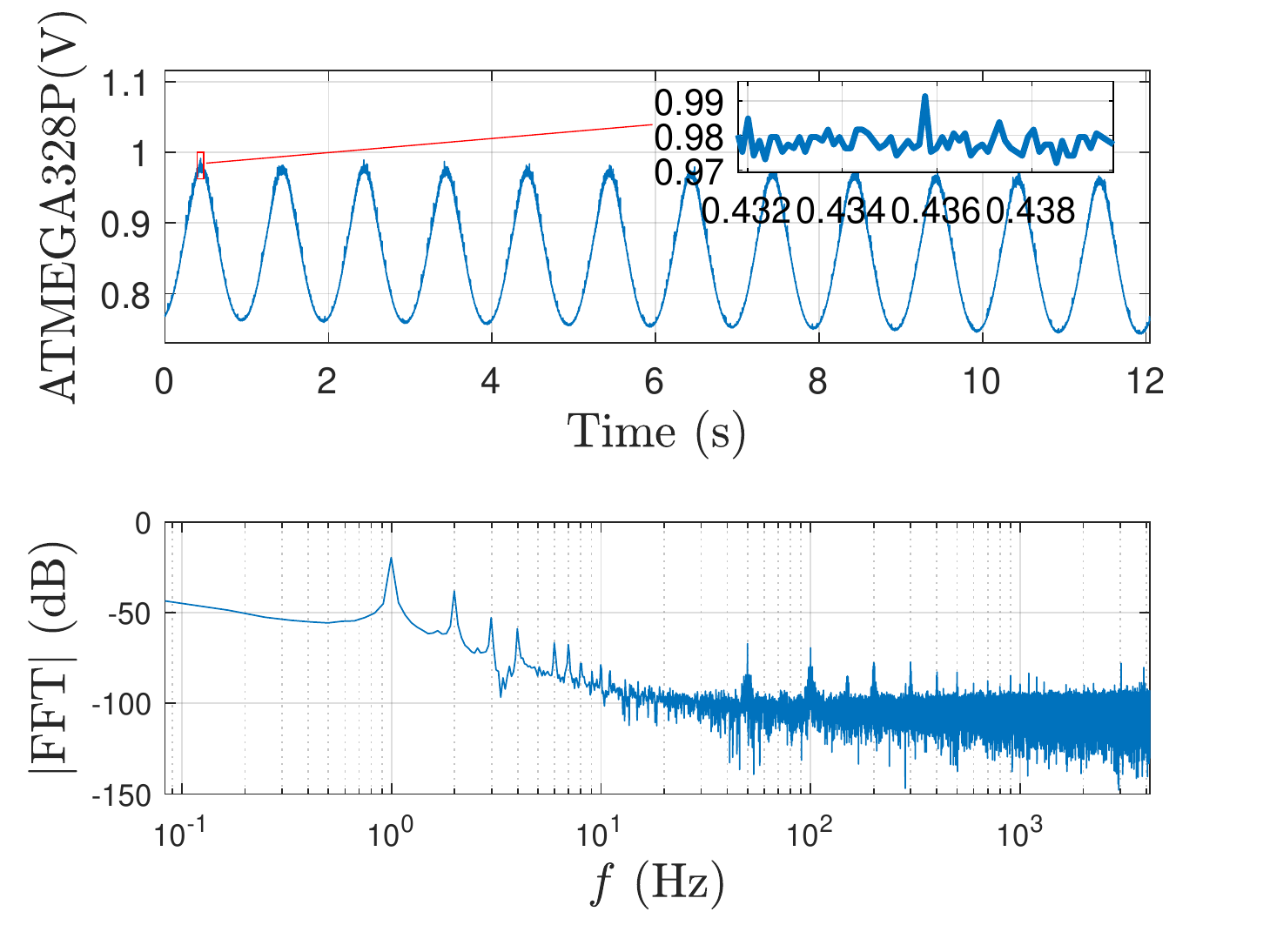}
          \caption{$f_c=\SI{144}{\mega\hertz}$}
  \end{subfigure}\hfill
  \begin{subfigure}[!t]{0.33\textwidth}
    \centering
            \includegraphics[width=\textwidth]{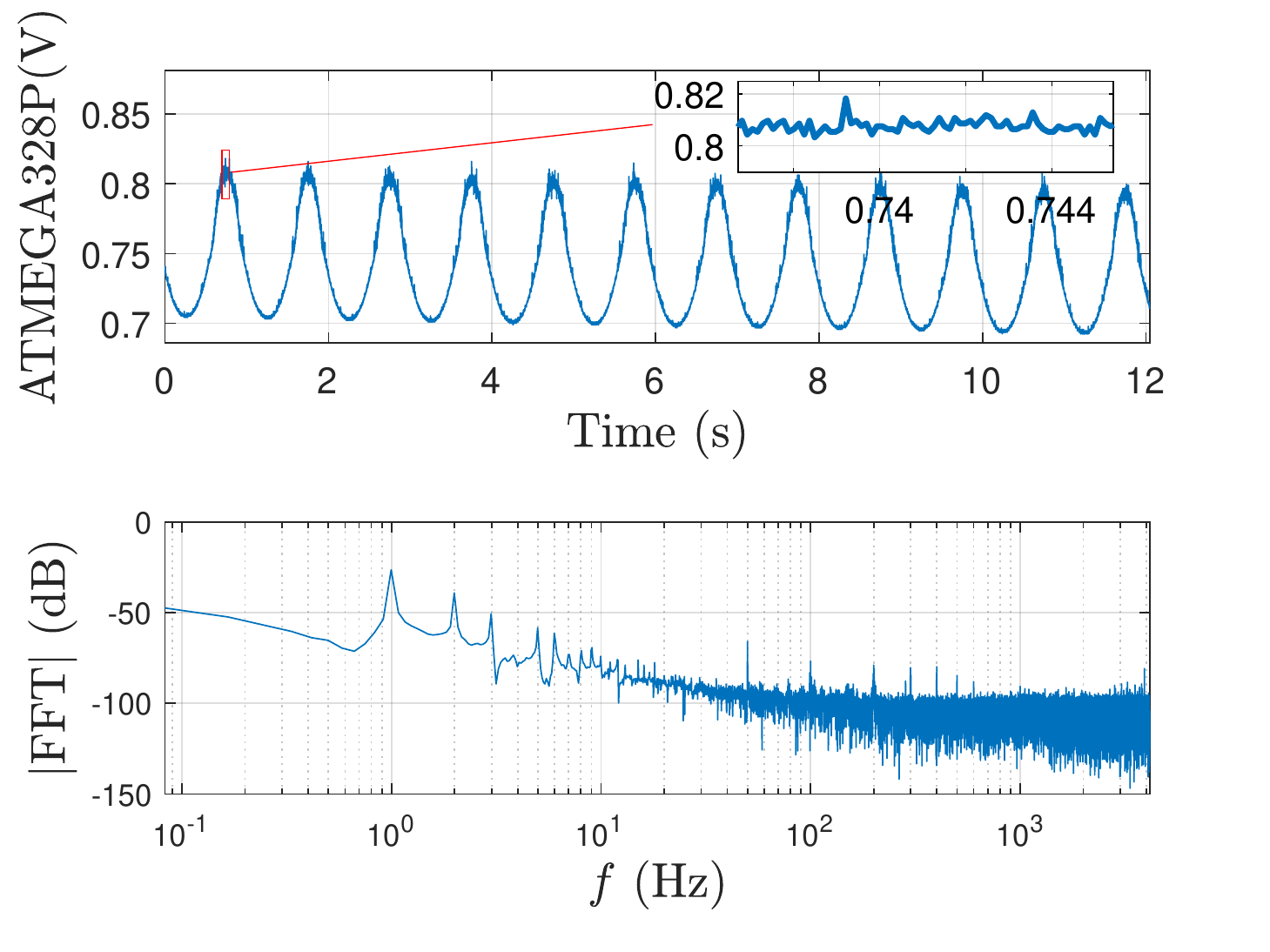}
            \caption{$f_c=\SI{400}{\mega\hertz}$}
  \end{subfigure}\hfill
  \begin{subfigure}[!t]{0.33\textwidth}
    \centering
            \includegraphics[width=\textwidth]{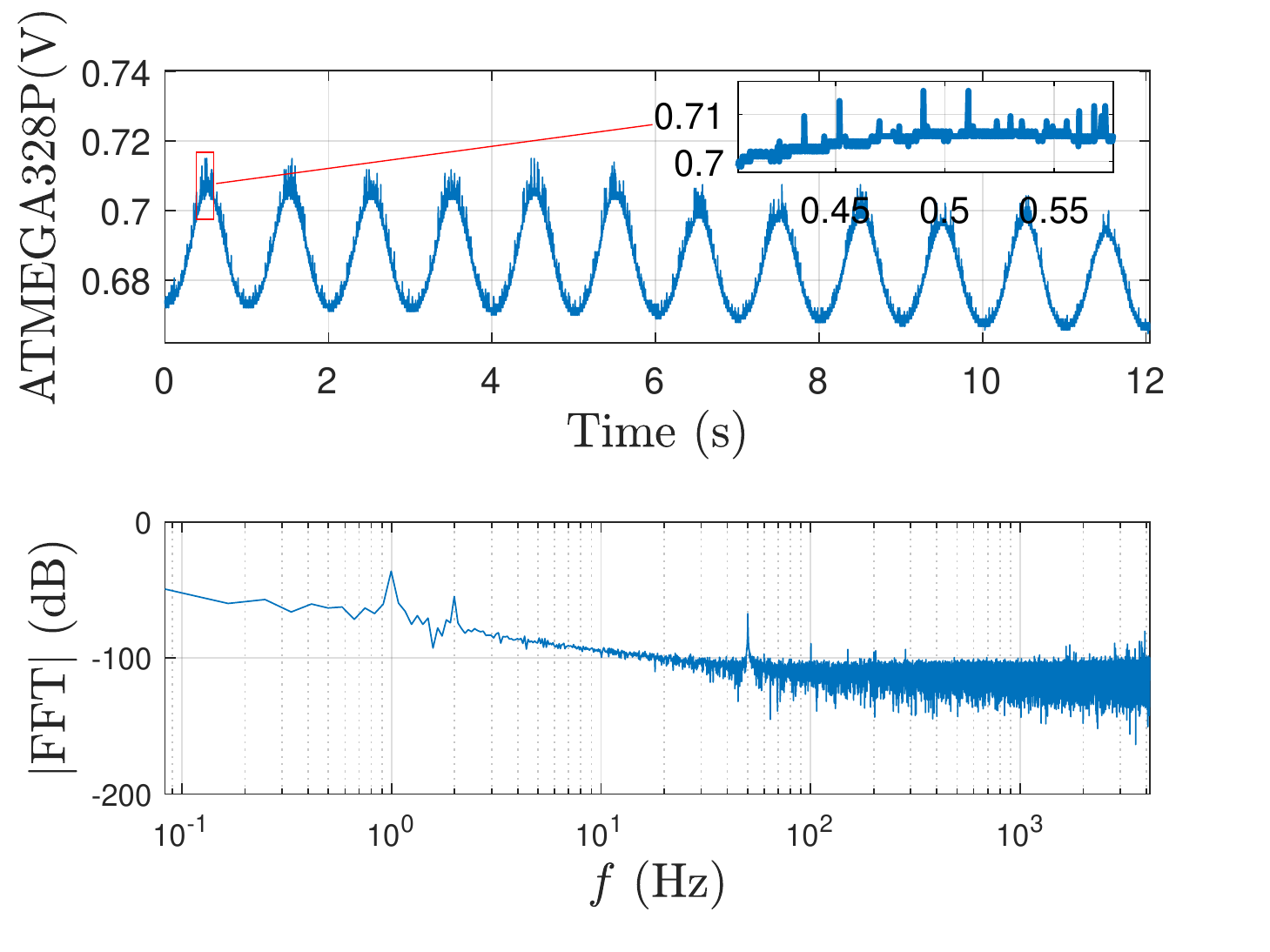}
            \caption{$f_c=\SI{1.2}{\giga\hertz}$}
  \end{subfigure}\hfill
  \caption{ATmega328P with amplifier and antenna output for power $P=\SI{0}{\decibel m}$, signal frequency $f_m=\SI{1}{\hertz}$,
          and modulation depth $\mu=0.5$.}
  \label{fig:app:atmega_ant}
\end{figure}

Figures~\ref{fig:app:atmega_full}--\ref{fig:app:atmega_ant} show
example outputs from the internal ATmega328P ADC for different carrier frequencies $f_c$,
powers $P$, and modulation depths $\mu$, with $f_m$ fixed at $\SI{1}{\hertz}$.
Figure~\ref{fig:app:atmega_full} first shows the results for the ADC
on its own, and complements Figure~\ref{fig:atmega_raw} of Section~\ref{sec:adcs}.
Although harmonics of the fundamental persist, the high-frequency component
becomes less pronounced as $f_c$ increases.

Figure~\ref{fig:app:atmega_amp} then shows output from the ATmega328P for two different
carrier frequencies $f_c$ when connected to an amplifier. The ADC no longer behaves
like a low-pass filter due to non-linearities, while harmonics of the fundamental
remain strong. Finally, Figure~\ref{fig:app:atmega_ant} shows output from the ATmega328P
for remote injections using the VERT400 antenna with the amplifier. As in the amplifier case,
carrier frequencies in the $\si{\giga\hertz}$ range are still demodulated, and
harmonics (and some high-frequency components) persist.

\subsection{Further ADC Demodulation Examples}
\label{sec:app:others}

This section contains additional examples of injections into the different ADCs.

\begin{figure}[!t]
    \centering
    \begin{subfigure}[!t]{0.49\textwidth}
    \centering
            \includegraphics[width=\textwidth]{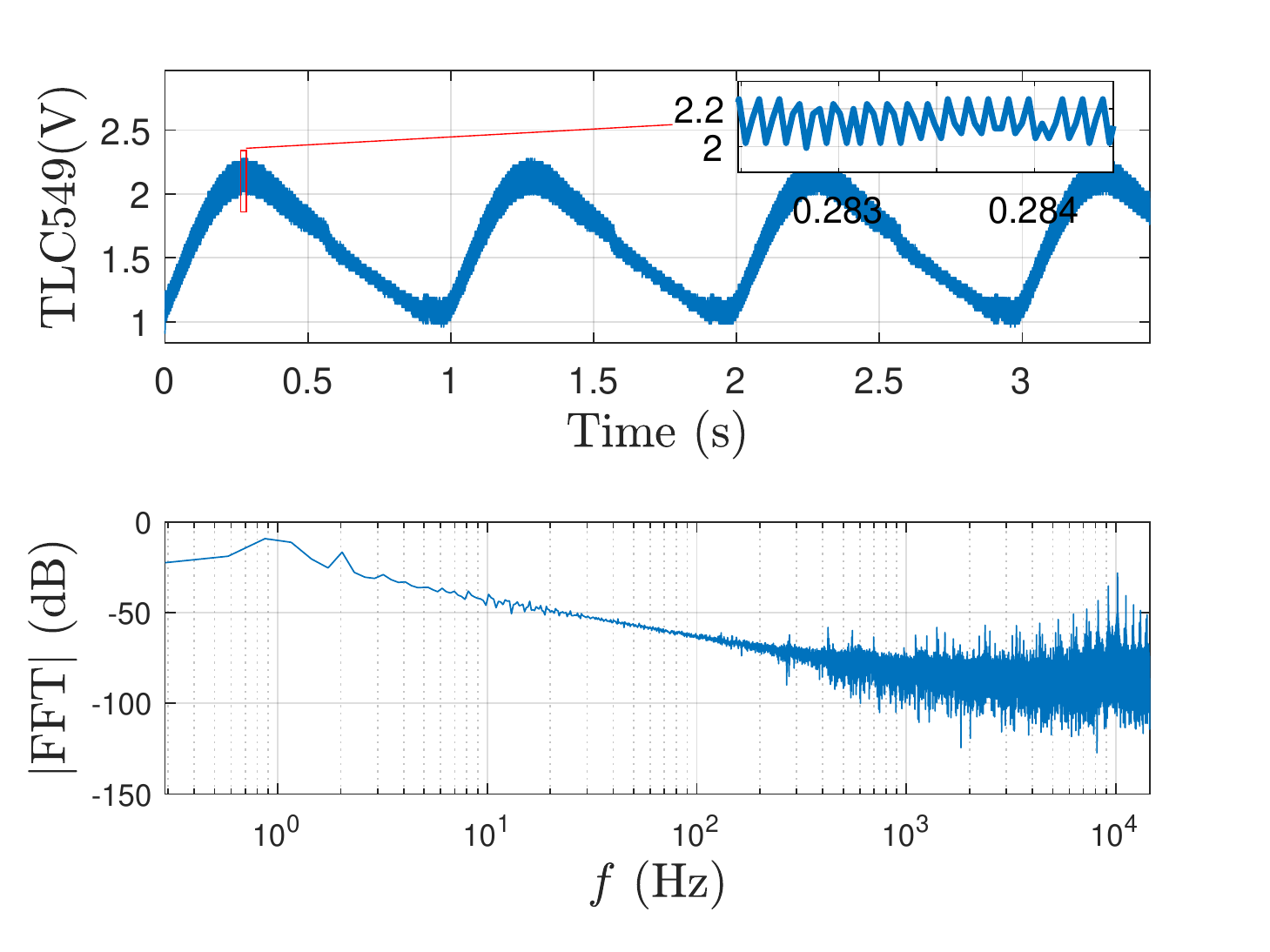}
            \caption{$f_c=\SI{20}{\mega\hertz}$}
    \end{subfigure}\hfill
      \begin{subfigure}[!t]{0.49\textwidth}
    \centering
            \includegraphics[width=\textwidth]{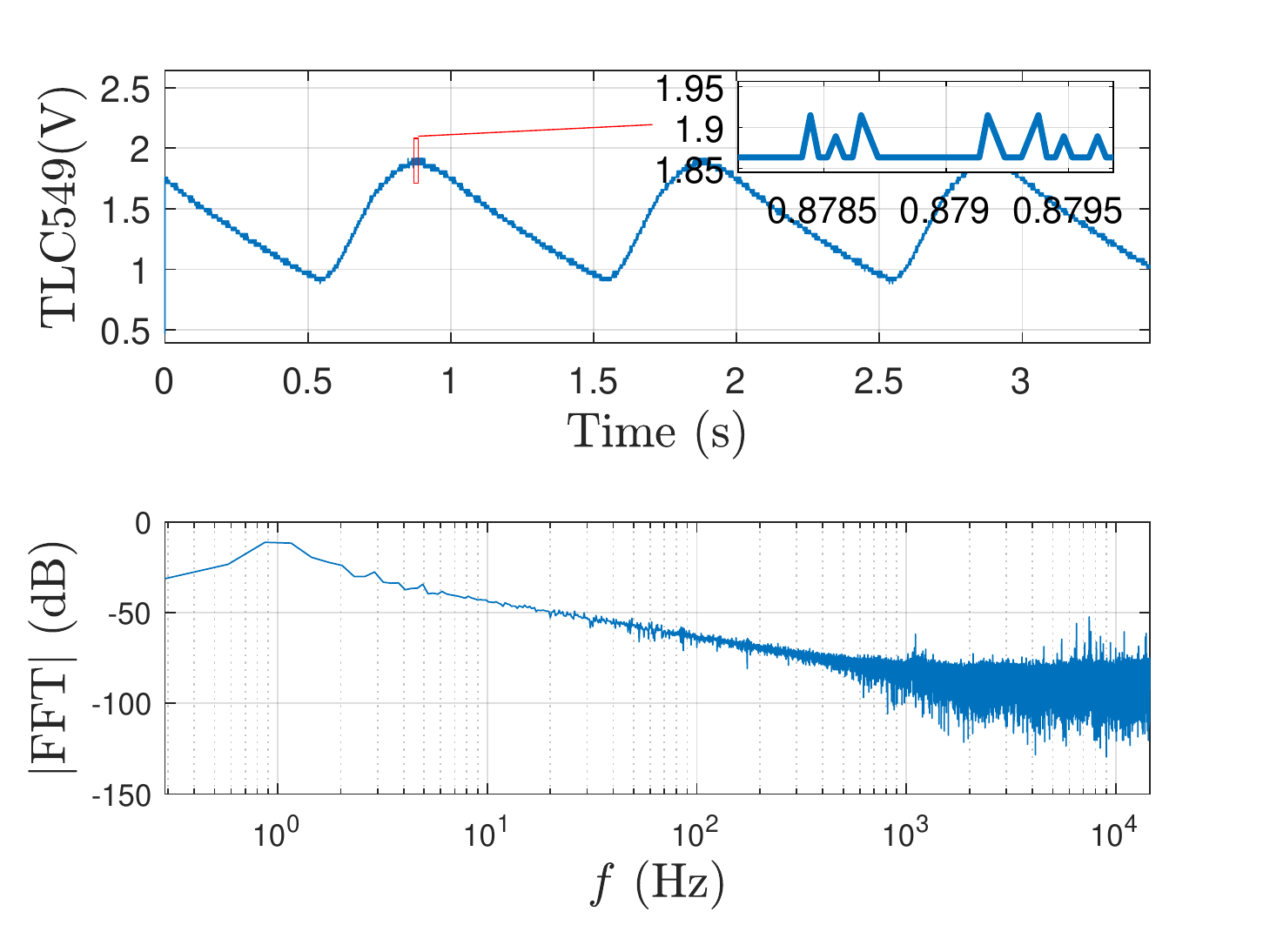}
            \caption{$f_c=\SI{120}{\mega\hertz}$}
    \end{subfigure}\hfill
    \caption{TLC549 output for power $P=\SI{5}{\decibel m}$, signal frequency $f_m=\SI{1}{\hertz}$,
    and depth $\mu=0.5$.}
    \label{fig:app:tlc549_full}
\end{figure}

\secpar{TLC549} Figure~\ref{fig:app:tlc549_full} shows example outputs from the
TLC549 ADC for two carrier frequencies $f_c$. Harmonics of the fundamental
are not pronounced, as the resolution is only 8 bits.

\begin{figure}[!t]
  \centering
  \begin{subfigure}[!t]{0.49\textwidth}
  \centering
          \includegraphics[width=\textwidth]{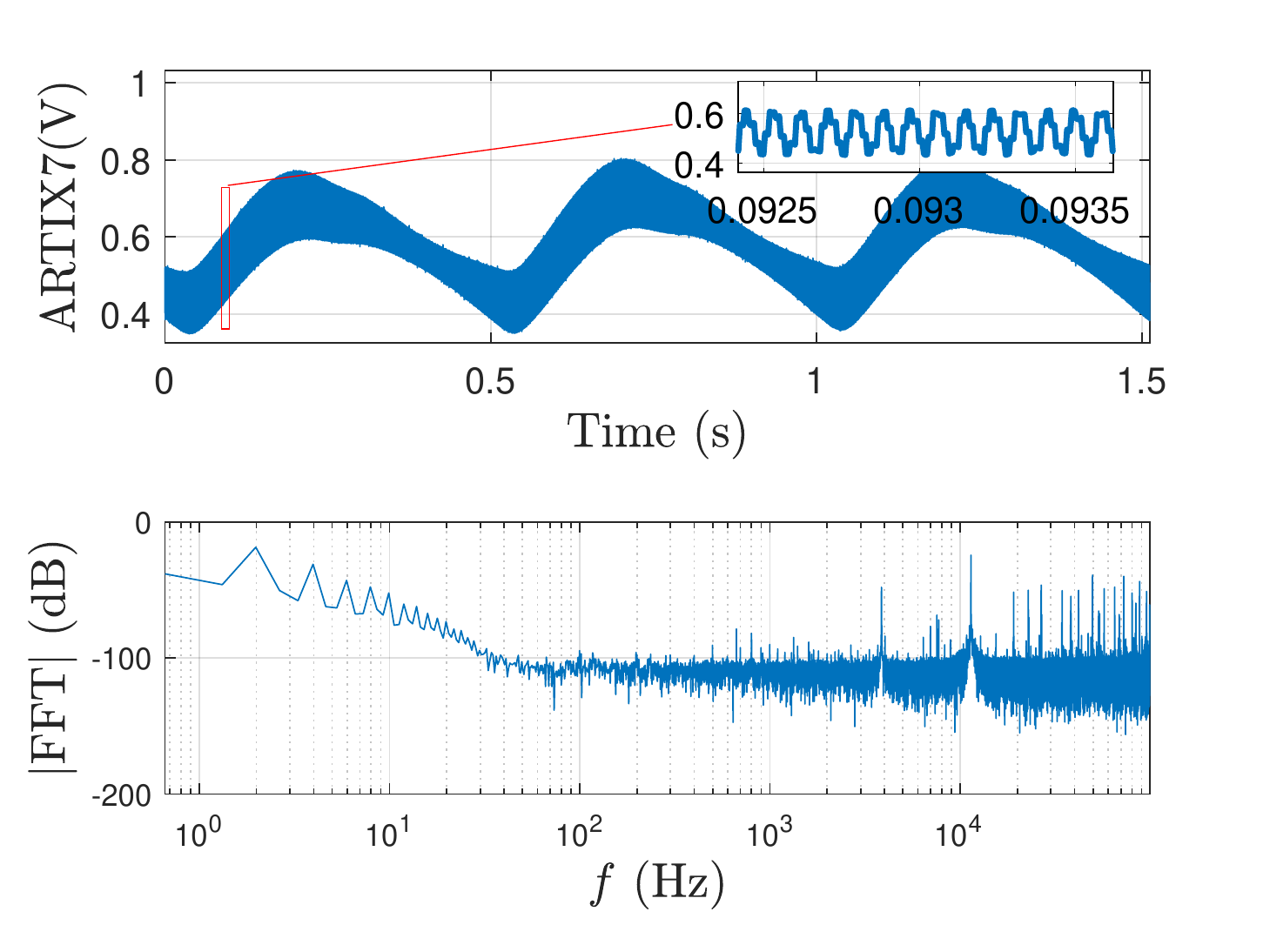}
          \caption{$f_c=\SI{80}{\mega\hertz}$}
  \end{subfigure}\hfill
  \begin{subfigure}[!t]{0.49\textwidth}
  \centering
          \includegraphics[width=\textwidth]{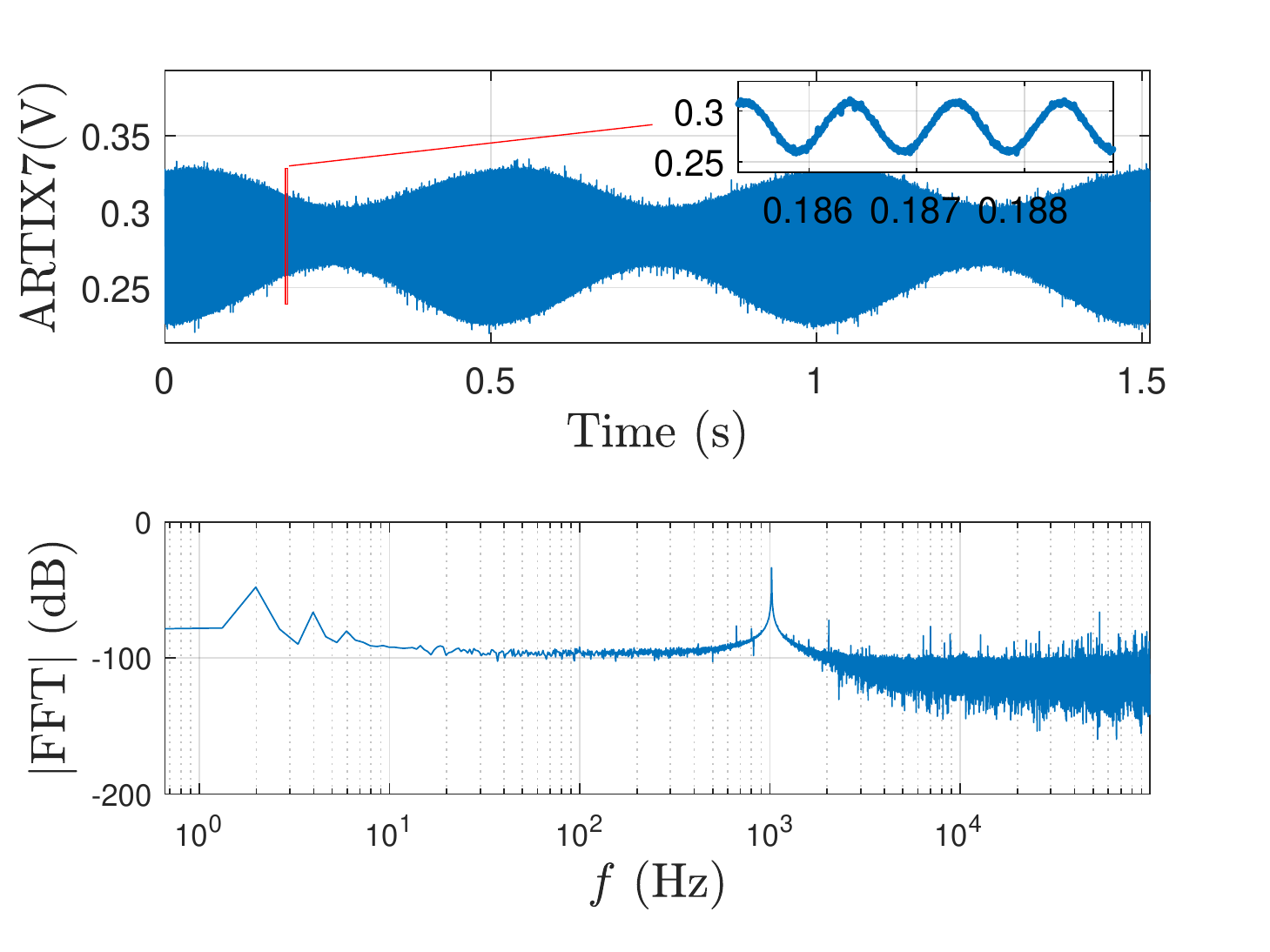}
          \caption{$f_c=\SI{100}{\mega\hertz}$}
  \end{subfigure}\hfill
  \caption{Artix~7 output for power $P=\SI{10}{\decibel m}$, signal frequency $f_m=\SI{1}{\hertz}$,
  and depth $\mu=0.5$.}
  \label{fig:app:artix7_full}
\end{figure}

\secpar{Artix~7} Figure~\ref{fig:app:artix7_full} shows example outputs from
the Artix~7 ADC for two carrier
frequencies $f_c$. The output contains high-frequency components which dominate
the target signal, forcing injections to require more fine-grained control over the
carrier frequency.

\begin{figure}[!t]
  \centering
  \begin{subfigure}[!t]{0.49\textwidth}
  \centering
          \includegraphics[width=\textwidth]{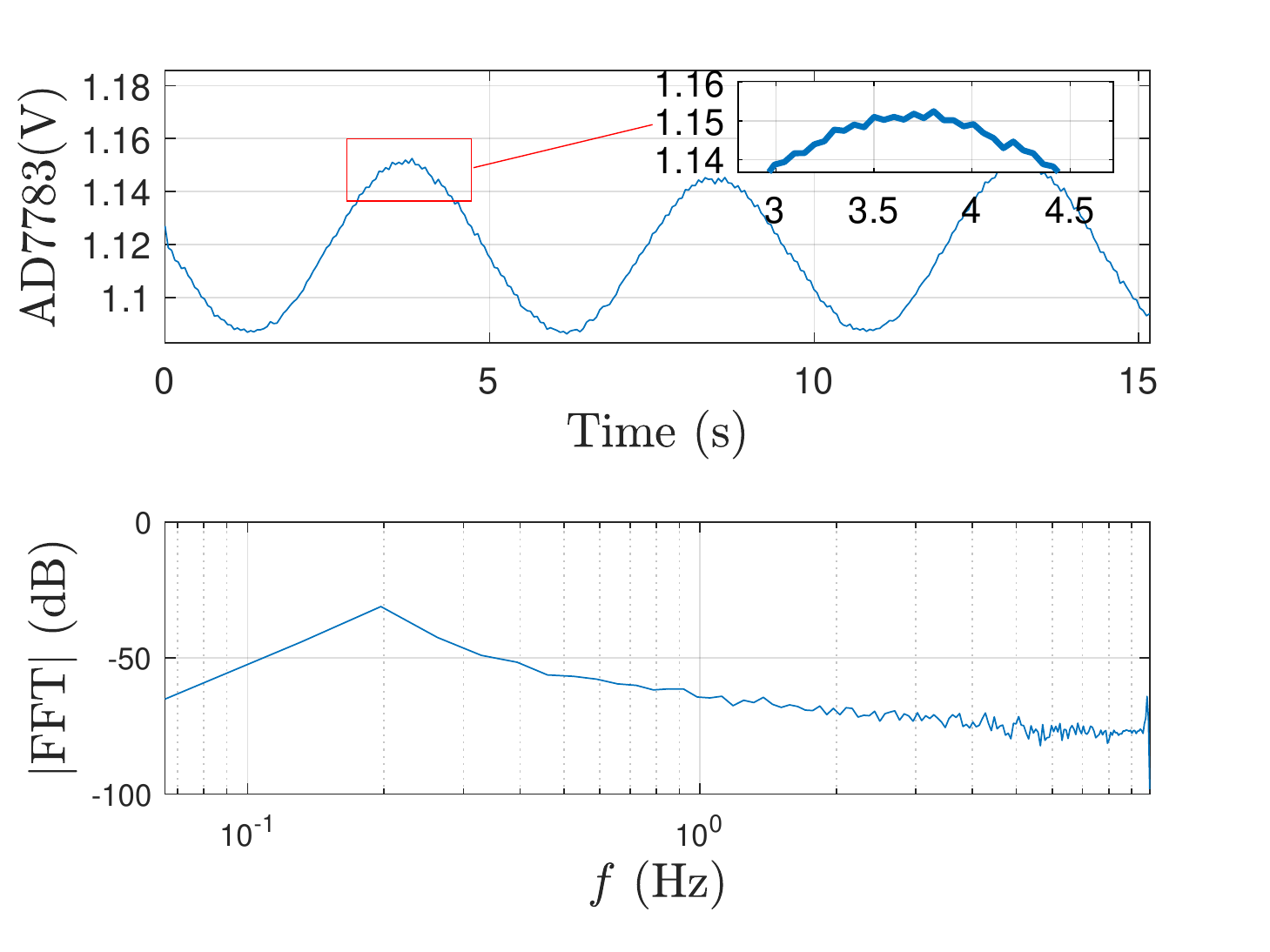}
          \caption{$f_c=\SI{40}{\mega\hertz}$, $\mu=0.5$}
  \end{subfigure}\hfill
  \begin{subfigure}[!t]{0.49\textwidth}
  \centering
          \includegraphics[width=\textwidth]{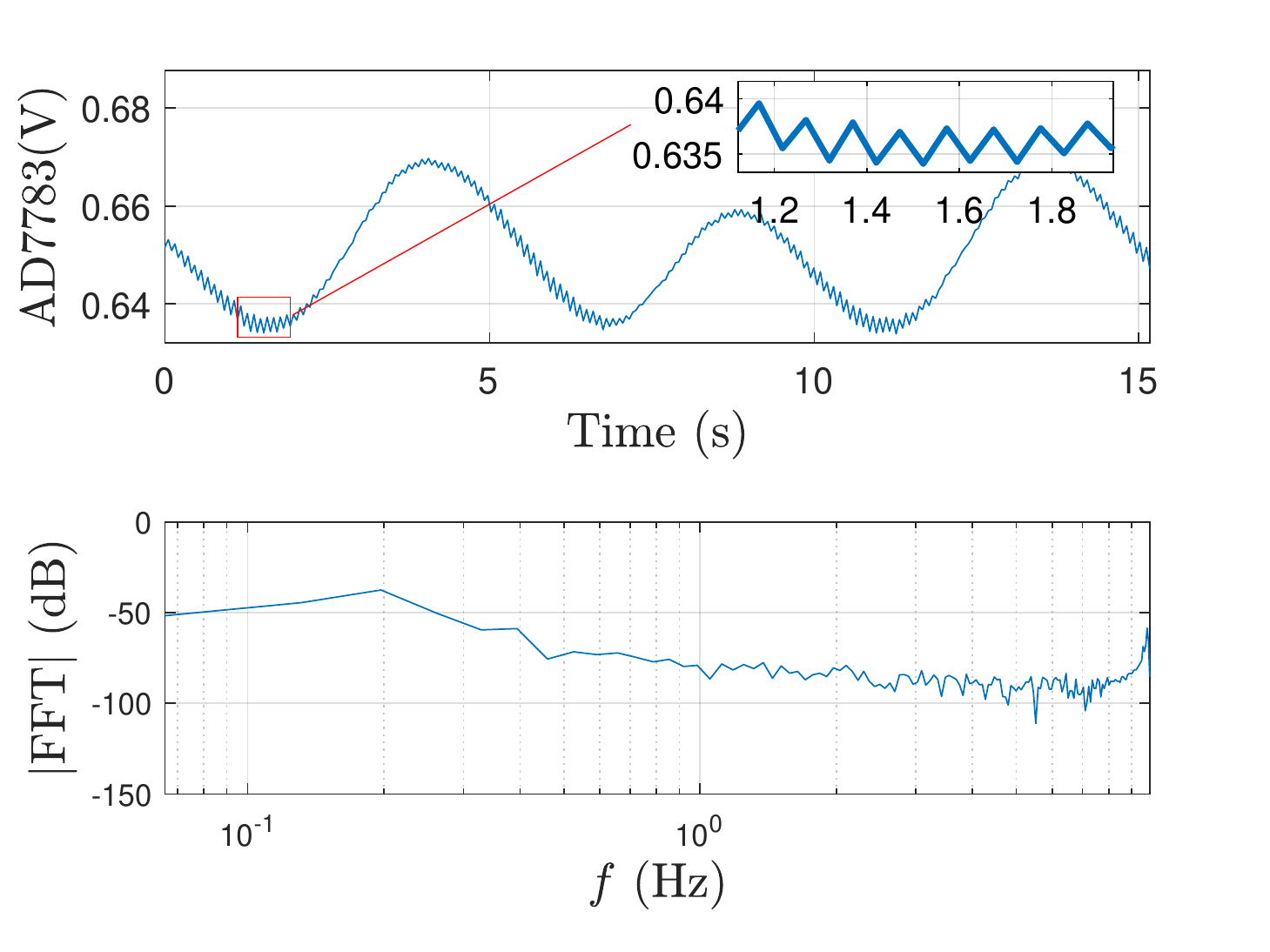}
          \caption{$f_c=\SI{80}{\mega\hertz}$, $\mu=1.0$}
  \end{subfigure}\hfill
  \caption{AD7783 output for power $P=\SI{5}{\decibel m}$ and signal frequency $f_m=\SI{10}{\hertz}$.}
  \label{fig:app:ad7783_full}
\end{figure}

\secpar{AD7783} Figure~\ref{fig:app:ad7783_full} shows the output from the (slow)
$\ds$ AD7783 ADC for
different carrier frequencies and modulation depths. As $f_m=\SI{10}{\hertz}$
is above the Nyquist frequency, aliasing occurs. The strongest frequency present is
$2f_m-f_s=\SI{0.21}{\hertz}$, while the high-frequency component is $f_s-f_m=\SI{9.79}{\hertz}$.

\begin{figure}[!t]
    \centering
    \begin{subfigure}[!t]{0.49\textwidth}
    \centering
      \includegraphics[width=\textwidth]{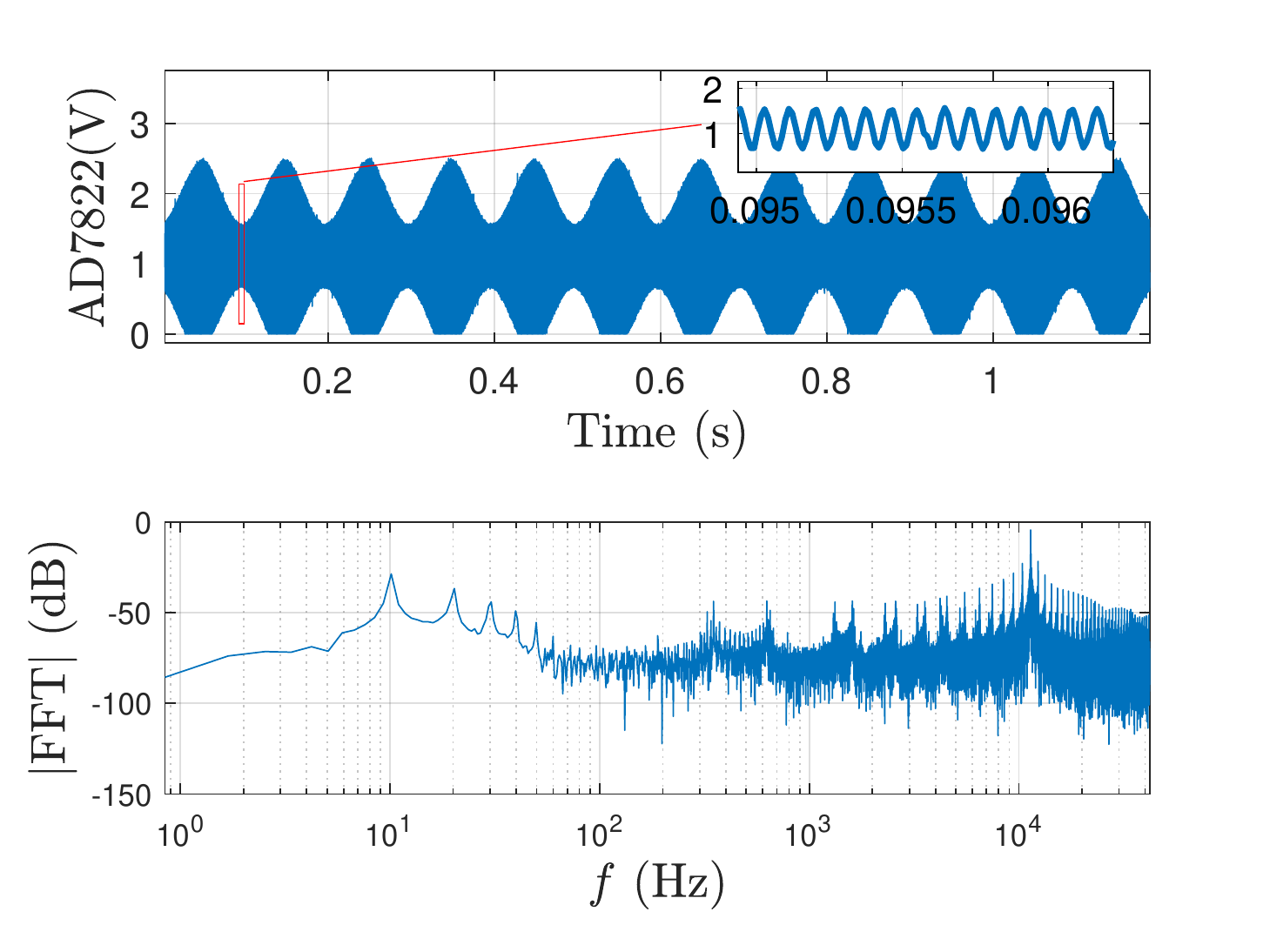}
      \caption{AD7822: $f_c=\SI{10}{\mega\hertz}$}
    \label{fig:app:ad7822_full}
    \end{subfigure}\hfill
    \begin{subfigure}[!t]{0.49\textwidth}
    \centering
      \includegraphics[width=\textwidth]{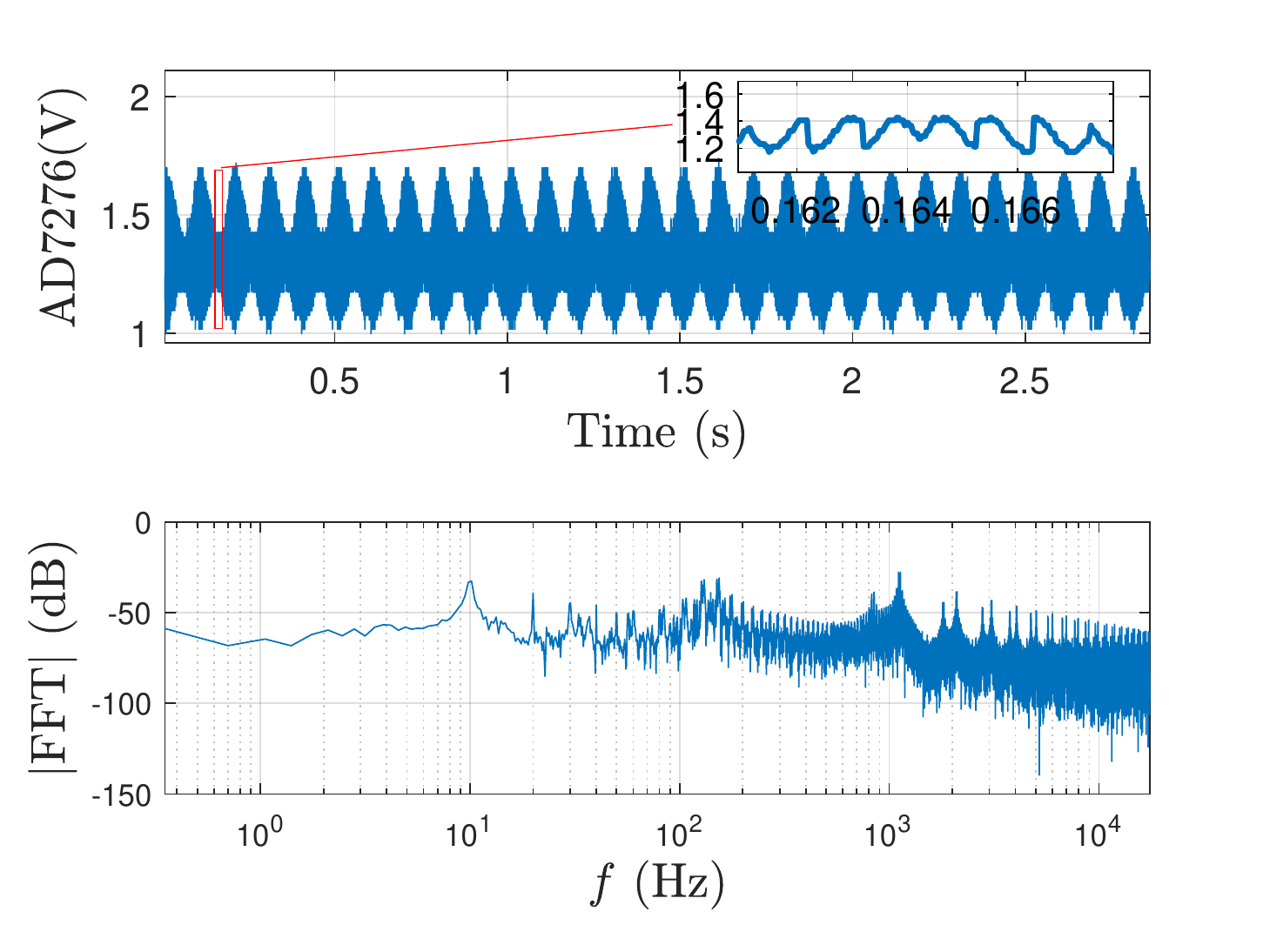}
      \caption{AD7276: $f_c=\SI{40}{\mega\hertz}$}
      \label{fig:app:ad7276_full}
    \end{subfigure}\hfill
    \caption{AD7822~(\subref{fig:app:ad7822_full}) and AD7276~(\subref{fig:app:ad7276_full})
    output for $P=\SI{-1}{\decibel m}$, $\mu=0.5$, and $f_m=\SI{10}{\hertz}$.}
  \end{figure}

\secpar{AD7822 \& AD7276} Figure~\ref{fig:app:ad7822_full} and Figure~\ref{fig:app:ad7276_full}
show example measurements for the AD7822 and the AD7276 ADCs respectively. As for
the Artix~7, high-frequency components dominate the output, and hence require
manual tuning to get a demodulated, low-frequency output.

\end{document}